\documentclass[12pt]{article}

\usepackage{epsfig}
\usepackage[height=10in, width=7in]{geometry}
\usepackage{graphics}
\usepackage{amsfonts}
\usepackage{amssymb}
\usepackage{amsmath}
\def\ri{\mathrm i}
\def\e{\mathrm e}
\def\C{\mathbb C}
\def\R{\mathbb R}
\newcommand{\be}{\begin{equation}}
\newcommand{\ee}{\end{equation}}
\newcommand{\bea}{\begin{eqnarray}}
\newcommand{\eea}{\end{eqnarray}}
\newcommand{\ba}{\begin{array}}
\newcommand{\ea}{\end{array}}
\newcommand{\rf}[1] {(\ref{#1})}
\newcommand{\half}{\mbox{$\frac{1}{2}$}}
\newcommand{\eps}{\epsilon}
\newcommand{\cn}{\mbox{cn}}
\newcommand{\sn}{\mbox{sn}}
\newcommand{\dn}{\mbox{dn}}
\newcommand{\sech}{\mbox{sech}}

\begin{document}

\title{On the stabilization of breather-type solutions of the  damped higher order nonlinear Schr\"odinger equation} 

\author{C.M. Schober and A. Islas\\
  Department of Mathematics\\
  University of Central Florida\\
  Orlando, FL 32816, USA}
\maketitle

\begin{abstract}
  Spatially periodic breather solutions (SPBs) of the nonlinear Schr\"o\-dinger  (NLS) equation are frequently used to model rogue waves and are typically unstable.
  In this paper  we study  the effects of dissipation and higher order nonlinearities 
  on the stabilization of both single and multi-mode SPBs in the framework of a damped higher order NLS (HONLS) equation.  We observe the onset of  novel  instabilities  associated with
 the development of critical  states which result from symmetry breaking  in the damped HONLS system.
 We broaden the Floquet characterization of  instabilities of solutions of  the NLS equation, using an even 3-phase solution of the NLS as an example,
 to show  instabilities are  associated with degenerate complex elements of both the
 periodic and continuous Floquet spectrum.
 As a result the
Floquet criteria for
the stabilization of a solution of the damped HONLS 
centers around
the elimination of all complex degenerate elements of the spectrum.
 For  an initial  SPB
 with a given  mode 
structure,  a perturbation analysis shows that for short time 
only  the  complex double points associated with resonant modes split under
the damped HONLS while those associated with  nonresonant modes remain
 effectively closed. The corresponding  damped HONLS 
  numerical
  experiments corroborate that
  instabilities associated with  nonresonant modes persist  on a longer time scale than the instabilities associated with resonant modes.
  \end{abstract}

\section{Introduction}
In one of his foundational studies, Stokes established the existence of
traveling nonlinear periodic wave trains in deep water \cite{stokes1847}.
The stability of these  waves  was resolved when
Benjamin and Feir proved  that in sufficiently deep water
the Stokes wave is modulationally unstable.
Small perturbations
of Stokes waves were found to lead to  exponential growth of the side bands \cite{bf67,z68}.
 More recently, modulational instability (MI) of the background state is considered to play a prominent role in the development of rogue  waves in oceanic sea states,
 nonlinear optics, and plasmas \cite{p83,dt99,kf10,kps09,orbma13,srkj07}.

 The
 nonlinear Schr\"odinger  (NLS) equation (when $\epsilon, \gamma = 0$ in
 Equation~\rf{dhonls}) is one of the simplest models for studying phenomena related to  MI; as such, special solutions 
 of the  NLS equation are regarded as prototypes of rogue waves. 
Amongst the more tractable ``rogue wave'' solutions of the NLS equation are  the rational solutions  (with
the Peregrine breather being the lowest order)
and the spatially periodic breathers
 (SPBs)  which are constructed as heteroclinic orbits of modulationally unstable
 Stokes waves \cite{oos00,cs02,asa09}.
 In the case of the Stokes waves with $N$ unstable modes (UMs), the associated  SPBs
 can be of dimension $M \le N$ and are referred to as $M$-mode SPBs; 
the single mode SPB
is  the Akhmediev breather \cite{aek87}.
 For more realistic sea states with non-uniform backgrounds,
 heteroclinic orbits of
unstable $N$-phase  solutions  have been used to describe rogue waves \cite{cs17,cp18}.

For theoretical and practical purposes it is important to understand the  stability of
the SPBs with respect to small variations in initial data and small
perturbations of the NLS
equation.
Using the squared eigenfunction connection between the Floquet spectrum of the NLS equation and the linear stability problem, the SPBs were shown to be typically unstable \cite{cs14}.
The effects of damping on deep water wave dynamics, even when weak, can be significant  and in many instances must  be included in  models to enable
accurate  predictions of laboratory and   field data  \cite{shch05,zc20,cha11,fetal19}.

In a recent study the authors examined the stabilization of 
symmetric SPBs using the linear damped NLS equation (a near-integrable system that  preserves even symmetry of solutions) \cite{si20}. 
The route to stability for these damped SPBs was determined  by appealing to
the Floquet spectral theory of the NLS equation.
Degenerate  complex elements of the periodic  spectrum (referred to as complex double points) are  associated with instabilities of the solution and
may split  under perturbation  to the system. 
In the restricted subspace of even solutions complex double points can
reform as time evolves. The damped solutions were found to be unstable
as long as complex double points were present in the spectral decomposition
of the data (either by persisting or reforming).
A key issue in analysing  the
route to stability  is determining which complex double points 
in the spectrum of the SPB are split by damping.
 For  an initial  SPB
 with a given  mode 
structure,  perturbation analysis showed that
only  the  complex double points associated with resonant modes split under damping while   those associated with  nonresonant modes remained
 effectively closed. The corresponding damped NLS 
  numerical
experiments corroborated the persistence  of
the nonresonant mode instabilities on a longer time scale than
resonant mode instabilities \cite{si20}.

In this paper we examine the competing effects of dissipation and higher order nonlinearities on the
routes to stability  of both single and multi-mode SPBs in the framework of  the linear damped  higher order 
NLS  (HONLS) equation over a spatially periodic domain:

\be
iu_t + u_{xx} + 2|u|^2 u 
+ i \eps\left(\half u_{xxx}
- 8|u|^2 u_x - 2ui\left[\mathcal{ H}\left(|u|^2\right)\right]_x\right) + i\gamma u = 0,
\label{dhonls}
\ee
where $u(x,t)$ is the complex envelope of the wave train, $\mathcal{ H}\{f(x)\} = \frac{1}{\pi}\int_{-\infty}^\infty \frac{f(\xi)}{x-\xi}d\xi$ is the Hilbert transform of $f$,
and $0 < \epsilon,  \gamma << 1$.
The initial data used in the numerical experiments is generated using  exact SPB solutions of the integrable NLS
equation. The SPBs are 
over Stokes
 waves with $N$ unstable modes  (referred to as the $N$-UM
regime) for $1 \le N \le 3$.
We interpret the damped HONLS (near-integrable) dynamics 
by appealing to the NLS Floquet spectral theory.

The higher order nonlinearities in Equation~\rf{dhonls} break the even symmetry of both the initial data and
the equation. This raises several interesting questions regarding the damped HONLS equation.  Which integrable instabilities are excited by the damped HONLS flow  and which  elements of the  Floquet spectrum are  associated with
these instabilities?  What are the routes to stability under  damping; i.e. what remnants of integrable NLS structures are detected in the damped HONLS evolution?

In the present study
we observe the onset of  novel  instabilities  as a result of symmetry breaking and
 the development of critical  states in the damped HONLS flow
which were
nonexistent in the previously examined damped NLS  system with even symmetry.
Significantly, we determine  these  
instabilities are associated with
complex degenerate elements of both the periodic  and continuous spectrum, i.e.
with both  complex ``double points'' and complex
``critical points'', respectively. This association was  not previously recognized. With regard to teminology, 
although double points are among the critical points of $\Delta$,
in this paper we exclusively call degenerate complex periodic spectrum
where $\Delta = \pm 2$ ``double points'' and reserve the term
``critical points'' for degenerate  complex spectrum where
$\Delta \neq \pm 2$.

The paper is organized as follows. In Section~2 we  present elements 
of  the NLS Floquet spectral theory which we use to distinguish instabilities in
the numerical experiments.
Whether  higher phase solutions,  such as the even 3-phase solution given in \rf{3phase},  are unstable with respect to
general  noneven perturbations and what the Floquet ``signature'' is of
the possible  instabilities, are open questions. The closest stability results
we are aware of are for the elliptic solutions of the focusing NLS equation \cite{ds17}.
We numerically assess  the stability of an even 3-phase solution of the NLS in order to develop  a broadened Floquet characterization of  instabilities of the NLS equation.

A brief overview of the SPB solutions of the NLS is provided at the end of Section 2 before numerically examining their stabilization under
the damped HONLS  flow in Section~3. 
The Floquet  decompositions of the numerical solutions 
are   computed for $0 \le t \le 100$. Complex double points are initially  present in the spectrum. If one of the complex double points present initially  splits due to the damped HONLS perturbation, the subsequent evolution involves repeated formation and splitting of complex critical points (not double points) which we correlate with the observed  instabilities. 
The Floquet spectral analysis is complemented by an  examination of the growth of small perturbations in the SPB initial data under the damped HONLS flow. We determine that the
instabilities  saturate and the solutions stabilize once all complex double points and 
complex critical points  vanish in the spectral decomposition of the perturbed flow. Variations in the spectrum 
under the HONLS flow are correlated with deformations of certain NLS solutions
to determine
the routes to stability for the damped HONLS SPBs. 

In Section~4, via perturbation analysis, we  examine  splitting of
the complex double points, present  in the SPB initial data, under the damped HONLS flow.
We  find that for short time, 
the double points
associated with  modes that resonate with the SPB structure
 split producing disjoint asymmetric bands,  while the 
complex double points associated with nonresonant modes remain effectively
closed,
 substantiating the spectral evolutions observed in the numerical experiments.
 The   nonresonant modes are observed to remain effectively closed for the duration of the experiments, even though the solution evolves as  a damped asymmetric multi-phase state. In this study resonances have a stabilizing effect; the 
instabilities of nonresonant modes persist  on a longer time scale than the instabilities associated with resonant modes.

\section{Analytical Framework}

\subsection{Floquet spectral characterization of instabilities}
The nonlinear Schr\"odinger equation (when $\epsilon, \gamma = 0$ in Equation \rf{dhonls}) arises as the solvability condition of the Zakharov-Shabat (Z-S) pair of linear systems \cite{zs72}:
  \bea
\mathcal{L}(u) \phi& = \lambda \phi,\qquad\mathcal{L}(u)=\begin{pmatrix} \ri{\partial_x} & u \\ -u^* & -\ri{\partial_x} \end{pmatrix},
\label{Lax}
 \\
  \phi_t & = \begin{pmatrix} -2\ri \lambda^2 +\ri |u|^2 & 2\ri \lambda u -u_x \\ 2 \ri \lambda u^* +u^*_x & 2\ri \lambda^2 -\ri |u|^2 \end{pmatrix} \phi,
\eea
where $\lambda$ is the spectral parameter, $\phi$ is a complex vector valued eigenfunction, and $u(x,t)$ is a solution of the NLS equation itself.
Associated with an $L-$periodic NLS
solution is it's Floquet spectrum
\be
\sigma(u) 
:= \left\{\lambda \in \C \, | \, \mathcal{L} \phi = \lambda \phi, |\phi|
\mbox{ bounded } \forall x\right\}.				
\ee

Given  a fundamental matrix solution of the Z-S system, $\Phi$,
one defines the Floquet discriminant as the trace
of the transfer matrix across one period $L$,
\(
\Delta(u,\lambda) = \mbox{Trace}\left[\Phi(x+L,t;\lambda) \Phi^{-1}(x,t;\lambda)\right]\).
The Floquet spectrum has an explicit representation in terms of the
 discriminant:
\be
\sigma(u) := \left\{ \lambda \in \C \, | \, \Delta(u,\lambda)\in\R,
 -2 \leq \Delta(u,\lambda) \leq 2 \right\}.
\ee
The Floquet discriminant  $\Delta(\lambda)$ is analytic and is a conserved functional of the NLS equation. As such, the spectrum 
$\sigma(u)$ of an NLS solution is invariant under the time evolution.

The spectrum consists of the entire real axis and curves or ``bands of spectrum'' in the complex $\lambda$ plane
($\mathcal{L}(u)$ is not self-adjoint). The periodic/antiperiodic points (abbreviated here as  periodic points) of the Floquet spectrum are
those at which $\Delta = \pm 2$. The endpoints of the bands of spectrum are
given by  the  simple points of the periodic spectrum 
  $\sigma^s(u) = \{\lambda_j^s\,|\, \Delta(\lambda_j) = \pm 2, \partial \Delta/\partial \lambda \neq 0\}$.
Located within the bands of spectrum  are two  important spectral elements:
\begin{enumerate}
\item Critical points of spectrum, $\lambda_j^c$, determined by the condition
$\partial \Delta/\partial \lambda = 0$. 
\item Double points of periodic spectrum
$\sigma^d(u) = \{\lambda_j^d \,|\,
\Delta(\lambda_j^d) = \pm 2, \partial \Delta/\partial \lambda = 0,\;
\partial^2 \Delta/\partial \lambda^2 \neq 0\}$.
\label{critical}
\end{enumerate}
Double points are among the critical points of $\Delta$.
However, in this paper, we exclusively call the degenerate periodic spectrum
where $\Delta = \pm 2$ ``double points'' and reserve the term
``critical points'' for degenerate elements of the spectrum where
$\Delta \neq \pm 2$.

The Floquet spectrum can be used to
represent a solution  in terms of a set of nonlinear modes
where the structure and stability of the modes are  determined by the
band-gap structure of the spectrum.
Simple periodic points  are associated with  
  stable active modes. 
The location of the double points is particularly important. Real double points correspond to zero amplitude inactive nonlinear modes. On the other hand, complex double points are associated with  degenerate, potentially  unstable, nonlinear modes with either positive or zero growth rate. 
When restricted to the subpace of even solutions, exponential instabilities of a
solution are  associated with complex double points in the spectrum \cite{efm90}.

A concrete example illustrating the correspondence between complex double points in the spectrum and linear instabilities  is the Stokes wave 
solution $u_a(t)=a \e^{\ri (2 a^2 t+\phi)}$.
For  small perturbations of the form 
$u(x,t) = u_a(t)(1 + \epsilon(x,t))$, $|\epsilon| <<1$, one finds 
$\epsilon$ satisfies the linearized NLS equation
\be
\ri\epsilon_{t} + \epsilon_{xx}  + 2|a|^2(\epsilon + \epsilon^*) = 0.
\label{lineqn}
\ee
Representing $\epsilon$ as a  Fourier series with modes $\epsilon_j \propto \e^{\ri\mu_j x + \sigma_j t}$, $\mu_j = 2\pi j/L$, gives 
$\sigma_j^2 = \mu_j^2\left(4|a|^2 - \mu_j^2 \right)$. As a result,
the $j$-th mode is unstable if
$0 < (j\pi/L)^2 < |a|^2$. The number of UMs is the largest integer M such that
$0 < M < |a|L/\pi$.

\begin{figure}[ht!]
\centerline{\includegraphics[width=.33\textwidth]{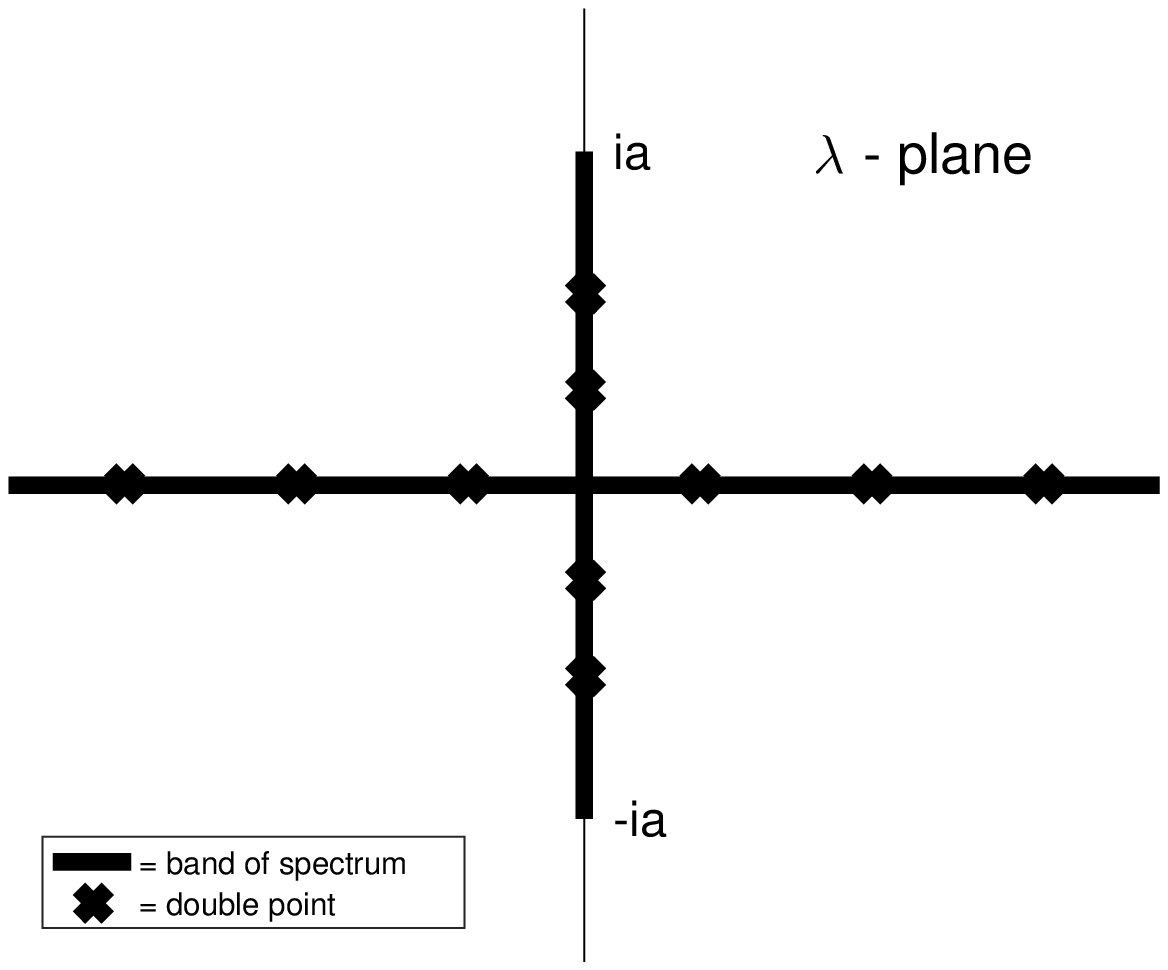}}
\caption{The Floquet spectrum of the Stokes wave with $\displaystyle \lfloor {aL}/{\pi} \rfloor=2.$}
\label{Figure1}
\end{figure}

The Floquet discriminant for the Stokes wave is  $\Delta = 2 \cos(\sqrt{a^2 + \lambda^2} L)$.
The Floquet spectrum consists of continuous bands $\R \bigcup$ $ [-\ri a,\ri a]$ and a discrete part containing $\lambda_0^s =  \pm \ri |a|$ 
and the infinte number of  double points 
\be
(\lambda_j^d)^2 = \left(\frac{j\pi}{L}\right)^2 - a^2,\quad j\in\mathbb{Z}, \quad j \ne 0,
\label{imdps}
\ee
as shown in Figure~\ref{Figure1}. Note that the condition for  $\lambda_j^d$ to be  complex is
precisely the condition  for the
$j$-th Fourier mode  $\epsilon_j$ to be unstable. The remaining
$ \lambda_j^d$ for $|j| >M$ are real double points. 

As  the NLS spectrum is symmetric
under complex conjugation, we subsequently 
only display the spectrum  in the upper half $\lambda$-plane.

{\sl Novel instabilities of noneven solutions:}
Earlier work on perturbations of the NLS equation dealt primarily with solutions with even symmetry
whose instabilities were identified solely via complex double points.
In general, imposing symmetry on a solution restricts it's dynamical behavior and may suppress instabilities.
In the current damped HONLS experiments we find 
instabilities  arise due to the asymmetry of the system  that are  not captured by complex double points.
Although complex double points, if present in the damped HONLS flow,  still identify instabilities, we need to develop a broader  Floquet spectral
characterization of  instabilities to capture all the instabilities of
noneven solutions.

A clue as to which new  spectral elements  are associated with instabilities in the full solution space of the NLS equation is provided by considering the following even 3-phase solution of the  NLS  \cite{aek87},
 \be 
  u_0(x,t) = a\e^{2\ri a^2 t}\frac{\sqrt{\frac{\kappa}{1+\kappa}}\,
    \cn\left(\frac{ax}{\sqrt\kappa},\sqrt{\frac{1-\kappa}{2}}\right)\,
    \dn\left(\frac{a^2t}{\kappa},\kappa\right) +
    i\kappa\,\sn\left(\frac{a^2t}{\kappa},\kappa\right)}
  {\sqrt 2\kappa\left[1 -
    \sqrt{\frac{\kappa}{1+\kappa}}\,
    \cn\left(\frac{ax}{\sqrt\kappa},\sqrt{\frac{1-\kappa}{2}}\right)
    \cn\left(\frac{a^2t}{\kappa},\kappa\right)\right]}.
  \label{3phase}
\ee
With respect to $t$ the solution has a double frequency; a frequency determined by the exponential function and a modulation frequency determined by the elliptic functions.
Formula \rf{3phase} describes an even  standing wave, periodic in space and time,
arising as the degeneration of a 3-phase solution due to symmetry in
it's spectrum.
The spatial period $L $ and temporal period $T$ are functions of
$\mathcal{ K}_x(\sqrt{\frac{1 - \kappa}{2}})$ and 
  $\mathcal{ K}_t(\kappa)$, respectively,  where $\mathcal{ K}$ is the complete elliptic integral of
the first kind.
 As $\kappa \rightarrow 1$ in \rf{3phase}, $T \rightarrow \infty$ and $u_0(x,t) \rightarrow U^{(1)}(x,t)$, the SPB given in Equation~\rf{SPB1} associated with one complex double point.
\begin{figure}[ht!]
  \centerline{
\includegraphics[width=.33\textwidth]{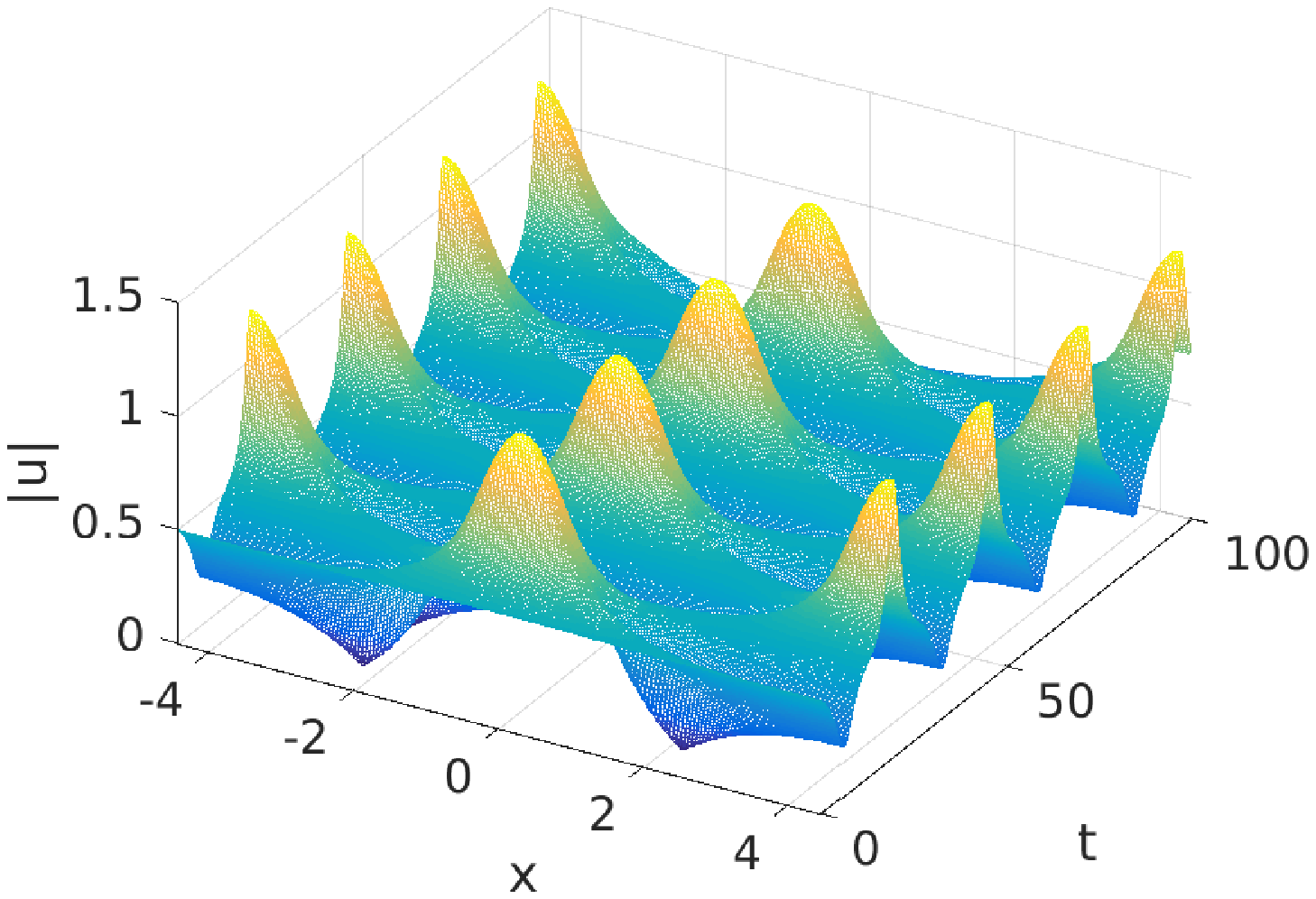}
\includegraphics[width=.33\textwidth]{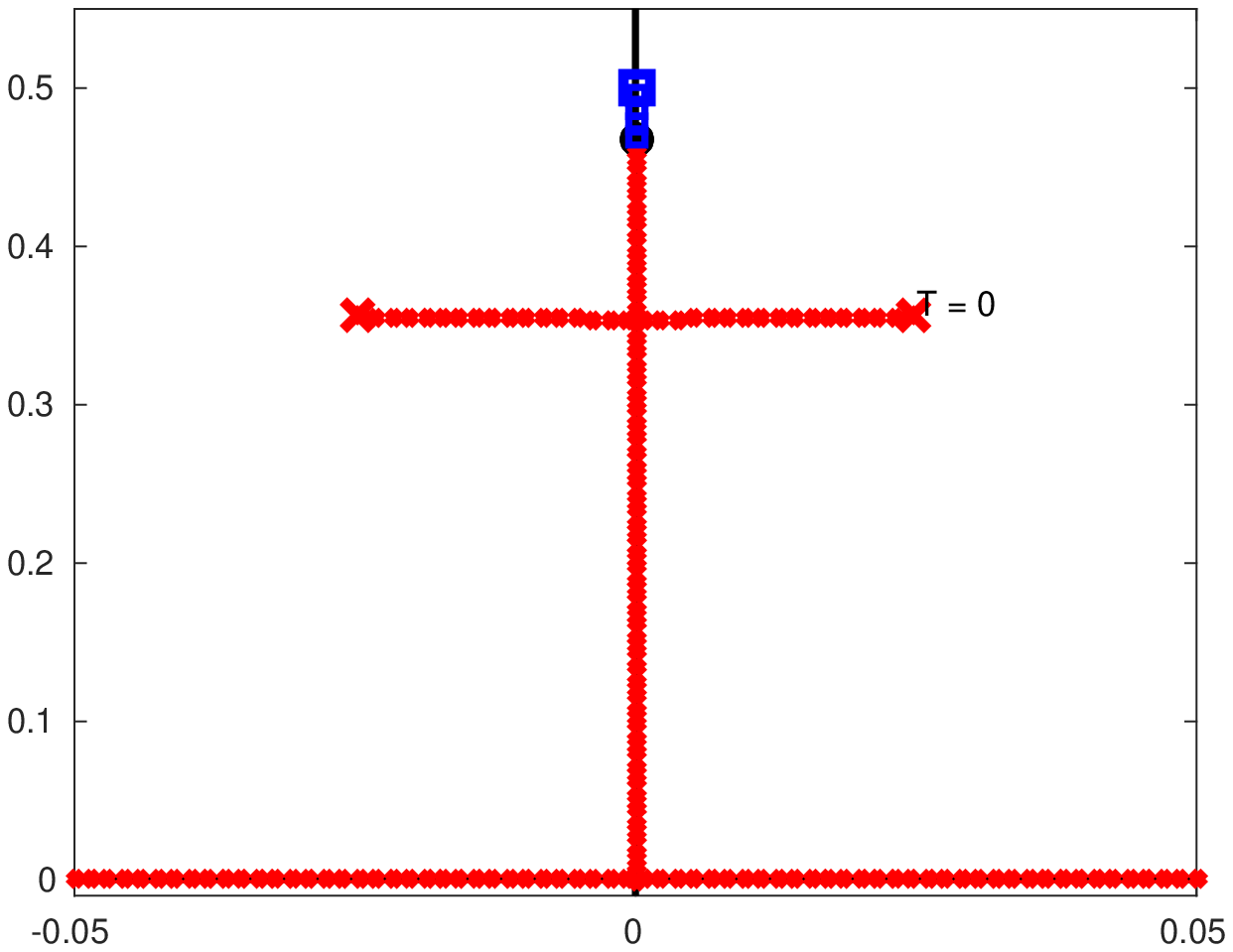}
\includegraphics[width=0.33\textwidth]{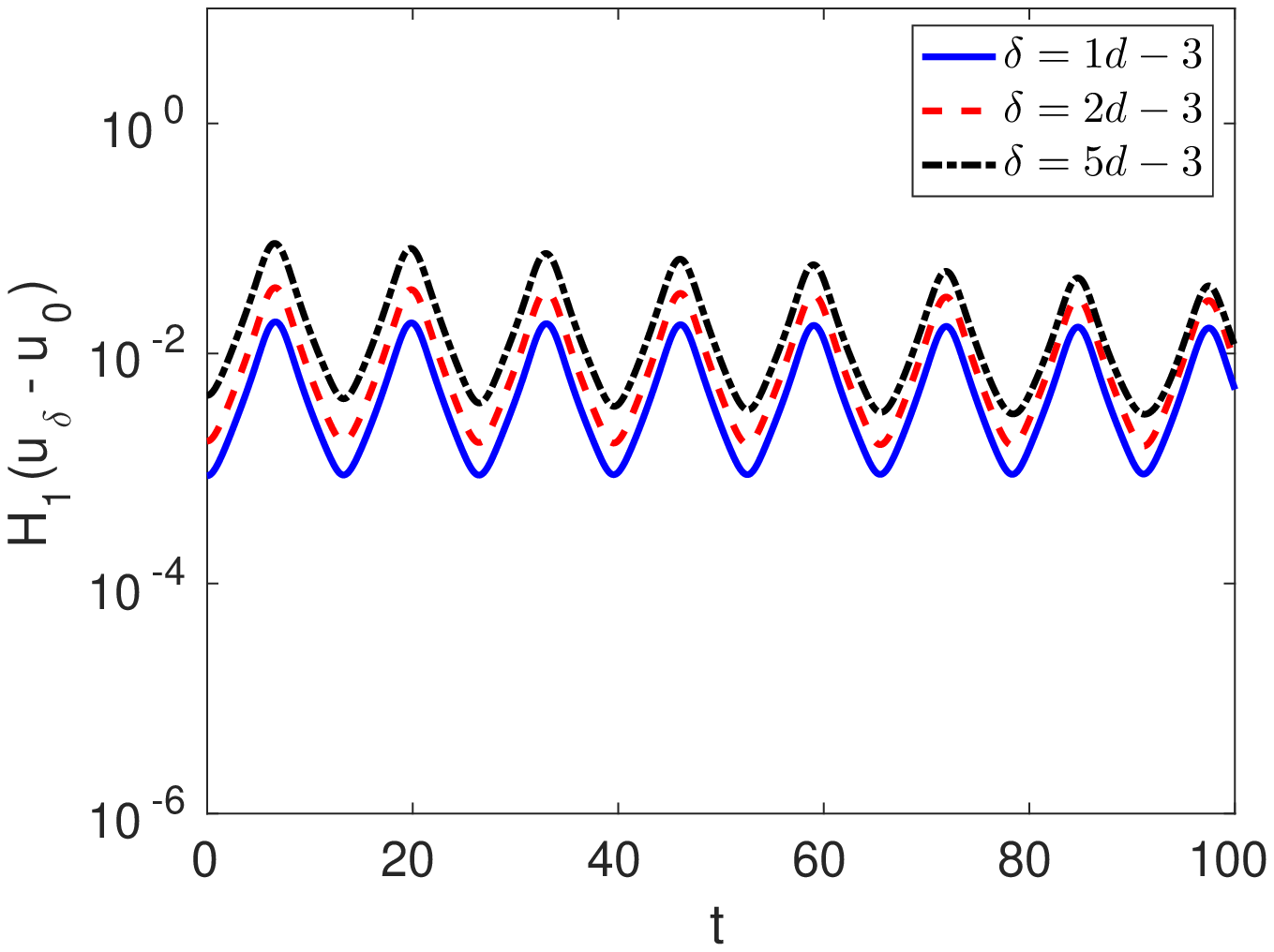}
}
  \caption{NLS cross state  a)  $|u_0(x,t)|$ for $0 \le t \le 100$, b) it's  spectrum at $t=0$, c) $\eta(t)$ for $u_{\delta}(x,0) = u_0(x,0) + \delta g_1(x)$.}
\label{even_cross}
\end{figure}

The surface  and  Floquet spectrum for $u_0(x,t)$ are shown in 
Figures~\ref{even_cross}A-B.
The spectrum forms an even  ``cross'' state  with two  bands of spectrum in the
upper half plane with endpoints given by the simple periodic spectrum $\lambda_0 = .5i$ and $\lambda_1^{\pm} = .35 i \pm \alpha$.
These two  bands intersect transversally at $\lambda^c$ on the
imaginary axis.
Since 
$\partial \Delta/\partial \lambda = 0$ at transverse intersections of bands of continuous spectrum,  $\lambda^c$ is a critical point.
There are no complex double points in $\sigma (u_0(x,t))$.

\begin{figure}[ht!]
  \centerline{
\includegraphics[width=.33\textwidth]{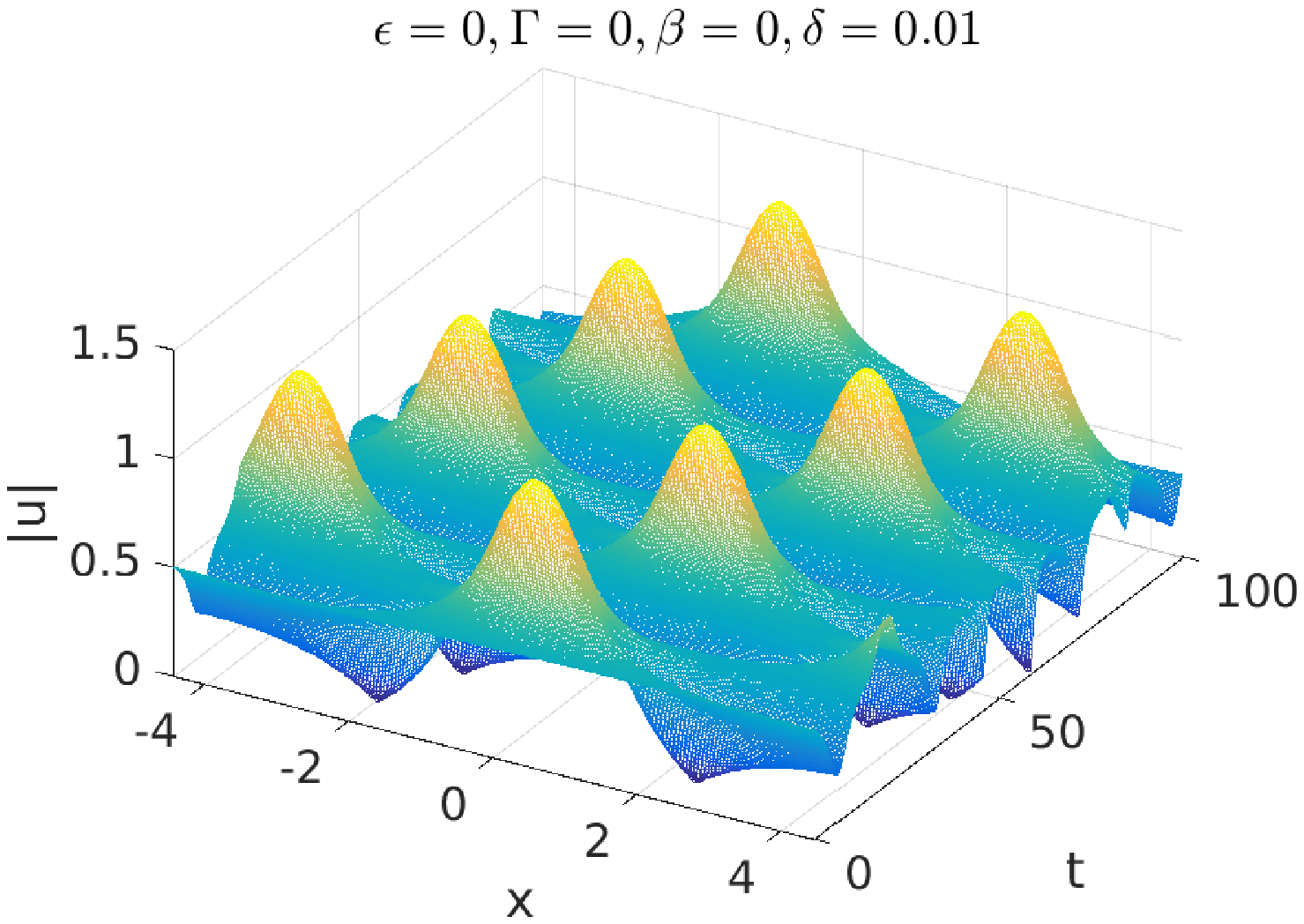}
\includegraphics[width=.33\textwidth]{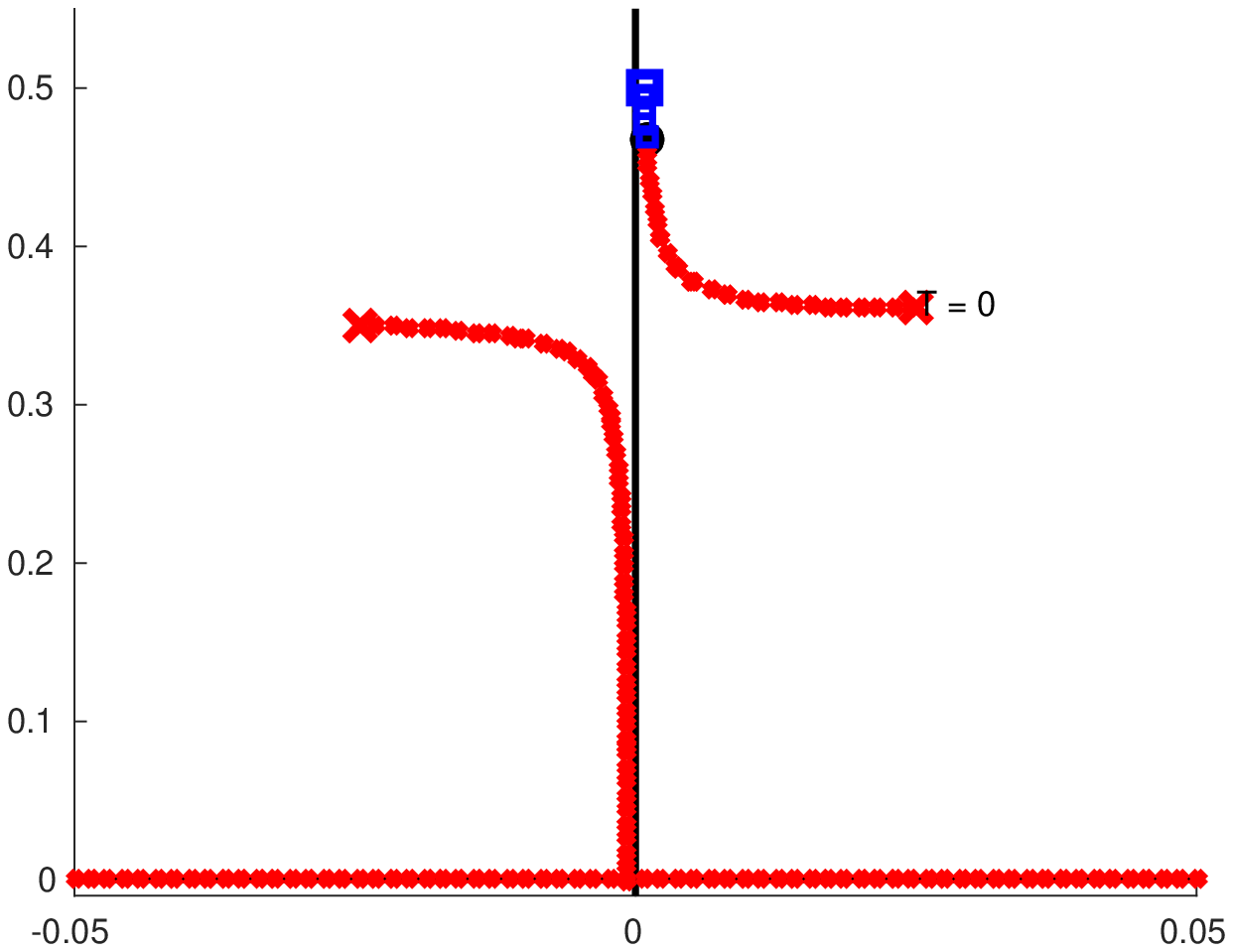}
\includegraphics[width=0.33\textwidth]{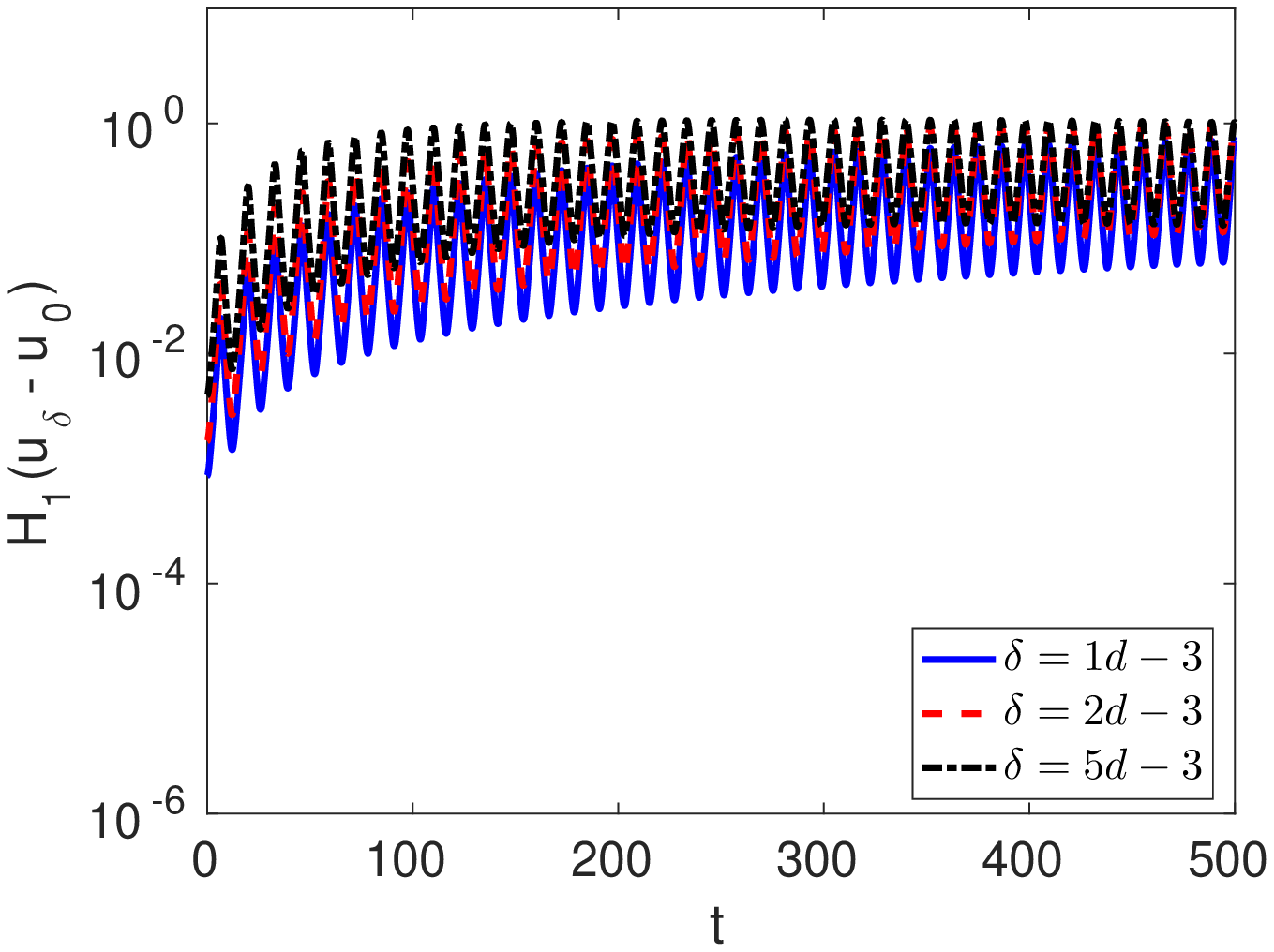}
}
  \caption{NLS noneven 3-phase solution a)  $|u_{\delta}(x,t)|$ for $0 \le t \le 100$, b) it's  spectrum at $t=0$, c) $\eta(t)$ for $u_{\delta}(x,0) = u_0(x,0)
    + \delta g_2(x)$.}
\label{noneven_cross}
\end{figure}

To  numerically address  the stability  of the cross state \rf{3phase} with respect to initial data we
consider small perturbations (both
 symmetric and asymmetric) of the following form:
\[u_{\delta}(x,0) = u_0(x,0) + \delta g_k(x),  \quad  k = 1,2  \] where
$g_1(x) = e^{i\phi} \cos \mu x$, $g_2(x) = e^{i\phi_1} \cos \mu x +  r e^{i\phi_2} \sin \mu x$, and $\delta = 10^{-3},\dots,  5 \times 10^{-3}$.
We examine i) the Floquet spectrum of $u_{\delta}(x,t)$  as compared with  $u_0(x,t)$  and ii) the  growth of
the perturbations as $\delta$ is varied. 
We consider a solution $u(x,t)$ of the NLS equation to be  stable if for every $\epsilon>0$ there exists a $\delta>0$ such that if
  $||u_{\delta}(x,0) - u(x,0)||_{H^1} < \delta$, then
$||u_{\delta}(x,t) - u(x,t)||_{H^1} < \epsilon$, for all $t$.
 Therefore, to determine whether $u$ and $u_{\delta}$ stay close as time evolves, we monitor
    the evolution of the difference
    \be
      \mathcal{ \eta}(t) = ||u_{\delta}(x,t) - u(x,t)||_{H^1}
      \label{H1Diff}
 \ee
where
$||f||^2_{H^1} = \int_{-L/2}^{L/2} \left(|f_x|^2 + |f|^2\right)\,dx.$

{\sl i) Symmetric perturbations of initial data:} 
As $\phi$ and $\delta$ are varied, the  surface and spectrum for $u_{\delta}(x,t)$ for perturbation $g_1(x)$ is qualitatively the same as in  Figures~\ref{even_cross}A-B.
 The endpoints of the band of spectrum,   $\sigma^s(u_{\delta}) $,  are
 slightly shifted maintaining even symmetry. Due to analyticity of $\Delta$,
 $\lambda^c$ does not split under even perturbations and
the  spectrum is not   topologically different.
Figure~\ref{even_cross}C shows the evolution of $\eta(t)$
for  even perturbations 
$u_{\delta}(x,0) = u_0(x,0) + \delta g_1(x)$.
The small osciallations in   $\eta(t)$, typical in Hamiltonian systems,
do not grow. The Floquet spectrum and the evolution of $\eta(t)$ show  that when
 restricted to the  subspace of even solutions, $u_0(x,t)$  is stable.
 
 {\sl ii) Asymmetric perturbations of initial data:}  The  surface and
 spectrum of  $u_{\delta}(x,t)$ for
 $u_{\delta}(x,0) = u_0(x,0) +  0.05 \sin \mu x$ are shown in
Figures~\ref{noneven_cross}A-B. A topologically different spectral configuration is obtained and the waveform is a modulated right traveling wave. 
The critical point  $\lambda^c$  has split into $\lambda_{\pm}^c  $ and the two disjoint bands of spectrum form a ``right'' state:  the upper band with endpoints  $\lambda_0^s $ and $\lambda_{1,\delta}^+$ in the right quadrant and the lower band with endpoint $\lambda_{1,\delta}^- $ extending to the real axis in the left quadrant.
   On the other hand, if e.g. 
   $u_{\delta}(x,0) = u_0(x,0) + 0.05 (\cos \mu x + e^{i\pi /3} \sin \mu x)$,
   the waveform is a modulated left traveling wave.
   In this case the orientation of the bands of spectrum is reversed,
   forming a ``left'' state with the upper band in the left  quadrant and the
   lower band in the right quadrant. As the parameters in $g_2(x)$  are varied these are the two possible noneven
   spectral configurations for $u_{\delta}(x,t)$.
   The perturbation analysis in Section~4 is related and shows noneven perturbations of the SPB  split complex double points into left and right sates.
   
Figure~\ref{noneven_cross}C shows  $\eta(t)$ grows to $\mathcal{ O}(1)$ for
   asymmetric perturbations
   $u_{\delta}(x,0) = u_0(x,0) + \delta g_2(x)$.
   Clearly  $u_{\delta}(x,t)$ does not remain close to $u_0(x,t)$
   for these perturbations.
 We associate the  transverse complex critical points with 
 instabilities arising from symmetry breaking which are not excited  when evenness is imposed.
The exact nature of the instability associated with complex critical points is under investigation.
 The current damped HONLS experiments in Section 3  corroborate the significance of
 transverse critical points $\lambda^c$ in identifying instabilities in the
 unrestricted solution space.

\subsection{Spatially periodic breather solutions of the NLS equation}
\begin{figure}[ht!]
  \centerline{
\includegraphics[width=.33\textwidth]{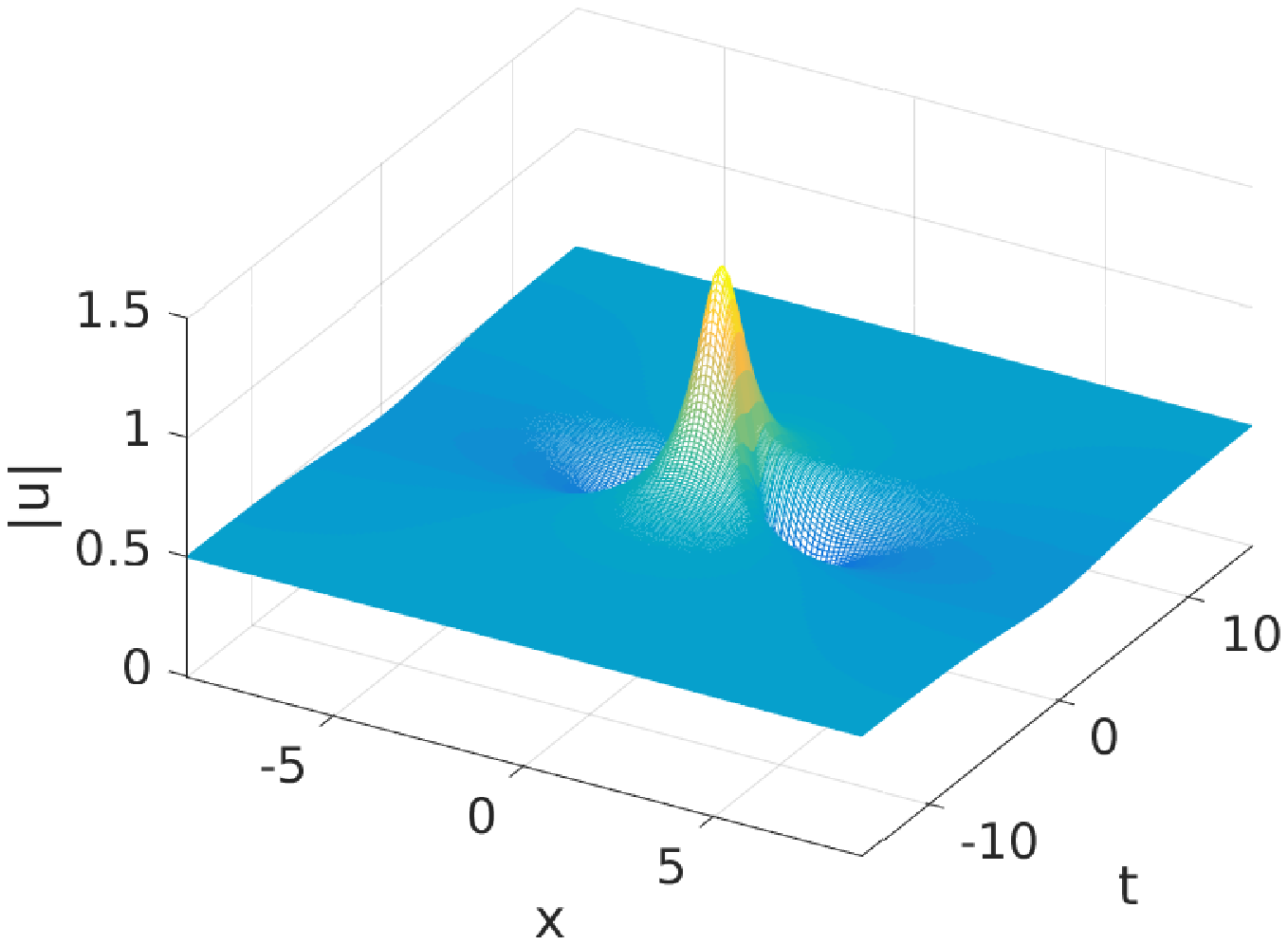}
\includegraphics[width=.33\textwidth]{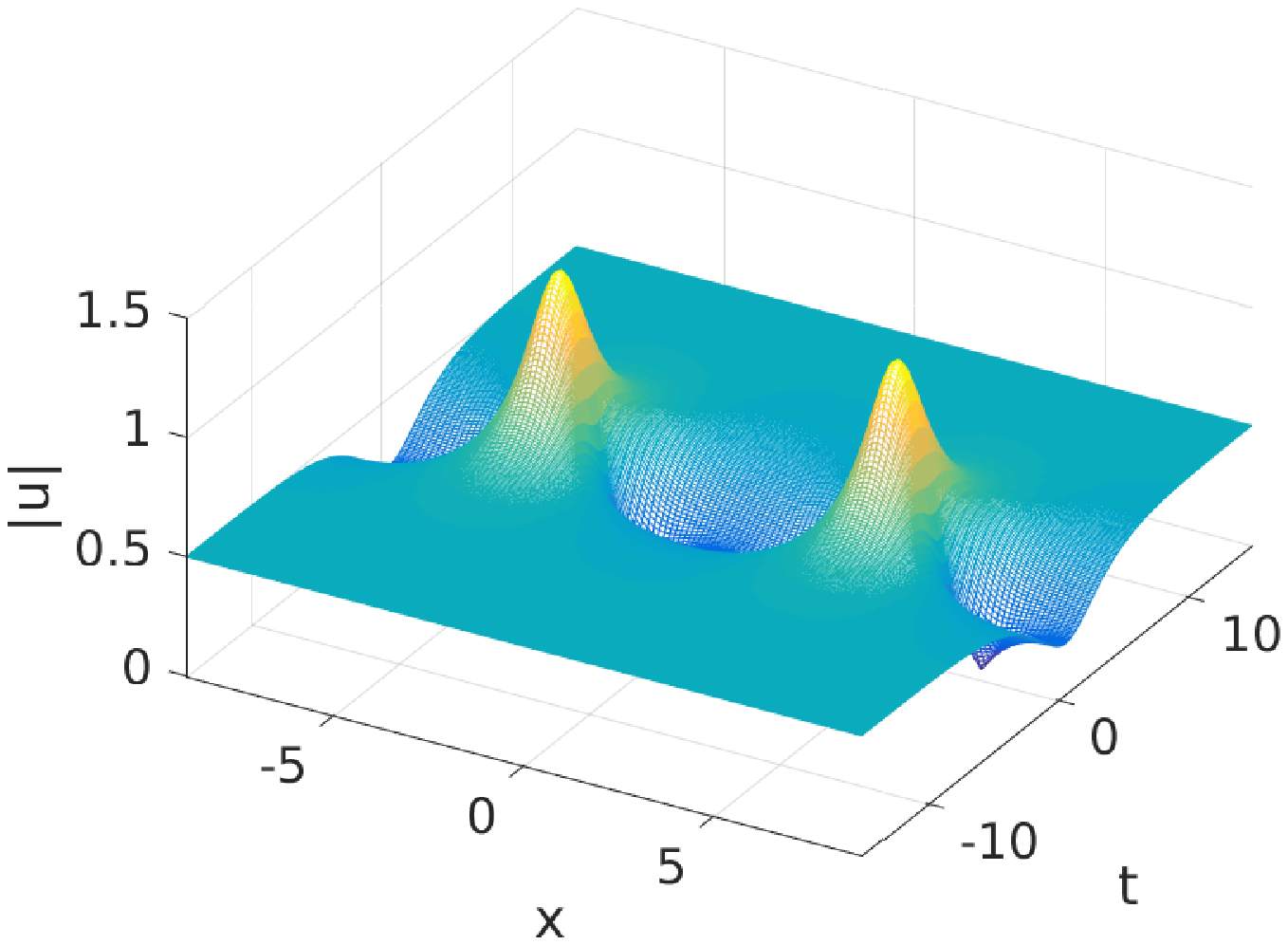}
\includegraphics[width=0.33\textwidth]{schober_fig4c}}
  \caption{SPBs over a Stokes wave with $N = 2$ UMs, $a = .5$,
    $L = 4\sqrt{2} \pi$: Amplitudes of (a-b) single mode SPBs $|U^{(1)}(x,t,\rho)|$ and $|U^{(2)}(x,t,\rho)|$ ($\rho = \beta = 0$) and (c) 2-mode SPB
  $|U^{(1,2)}(\rho,\tau)|$.}
\label{Figure_SPB}
\end{figure}

The heteroclinic orbits of a  spatially periodic unstable NLS potential $u(x,t)$ with complex double points, $\lambda^d_j$, in its Floquet spectrum can be derived using the
B\"acklund-gauge transformation (BT)  for the NLS equation \cite{sz87}.
Given a Stokes wave $u_a(t)$ with $N$ complex double points,
a single BT of $u_a(t)$ at  $\lambda^d_j$
 yields the one mode SPB, $U^{(j)}(x,t)$, associated with the $j$-th UM, $1 \le j \le N$.
Introducing $\mu_j =  2\pi j/L$, $\sigma_j = -2\ri \mu_j\lambda_j$, $\cos p_j = \mu_j/2a$,
and $\tau_j = (\rho - \sigma_j t)$ one obtains \cite{aek87,mo95}:
\be
U^{(j)}(x,t) = 
 a\e^{2\ri a^2t}\left( \frac{\ri\sin 2p_j\tanh\,\tau_j+\cos 2p_j-\sin p_j\cos\left(\mu_jx+\beta\right)\sech \tau_j}{1+\sin p_j\cos\left(\mu_jx+\beta\right)\mbox{ sech }\tau_j} \right).\label{SPB1}
\ee
 $U^{(j)}(x,t)$  exponentially approaches a phase shift of the Stokes wave, 
$\displaystyle\lim_{t \rightarrow \pm\infty} U^{(j)}(x,t) = ae^{2\ri a^2t + \alpha_{\pm}}$,
at a rate depending on $\lambda^d_j$.
Figures~\ref{Figure_SPB}A-B show the amplitudes of two distinct
single mode SPBs, $U^{(1)}(x,t)$ and $U^{(2)}(x,t)$   over a Stokes waves with $N = 2$ UMs.
 $U^{(1)}(x,t)$ and $U^{(2)}(x,t)$ are both unstable as the  BT based
at $\lambda_j$ saturates the instability of the
$j$-th UM while the other instabilities of the background  persist.

When the Stokes wave  posesses two or more UMs, the BT can be iterated to obtain  multi mode SPBs. For example,
the two-mode SPB with wavenumbers $\mu_i$ and $\mu_j$, obtained by applying the BT successively at complex $\lambda_i^d$ and $\lambda_j^d$, is of the following form:
\be
  U^{(i,j)}(x,t;\rho,\tau) =
a\e^{2\ri a^2t} \frac{N(x,t;\rho,\tau)}{D(x,t;\rho,\tau)}.
\label{SPB2}
\ee
The exact formula is provided in  \cite{cs02}. The parameters $\rho$ and $\tau$ determine the time
at which the first and second modes become excited, respectively.
Figure~\ref{Figure_SPB}C shows the amplitude of the ``coalesced'' two mode  SPB
$U^{(1,2)}(x,t;\rho,\tau)$ ($\rho = -2$, $\tau = -3$) over a Stokes waves with $N = 2$ UMs where 
the  two modes are excited simultaneously.

{\sl An important property of the B\"acklund transformation:}  The  BT is isospectral,
i.e.  $\sigma(u_a(t)) = \sigma(U^{(j)}(x,t)) = \sigma(U^{(i,j)}(x,t))$.
For example, the Stokes wave with $N = 2$ UMs (given in Figure~\ref{Figure1}) and 
each of the SPBs shown in  Figures~\ref{Figure_SPB}A-B-C share the same Floquet spectrum.

\section{Numerical investigation of  routes to stability of SPBs in the damped higher order NLS equation}

In our examination of the even 3-phase solution in Section 2 we find 
that when evenness is relaxed  novel instabilities arise
which  are  associated with complex critical points.
Armed with this result, in this section 
we return to the questions posed at the outset of our study:
i) Which  integrable instabilities are excited by the damped HONLS flow and
what is the Floquet criteria for their  saturation?
ii) What remnants of integrable NLS structures are detected in the damped HONLS evolution? 

The notation used in this section is as follows: 
i) The ``$N$ UM
regime'' refers to the range of parameters $a$ and $L$  for which 
the underlying Stokes wave  initially has $N$ unstable modes.
ii) The initial data used in the numerical experiments is generated using  exact SPB solutions of the integrable NLS equation. 
The perturbed SPBs are indicated with subscripts: $U^{(j)}_{\epsilon,\gamma}(x,t)$  refers to the  solution of
the damped HONLS  Equation~\rf{dhonls}  for one-mode SPB initial data  $ U^{(j)}(x,0) $.
Likewise $U^{(i,j)}_{\epsilon,\gamma}(x,t)$
is the solution to (\ref{dhonls}) for iterated SPB initial data.

The damped HONLS equation  is solved numerically 
using a high-order spectral method due to Trefethen \cite{t2000}.
The integrator uses a Fourier-mode decomposition  in space
with a fourth-order Runge-Kutta discretization in time.
The number of Fourier modes and the time step used depends on the complexity of the solution. For example, for initial data in the three UM regime,
$N = 1024$  Fourier modes are used with time step $\Delta t = 7.5 \times 10^{-5}$.  As a benchmark the first three global invariants of the HONLS equation, the energy $ E = \int_0^L|u|^2\,dx $, momentum $ P = i\int_0^L\left(u^* u_x - uu_x^*\right)dx$, and 
Hamiltonian
\[H =  \int_0^L \left\{ -i | u_x |^2 + i | u |^4 - \frac{\epsilon}{4}
\left( u_x u_{xx}^* - u_x^* u_{xx} \right) 
+ 2 \epsilon | u |^2\left(u^* u_x - u u_x^*\right) + i\epsilon |u|^2
\left[\mathcal{ H} \left( | u |^2 \right) \right]_x\right\}\,dx
\]
are preserved 
with an accuracy of $\mathcal{ O} (10^{-12})$ for $0 \le t \le 100$.  
The  invariant for the damped HONLS system, the spectral center $k_m = -P/2E$,  is preserved  with an accuracy of at least $\mathcal{ O} (10^{-12})$
in the  experiments.

{\sl Nonlinear mode decomposition of the damped HONLS flow:}
At each time $t$  we compute the spectral decomposition of the damped HONLS data using the numerical procedure developed by Overman et. al. \cite{omb86}.  After solving 
system \rf{Lax}, the  discriminant $\Delta$ is constructed.
The zeros of $\Delta \pm 2$ are determined using a root solver 
based on M\"uller's method and then the curves of spectrum filled in.
The spectrum is calculated with an accuracy of ${\mathcal O}(10^{-6})$
which is sufficient given the perturbation parameters $\epsilon$ and $\gamma$
used in the numerical experiments are ${\mathcal O}(10^{-2})$.

{\sl Notation used in the spectral plots:} The periodic spectrum is indicated with
a large $\times$ when $\Delta = -2$ and a large box when $\Delta = 2$.
The continuous spectrum is indicated with small $\times$ when the $\Delta$ is
negative and a small box when $\Delta$ is positive.

{\sl Interpreting the damped HONLS flow via the NLS spectral theory:}
A tractable example which illustrates the use of the Floquet spectrum to interpret near integrable dynamics is the spatially  uniform solution (there is no depenence on $\epsilon$) of the damped HONLS Equation~\rf{dhonls}, 
\begin{align}
  u_{a,\gamma}(t) = a \e^{-\gamma t} \e^{\ri (|a|^2\frac{(1 - \e^{-2\gamma t})}{\gamma})}.
  \label{pwdhonls}
\end{align}
At a given time $t = t_*$ the nonlinear spectral decomposition of \rf{pwdhonls} can be explicitly determined
by substituting $  u_{a,\gamma}(t_*)$  into
$\mathcal{L}(u) \phi = \lambda \phi$.
We find the periodic Floquet spectrum   
consists of  
$\lambda_0^s =   \pm \ri a \e^{-\gamma t_*}$
and infinitely many double points 
\be
(\lambda_j^d)^2 = \left(\frac{j\pi}{L}\right)^2 - \e^{-2 \gamma t_*} a^2,
\quad j\in\mathbb{Z}, \quad j \ne 0,
\label{imdps1}
\ee
where $\lambda_j^d$ is complex if $\left(\frac{j\pi}{L}\right)^2 <\e^{-2 \gamma t_*} a^2$.
Under the damped HONLS evolution the endpoint of the band of spectrum
$\lambda^s_0$  and  the complex double points $\lambda^d_j$ move down the imaginary
axis and then onto the real axis.
Similarly, at  $t = t_*$, a linearized stability analysis about
$u_{a,\gamma}(t_*)$
shows the growth rate of the $j$-th mode
$\sigma_j^2 = \mu_j^2\left(4\e^{-2\gamma t_*}a^2 - \mu_j^2 \right)$.
Thus the 
number of complex double points and the number of unstable modes
diminishes in time due to damping.
As a result, $u_{a,\gamma}(t)$
stabilizes when the growth rate $\sigma_1 = 0$, i.e. when
$\lambda_1^d = 0$, giving
$t_* = \frac{\ln(aL/\pi)}{\gamma}$.

{\sl Saturation time of the instabilities:} Since the association of complex critical points with instabilities is a new result, we supplement the
spectral analysis with an examination of the saturation time of the instabilities for the damped SPBs 
$U^{(j)}_{\epsilon,\gamma}(x,t)$ and $U^{(j,k)}_{\epsilon,\gamma}(x,t)$
as follows:  we  examine the growth
of small asymmetric perturbations in
the initial data of the following form,
\[U^{(j)}_{\epsilon,\gamma,\delta}(x,0) = U^{(j)}(x,0) + \delta f_k(x)\] and
\[U^{(i,j)}_{\epsilon,\gamma,\delta}(x,0) = U^{(i,j)}(x,0) + \delta f_k(x)\] where

{\bf i.}  $f_k(x) = \cos \mu_k x + r_k e^{i\phi_k} \sin \mu_k x$, $\mu_k = 2\pi k/L$, $1 \le k \le 3$, 

{\bf ii.} $f_4(x) = r(x)$, $r(x) \in [0,1]$ is random noise.

To determine the closeness of  $ U^{(j)}_{\epsilon,\gamma}(x,t)$ and 
 $U^{(j)}_{\epsilon,\gamma,\delta}(x,t)$ as time evolves we monitor 
the evolution of $\eta(t)$, as given by Equation~\rf{H1Diff}.
We consider the solution to have stabilized under the damped HONLS flow once $\eta(t)$ saturates.

In the damped HONLS numerical experiments
we obtain a new criteria for the saturation of
instabilities:   $\eta(t)$ saturates and  the SPB stabilizes once damping eliminates {\sl all}  complex double points and complex critical points in the spectrum.

\subsection{Damped HONLS SPB in the one unstable mode regime}
\begin{figure}[ht!]
  \centerline{
    \includegraphics[width=.5\textwidth,height=1.5in]{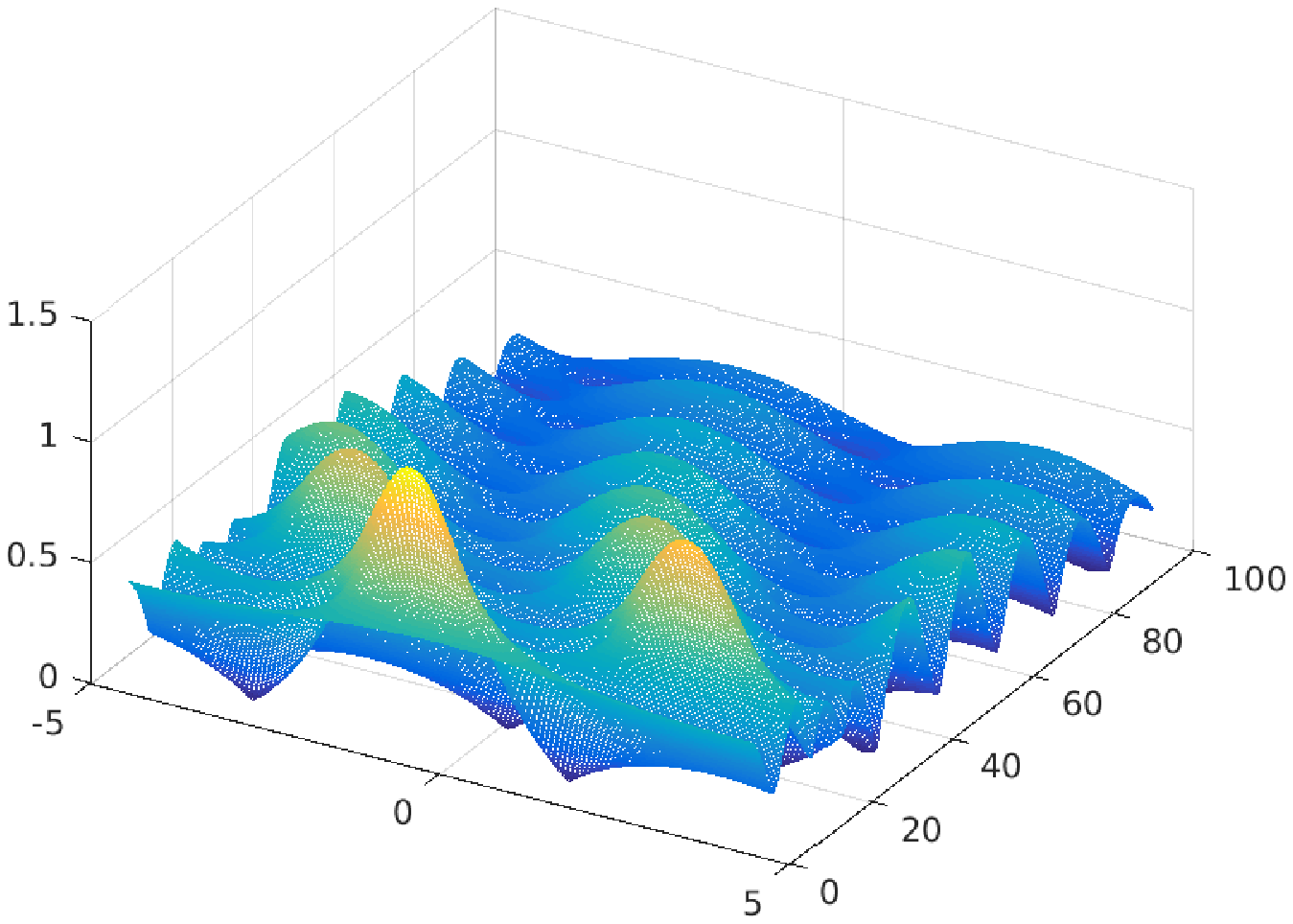}
  \hspace{12pt}
  \includegraphics[width=.33\textwidth,height=1.125in]{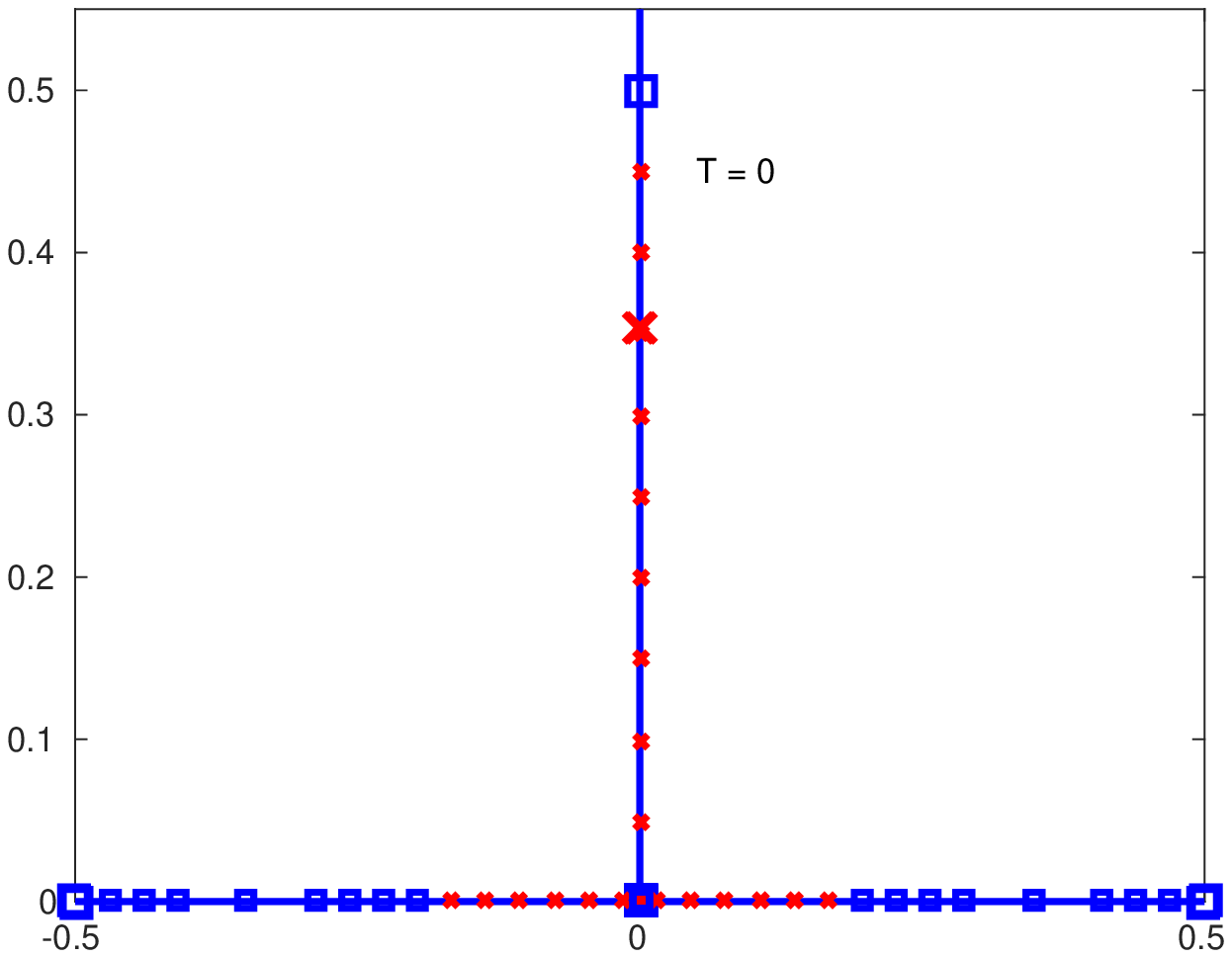}
  }
  \centerline{\hspace{.1in} A\hspace{2.5in} B}
  \vspace{12pt}
  \centerline{
    \includegraphics[width=.33\textwidth,height=1.125in]{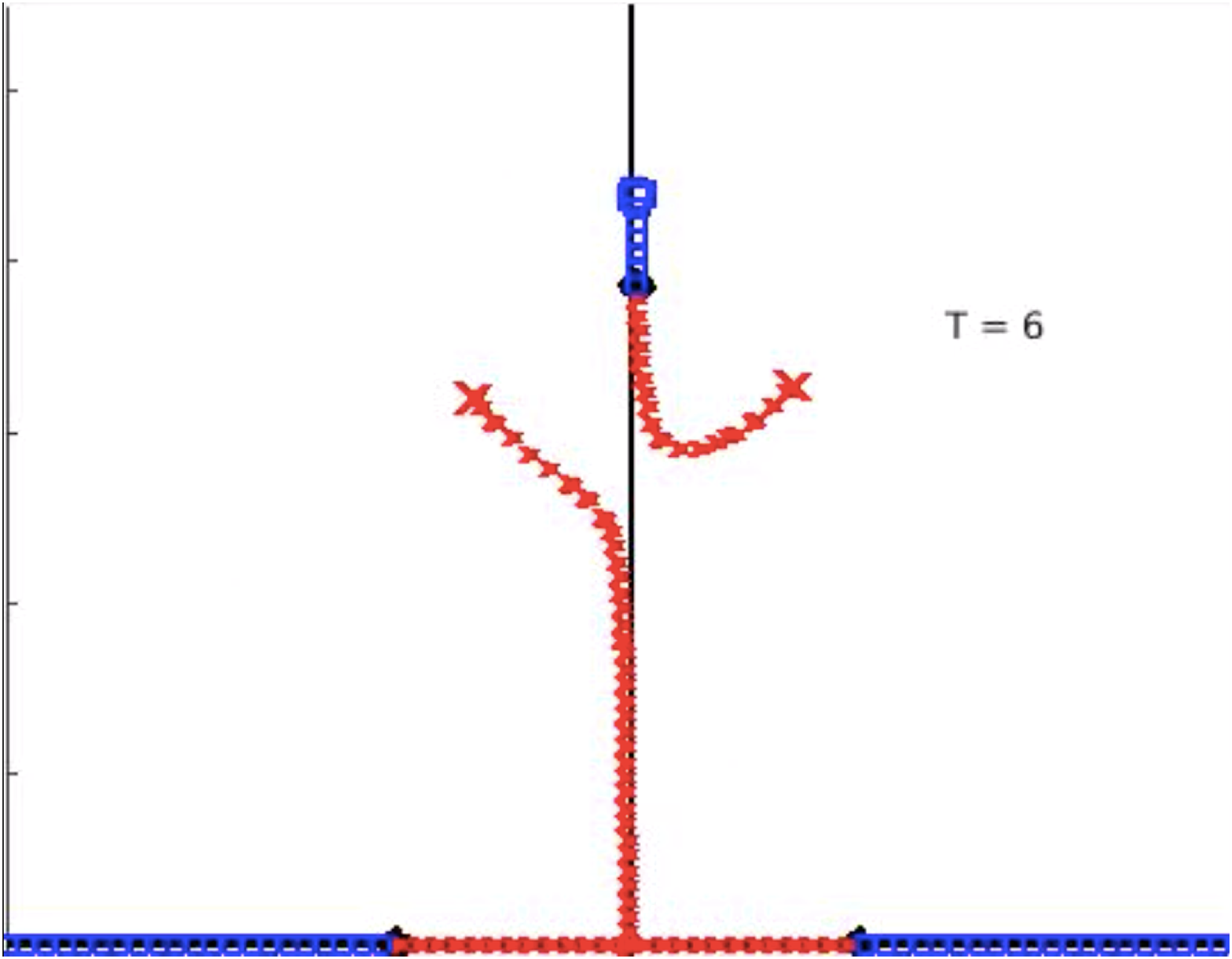}
    \includegraphics[width=.33\textwidth,height=1.125in]{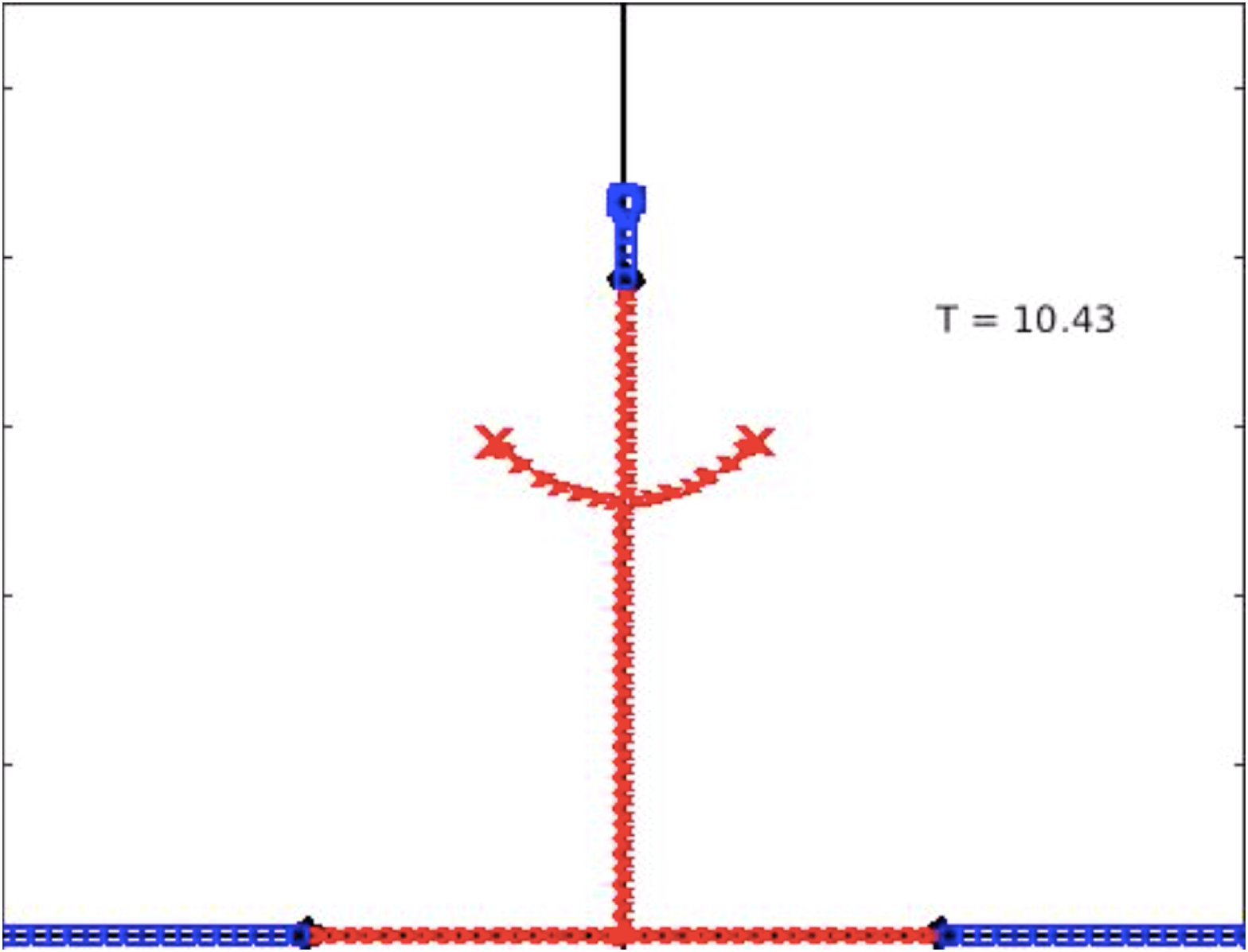}
    \includegraphics[width=.33\textwidth,height=1.125in]{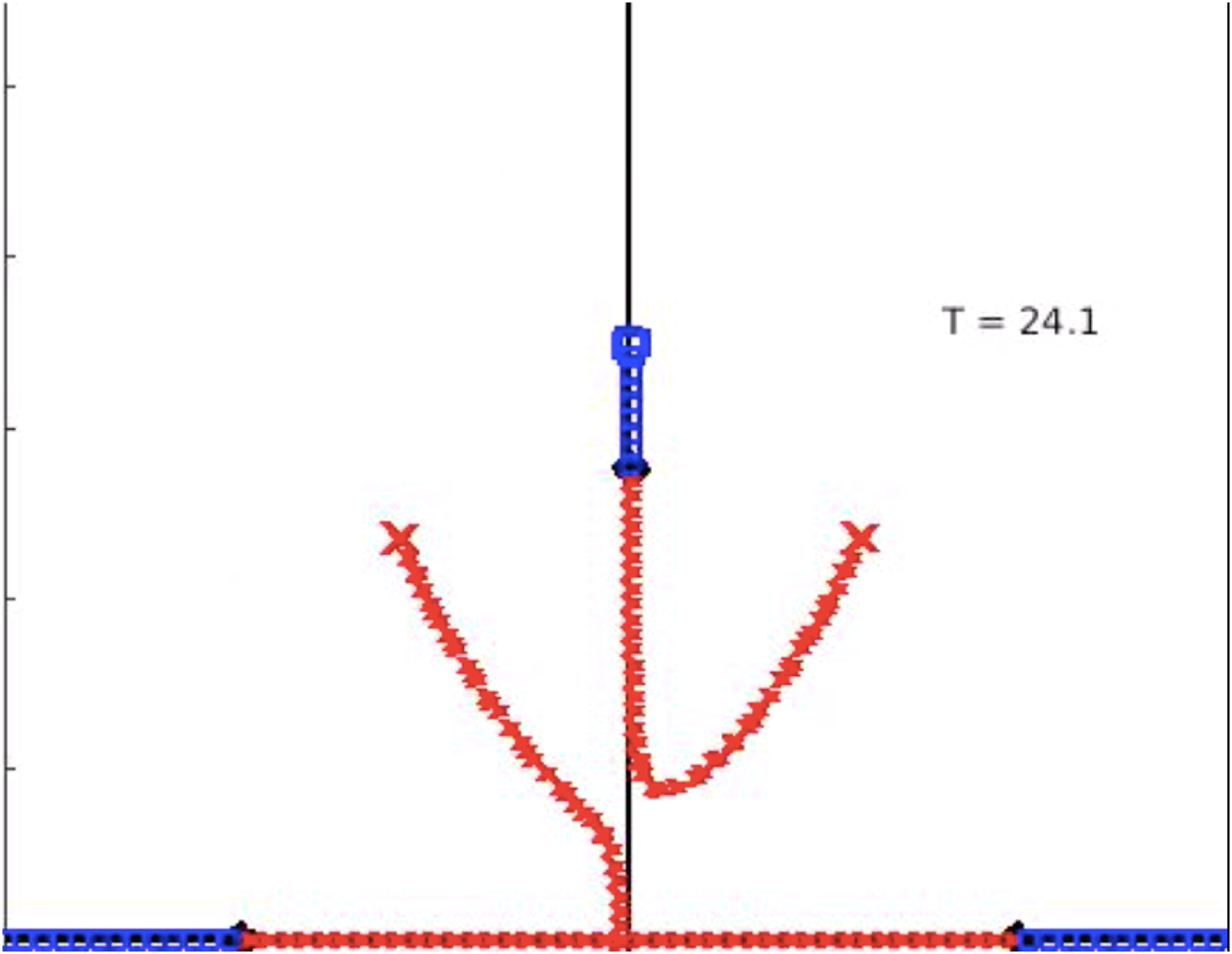}
    }
  \centerline{\hspace{.1in} C\hspace{2.5in} D\hspace{2.5in} E}
  \vspace{12pt}
  \centerline{
        \includegraphics[width=.33\textwidth,height=1.125in]{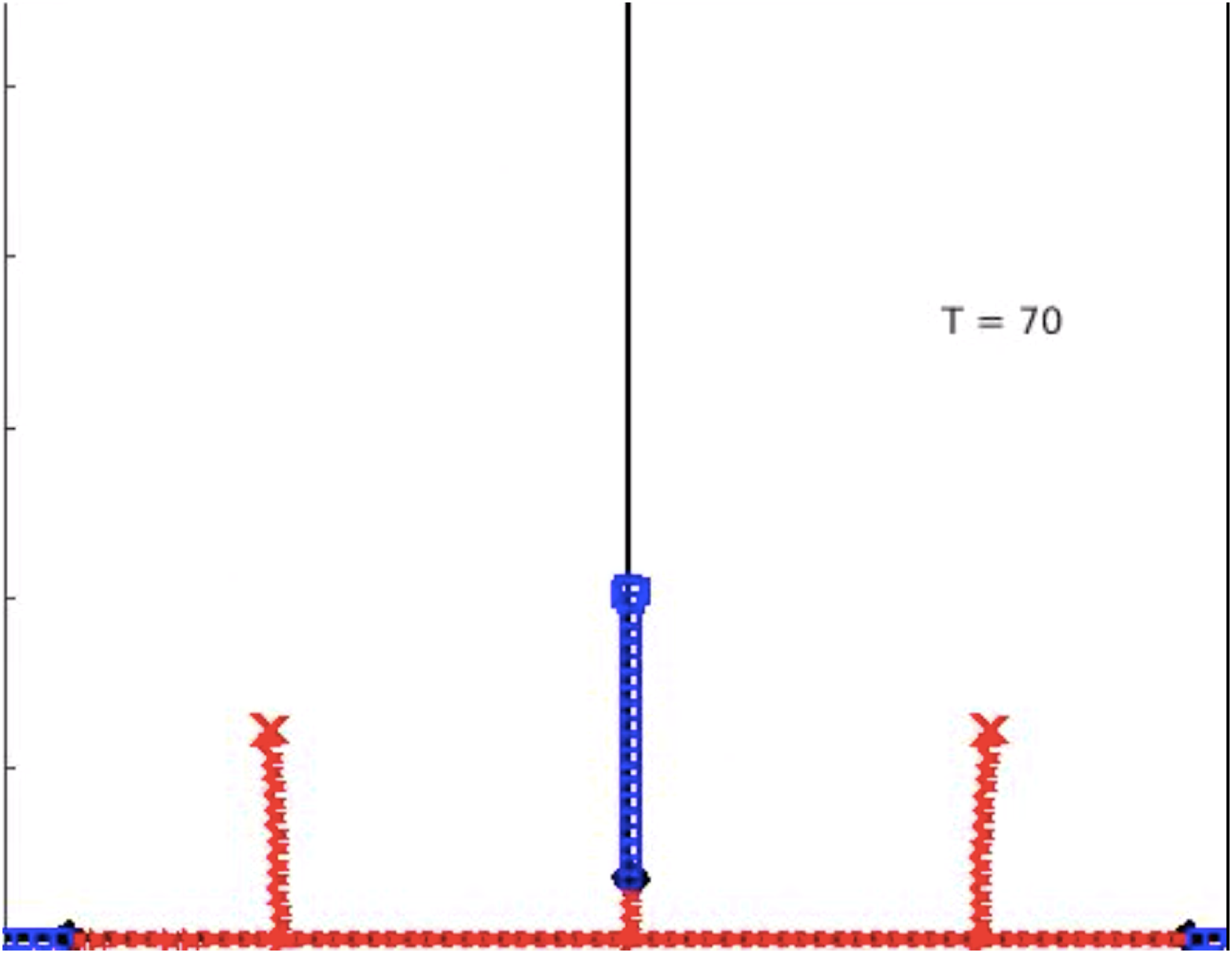}
  \includegraphics[width=.33\textwidth,height=1.125in]{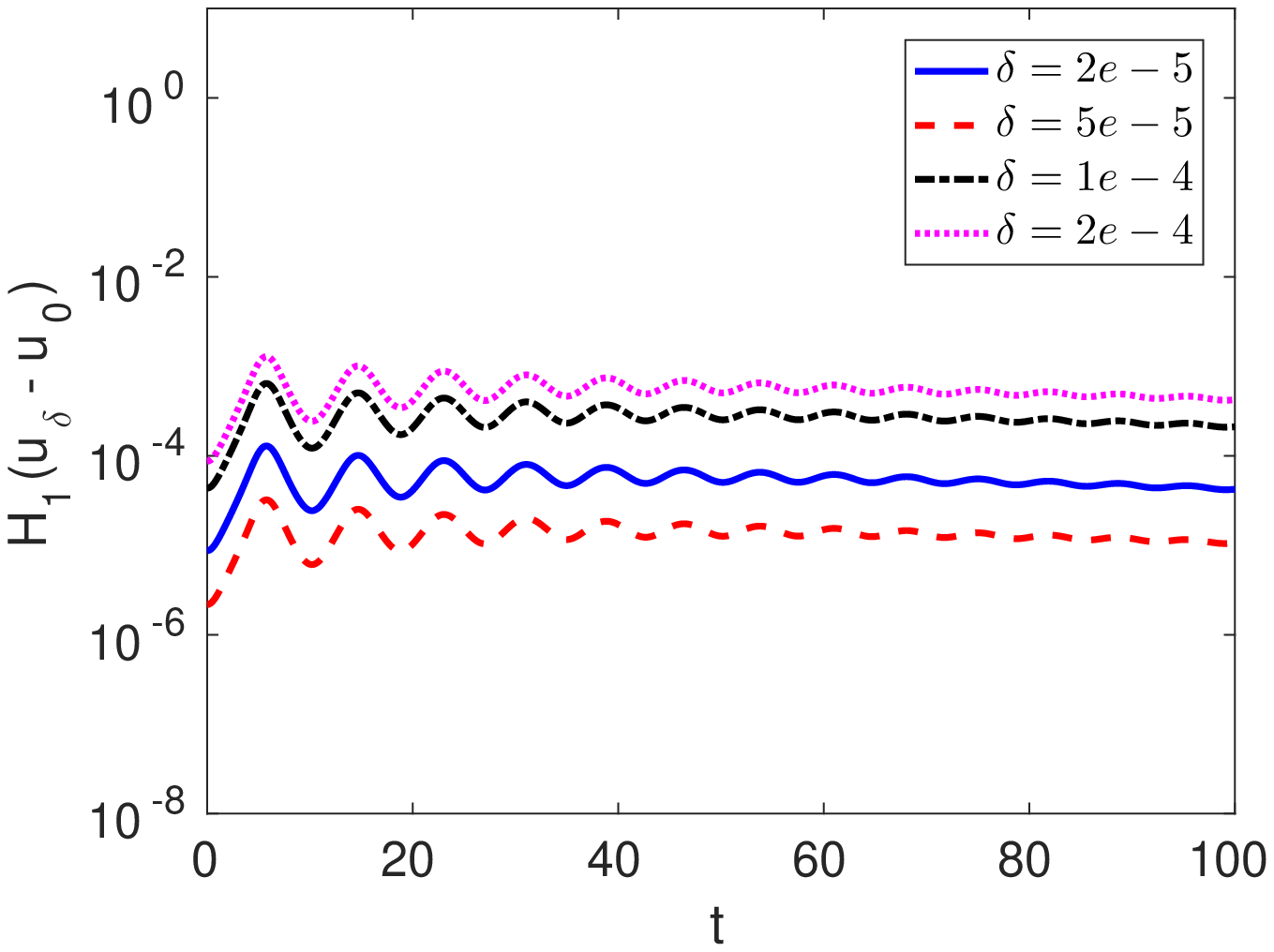}
}
  \centerline{\hspace{.1in} F\hspace{2.5in} G}
  \caption{One UM regime: A)  $|U_{\epsilon,\gamma}^{(1)}(x,t)|$ for $0 \le t \le 100$ and the  spectrum at B) $t=0$, C) $t = 6$, D) $t = 10.43$, E) $t = 24.1$,
  F) $t = 70$ and G) $\eta(t)$ for $f_4(x,t)$,  $\delta = 10^{-5},\dots,  10^{-4}$ and $\epsilon = .05$, $\gamma = 0.01$.}
\label{fig:5}
\end{figure}

{\bf 1. $\bf U_{\epsilon,\gamma}^{(1)}(x,t)$ in the one UM regime:} We begin by considering  $U_{\epsilon,\gamma}^{(1)}(x,t)$  for $\epsilon = 0.05$ and  $\gamma = 0.01$
in the one UM regime with initial data generated using  Equation~\rf{SPB1} with  $j = 1$, $a = 0.5$ and $L = 2\sqrt{2}\pi$.
Figure~\ref{fig:5}A shows the surface
 $|U_{\epsilon,\gamma}^{(1)}(x,t)|$  for $ 0 < t < 100$.

The evolution of the Floquet spectrum for $U_{\epsilon,\gamma}^{(1)}(x,t)$ is
as follows:
At $t = 0$, the spectrum in the upper half plane consists of a single band of spectrum with
end point at $\lambda_0^s = 0.5\ri$, indicated by a large ``box'',  and
one imaginary double point  at $\lambda_1^d = 0.3535 \ri$,
indicated by a large  ``$\times$'' (Figure~\ref{fig:5}B).
Under the damped HONLS
$\lambda_1^d$ splits asymmetrically forming a right state, with the upper band
of  spectrum in the right quadrant and the lower band in the left quadrant,
consistent with the short time perturbation analysis in Section~4.
The right state is clearly visible at $t = 6$ in Figure~\ref{fig:5}C
with the waveform  characterized by a single damped modulated mode
traveling to the right.
The spectrum persists in a right state with the  separation distance between the two bands varying until $t = 10.43$ when a cross state forms with an embedded critical point, Figure~\ref{fig:5}D, indicating an instability. Subsequently the critical point splits into a right state.  As damping continues the band in the right quadrant widens, Figure~\ref{fig:5}E,  and the vertex of the loop  eventually touches the origin at $t \approx 27.5.$
The spectrum  then has three  bands emanating off the real axis,
 with endpoints $\lambda_1^{-}, \lambda_0^{s}, \lambda_1^{+}$ which, as damping
continues,   diminish in amplitude and 
 move away from the imaginary axis, Figure~\ref{fig:5}F.

Figure~\ref{fig:5}G shows the evolution of $\eta(t)$
for $U_{\epsilon,\gamma}^{(1)}(x,t)$ 
  using  $f_4$ and $\delta = 10^{-5},\dots,  10^{-4}.$
  The saturation of $\eta(t)$ at $t \approx 16$ is consistent with the Floquet
  criteria that the solution stabilizes after complex double points and complex critical points are eliminated in the damped HONLS flow.
  
From the spectral analysis of $U_{\epsilon,\gamma}^{(1)}(x,t)$  in the one UM regime, we find it may be
characterized as a continuous deformation of a noneven generalization of
the  3-phase solution, i.e. the right state \rf{3phase}.
The  amplitude of the oscillations of $U_{\epsilon,\gamma}^{(1)}(x,t)$ decreases and the frequency increases until small fast oscillations about the damped
Stokes wave, visible in Figure~\ref{fig:5}A, are obtained.

 \subsection{Damped HONLS SPBs in the two unstable mode regime}
 For the two UM  regime we let $a = 0.5$,
$L = 4\sqrt{2}\pi$ 
($N = \displaystyle \lfloor {aL}/{\pi} \rfloor = 2$) and
 consider the two distinct perturbed single mode
 SPBs $U_{\epsilon,\gamma}^{(1)}(x,t)$ and $U_{\epsilon,\gamma}^{(2)}(x,t)$ and the perturbed iterated
 SPB $U_{\epsilon,\gamma}^{1,2}(x,t)$. The damped HONLS perturbation parameters are
 $\epsilon = 0.05$ and  $\gamma = 0.01$.

\begin{figure}[ht!]
  \centerline{
    \includegraphics[width=.5\textwidth,height=1.5in]{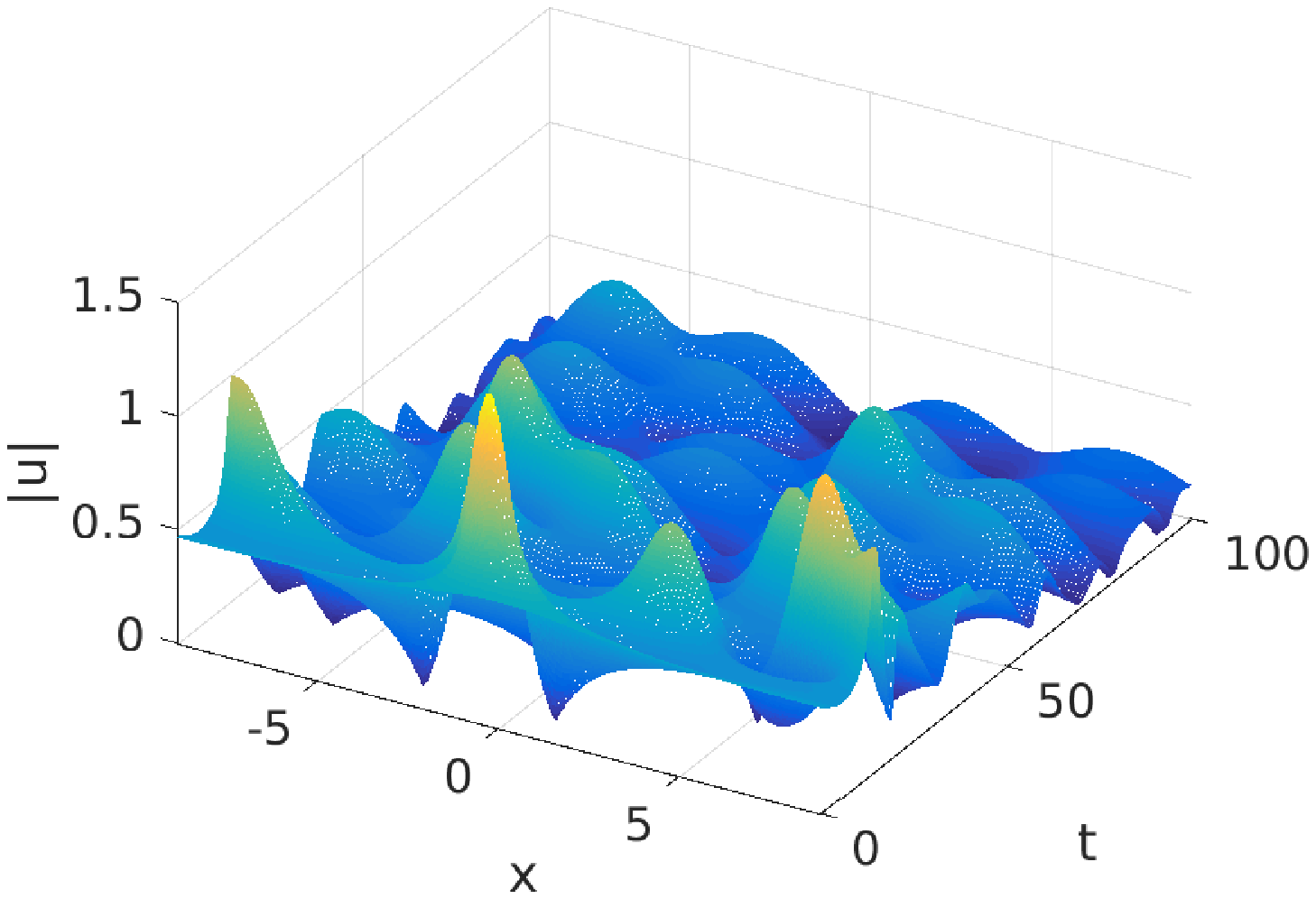}
    \hspace{12pt}
    \includegraphics[width=.245\textwidth,height=1.125in]{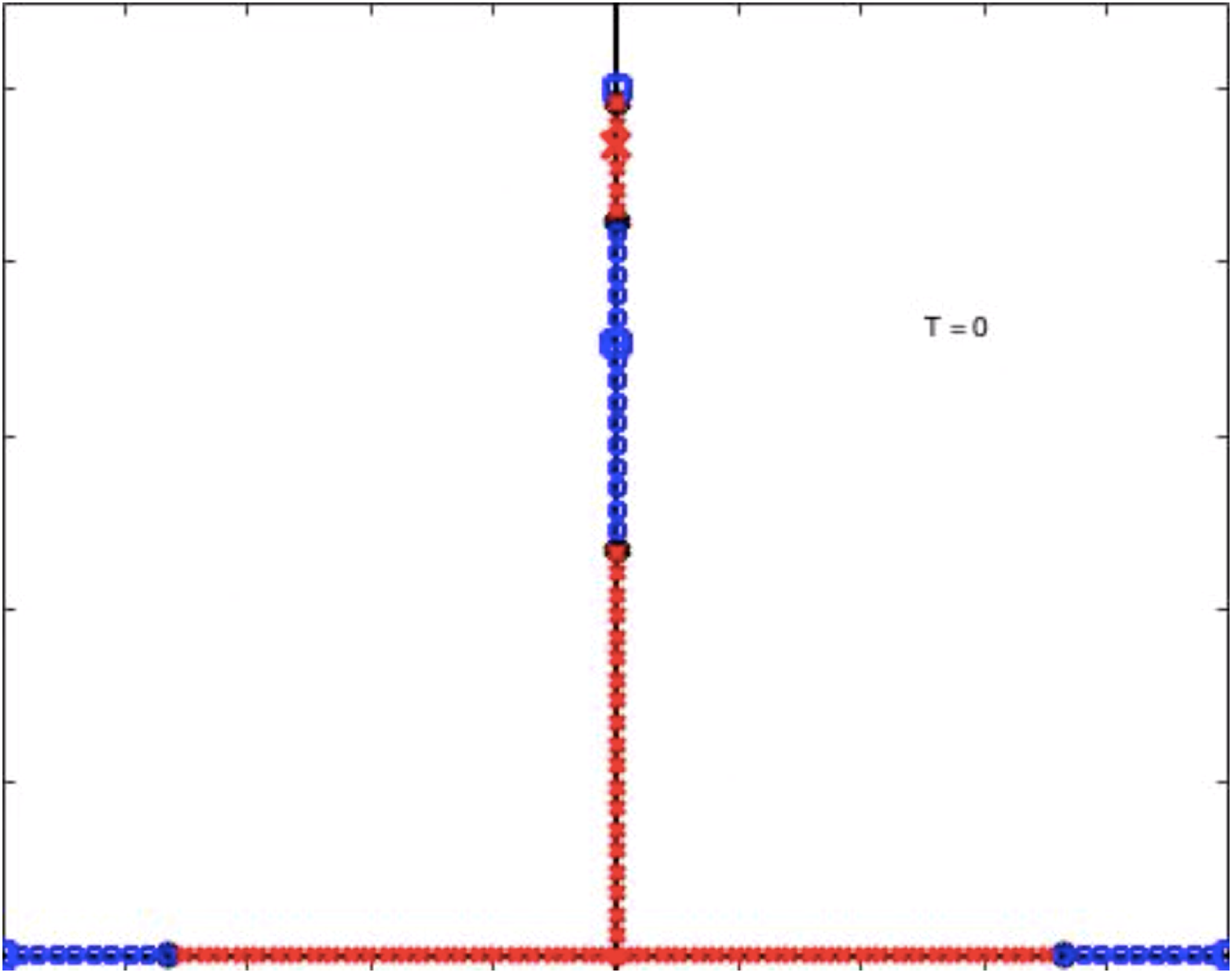}
  }
  \centerline{A\hspace{3.5in} B}
  \vspace{12pt}
  \centerline{
\includegraphics[width=.33\textwidth,height=1.125in]{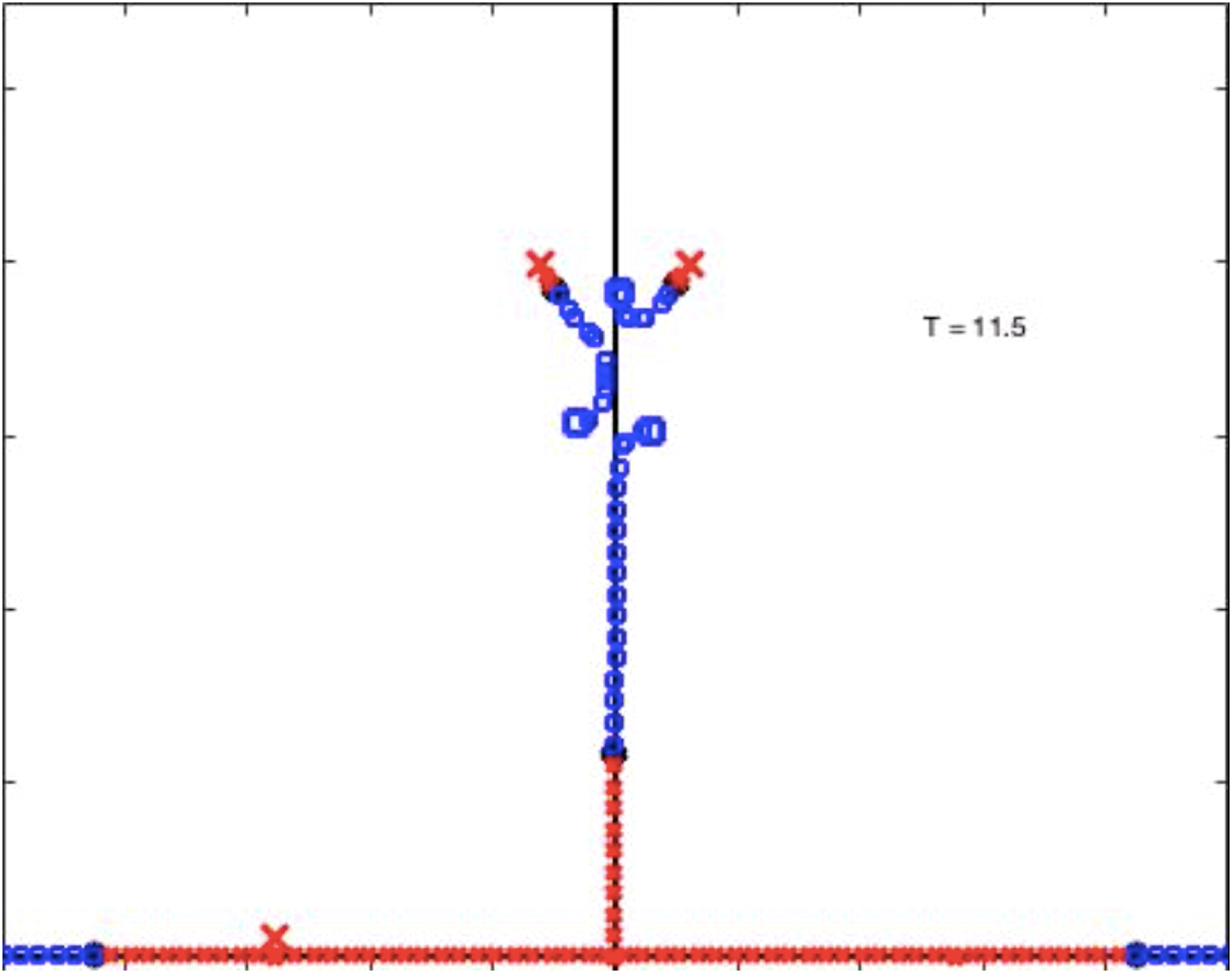}
\includegraphics[width=.33\textwidth,height=1.125in]{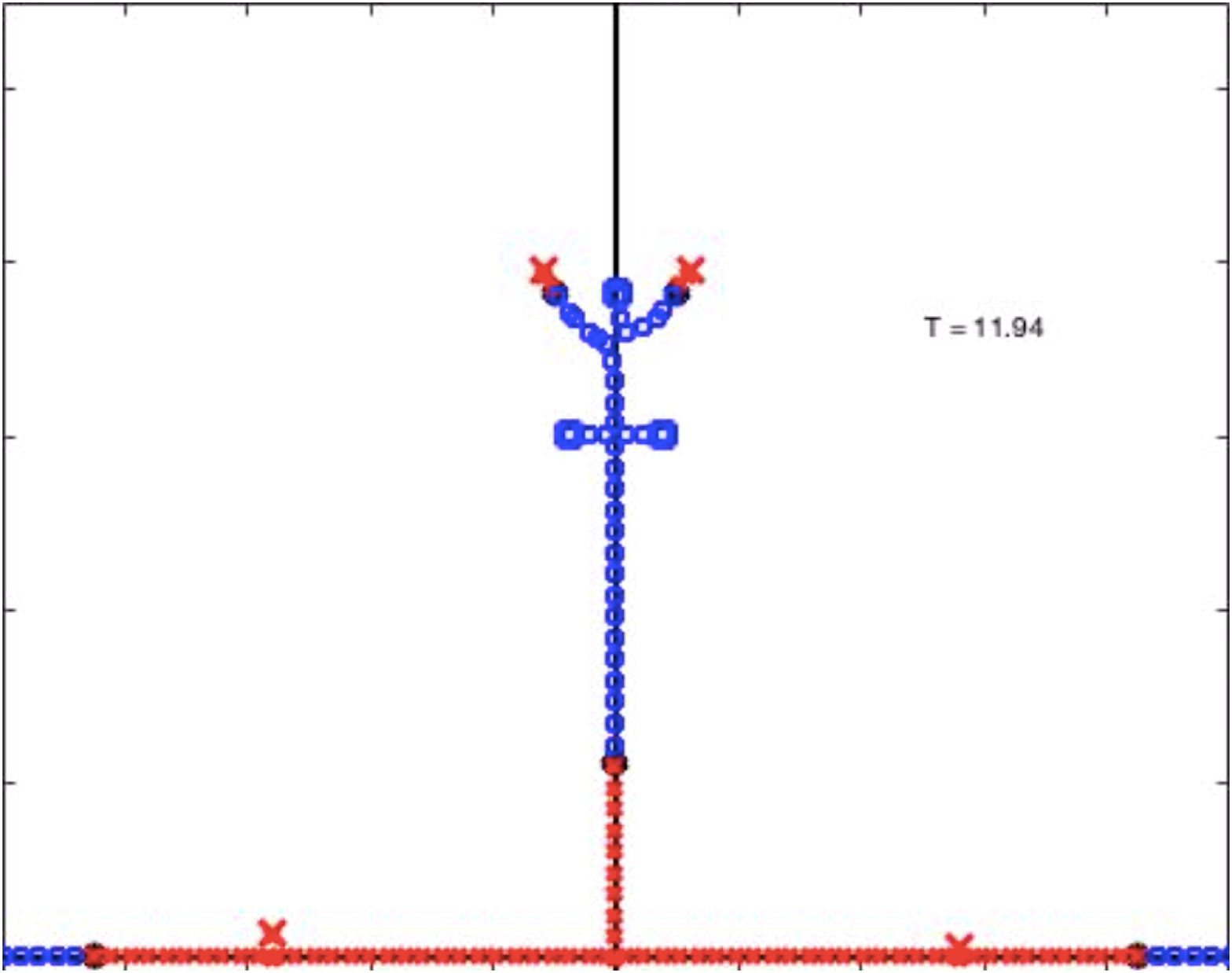}
\includegraphics[width=.33\textwidth,height=1.125in]{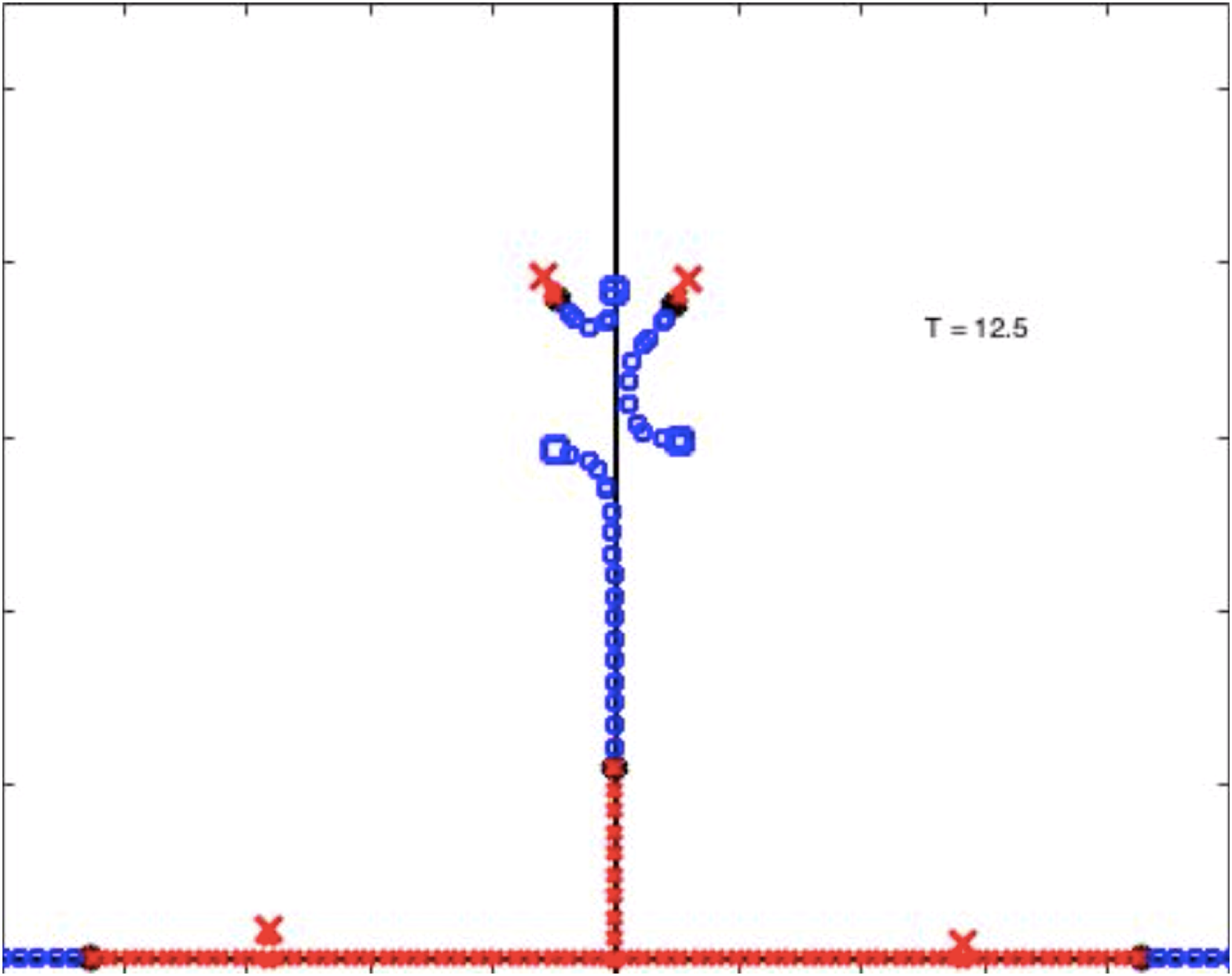}
}
 \centerline{C\hspace{2.5in} D\hspace{2.5in} E}
  \vspace{12pt}
  \centerline{
    \includegraphics[width=.3\textwidth,height=1.125in]{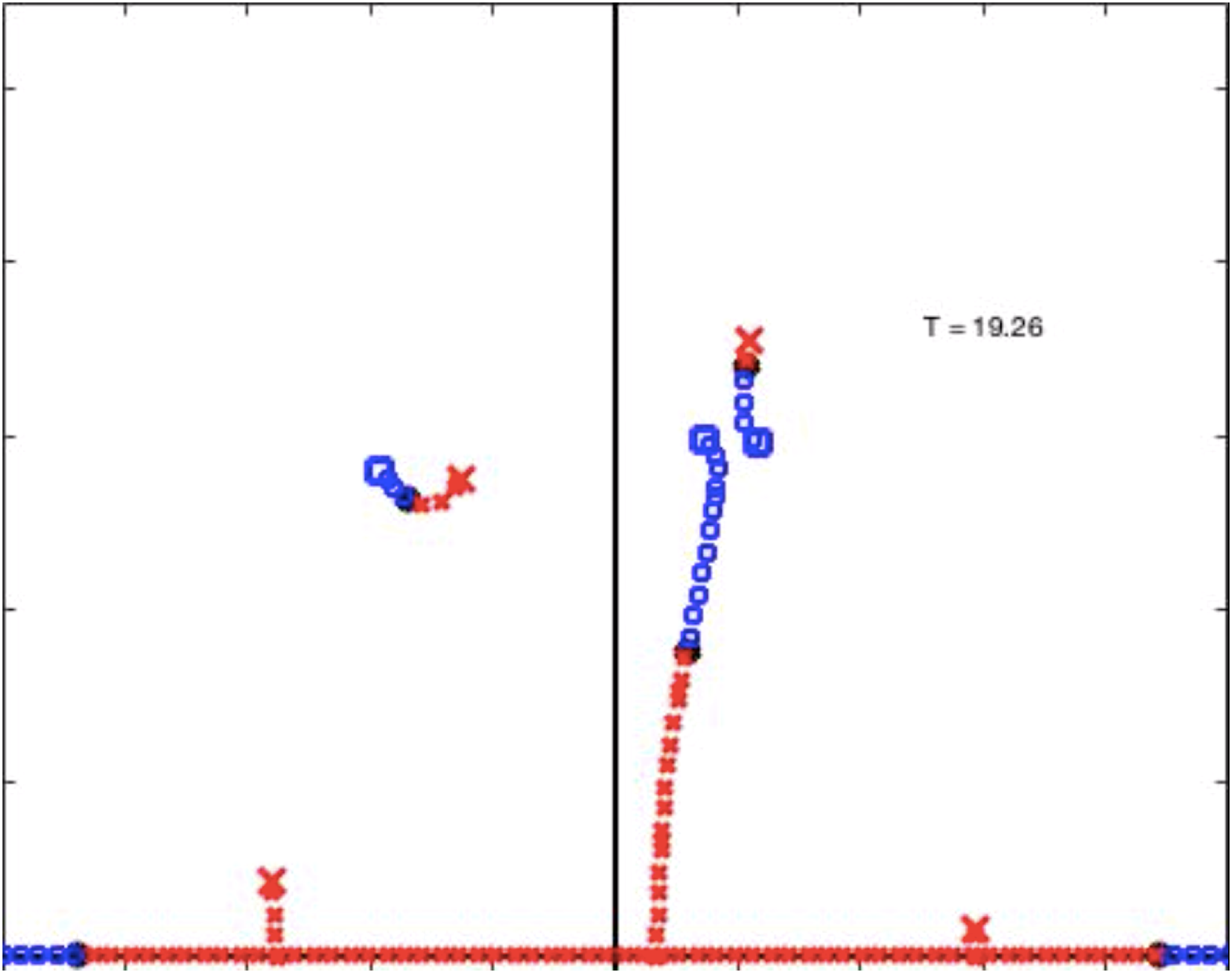}
  \includegraphics[width=.3\textwidth,height=1.125in]{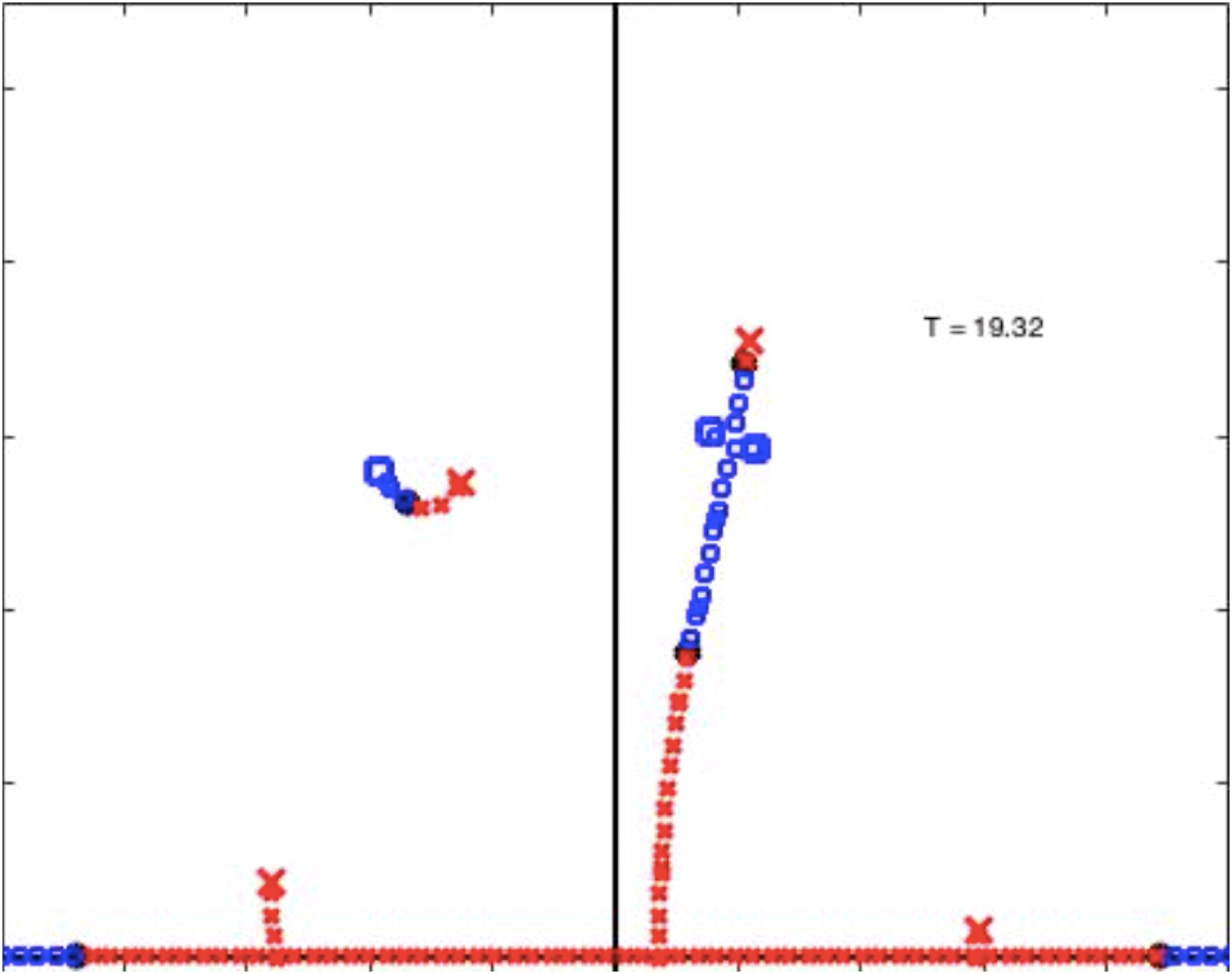}
  \includegraphics[width=.3\textwidth,height=1.125in]{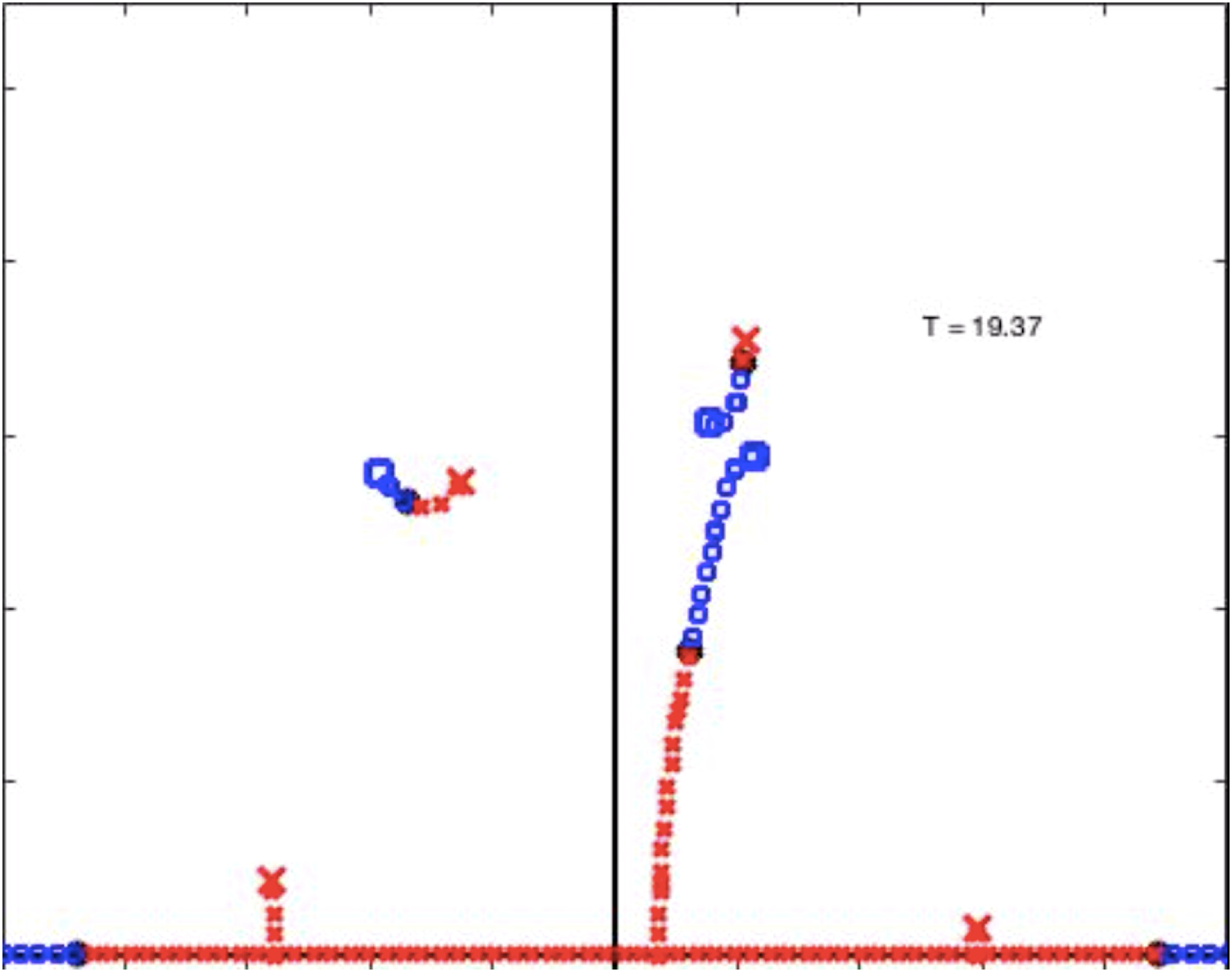}
}
 \centerline{F\hspace{2.5in} G\hspace{2.5in} H}
  \vspace{12pt}
\centerline{
    \includegraphics[width=.33\textwidth,height=1.125in]{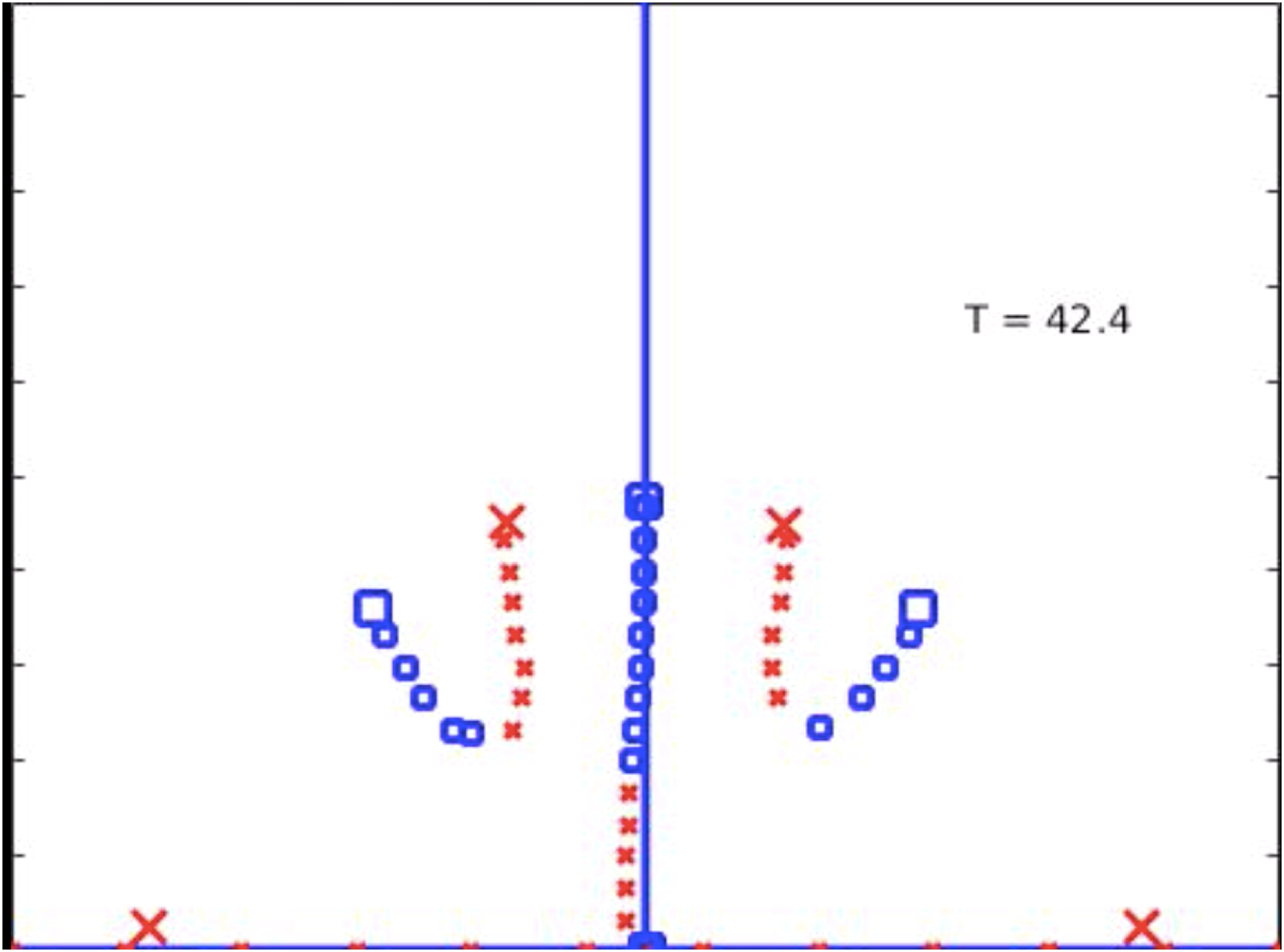}
\includegraphics[width=.33\textwidth,height=1.125in]{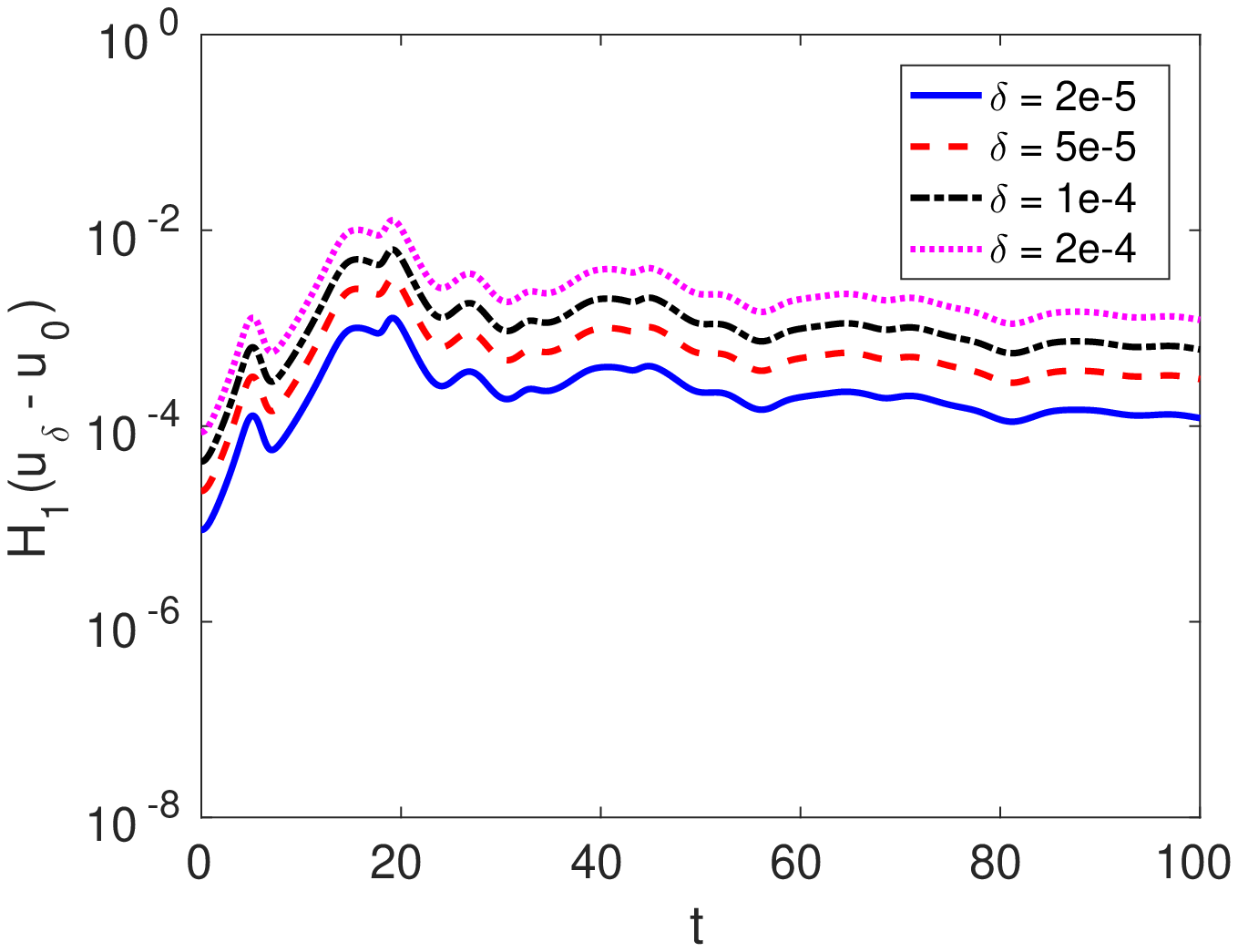}
}
 \centerline{I\hspace{2.5in} J}
 \caption{Two UM regime: A) $|U_{\epsilon,\gamma}^{(1)}(x,t)|$ for $0 \le t \le 100$, and spectrum at B) $t=0$,
   C) $t = 11.5$, D) $t = 11.94$, E) $t=12.5$,
   F) $t=19.26$, G) $t=19.32$, and H) $t=19.37$,
I) $t=42.4$, and J) $\eta(t)$ for $f_1(x,t)$, $\delta = 10^{-5},\dots, 10^{-4}$.}
 \label{fig:6}
\end{figure}

 {\bf 1.  $ \bf U_{\epsilon,\gamma}^{(1)}(x,t)$ in the two UM regime:}
  Figure~\ref{fig:6}A shows the surface  $|U_{\epsilon,\gamma}^{(1)}(x,t)|$
  for $ 0 < t < 100$ for initial data given by Equation~\rf{SPB1} with $j =1$.
The Floquet spectrum  at $t = 0$ is given in Figure~\ref{fig:6}B.
The end point of the band of spectrum at $\lambda_0^s = 0.5\ri$ is indicated by a ``box''.
There are two complex double points  at $\lambda_1^d = 0.4677 \ri$ and $\lambda_2^d = 0.3535 \ri$,
indicated by a ``$\times$'' and ``box'', respectively.
For $t >0$ both double points split asymmetrically: $\lambda_1^d$,
the complex double point at which $U^{(1)}(x,t)$ is constructed,  splits at leading order  into $\lambda_1^{\pm}$  such that a right state forms with the first mode
traveling to the right.
The second double point $\lambda_2^d$  splits at higher order  
into $\lambda_2^{\pm}$ such that a left state forms with  the second mode traveling to  the left.
These disjoint asymmetric bands of  spectrum  are consistent with the short time perturbation analysis in Section~4 for damped HONLS data of the form
(\ref{uij2}).

This spectral configuration is representative of the spectrum during the initial stage of it's evolution and is   still observable  in
Figure~\ref{fig:6}C at $t = 11.5$.
A bifurcation occurs  at  $t = 11.94$, Figure~\ref{fig:6}D,
when a double cross state forms and two complex critical points
emerge in the spectrum, indicating instabilities associated with both nonlinear modes. 
Subsequently both complex critical points split, Figure~\ref{fig:6}E,
with  the upper band in the left quadrant  
and the second band in the right quadrant
corresponding to a damped waveform  with the first mode traveling to the
left and the second mode traveling to  the right.
  
The bifurcation  at $t= 11.94$ corresponds to transitioning through the remnant of an unstable  5-phase solution (with two instabilities) of the NLS equation. 
The bands eventually become completely detached
from the imaginary axis and 
a second complex critical point forms
at $t = 19.32$. The bifurcation sequence is shown in
Figures~\ref{fig:6}F-G-H.  The main band emanating from the real axis
  then reestablishes itself  close to the imaginary axis,
  Figure~\ref{fig:6}I, and the spectrum settles into a configuration corresponding to a stable 5 phase solution. 
The bands move
apart and downwards and hit the real axis
with no further development of complex critical points.
From the Floquet spectral perspective,  once damping 
eliminates complex critical  points in the spectrum
at approximately $t \approx 20$, $U_{\epsilon,\gamma}^{(1)}(x,t)$ 
stabilizes. 

Figure~\ref{fig:6}J shows the evolution of $\eta(t)$
for $U_{\epsilon,\gamma}^{(1)}(x,t)$ 
with $f_2$ for  $\delta = 10^{-5},\dots,  10^{-4}.$
The perturbation $f_2$  is chosen in the direction of the
unstable mode associated with $\lambda_2^d$.
$\eta(t)$  stops growing by  $t\approx 20$, confirming  the
instabilities associated with the complex critical ponts and 
time of 
  stabilization obtained 
  from  the nonlinear spectral analysis.

For $U_{\epsilon,\gamma}^{(1)}(x,t)$  in the 2-UM regime, both
$\lambda_1^d$ and $\lambda_2^d$ resonate with the perturbation.
The route to stability is characterized by the appearance of the double cross state of the NLS  and the proximity to this state is significant in organazing
the damped HONLS dynamics.
  Once stabilized, $U_{\epsilon,\gamma}^{(1)}(x,t)$ may be characterized as a
  continuous deformation of a stable 5-phase solution.

\begin{figure}[ht!]
  \centerline{
    \includegraphics[width=.5\textwidth,height=1.5in]{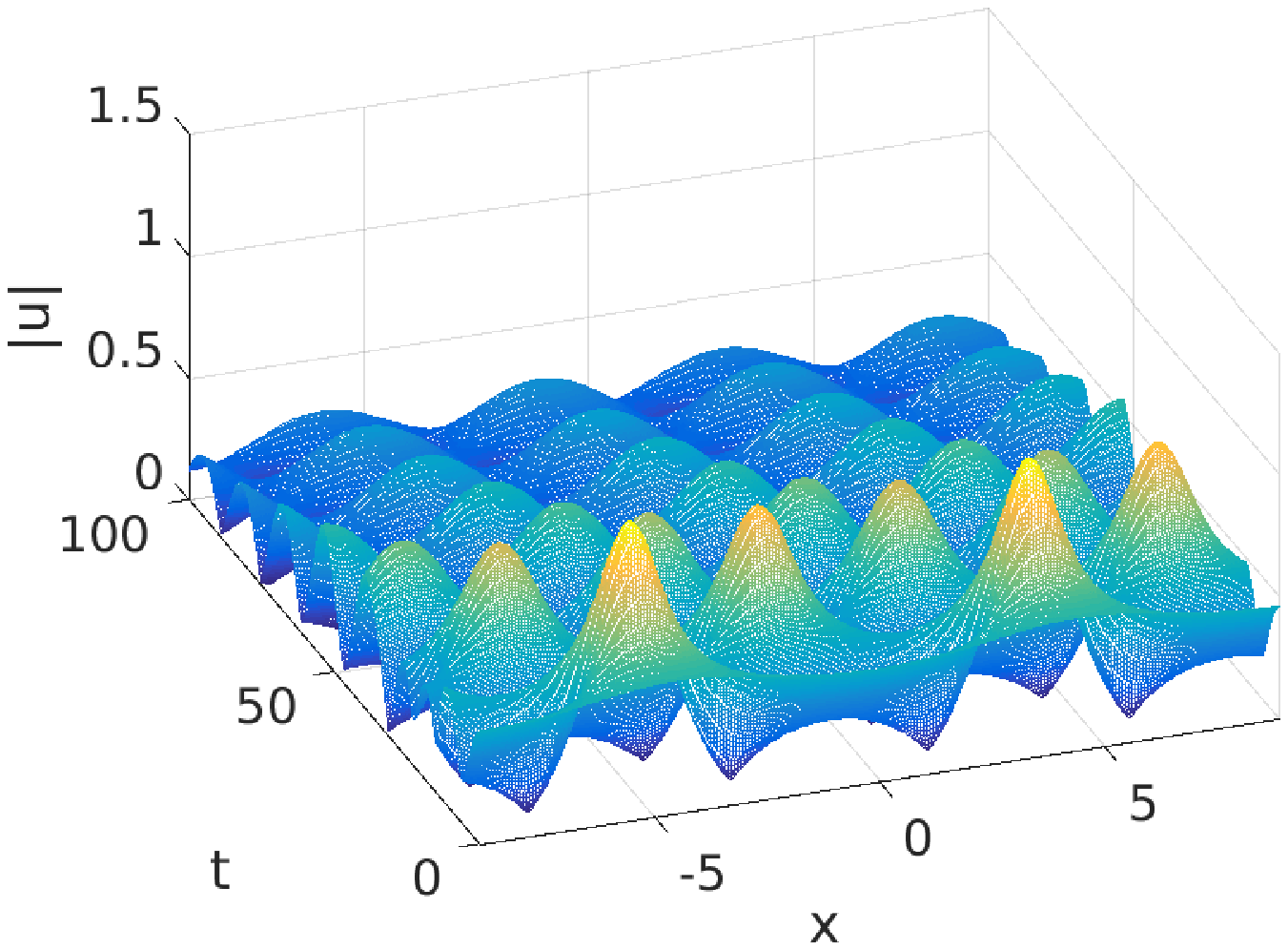}
    \hspace{12pt}
    \includegraphics[width=.33\textwidth,height=1.125in]{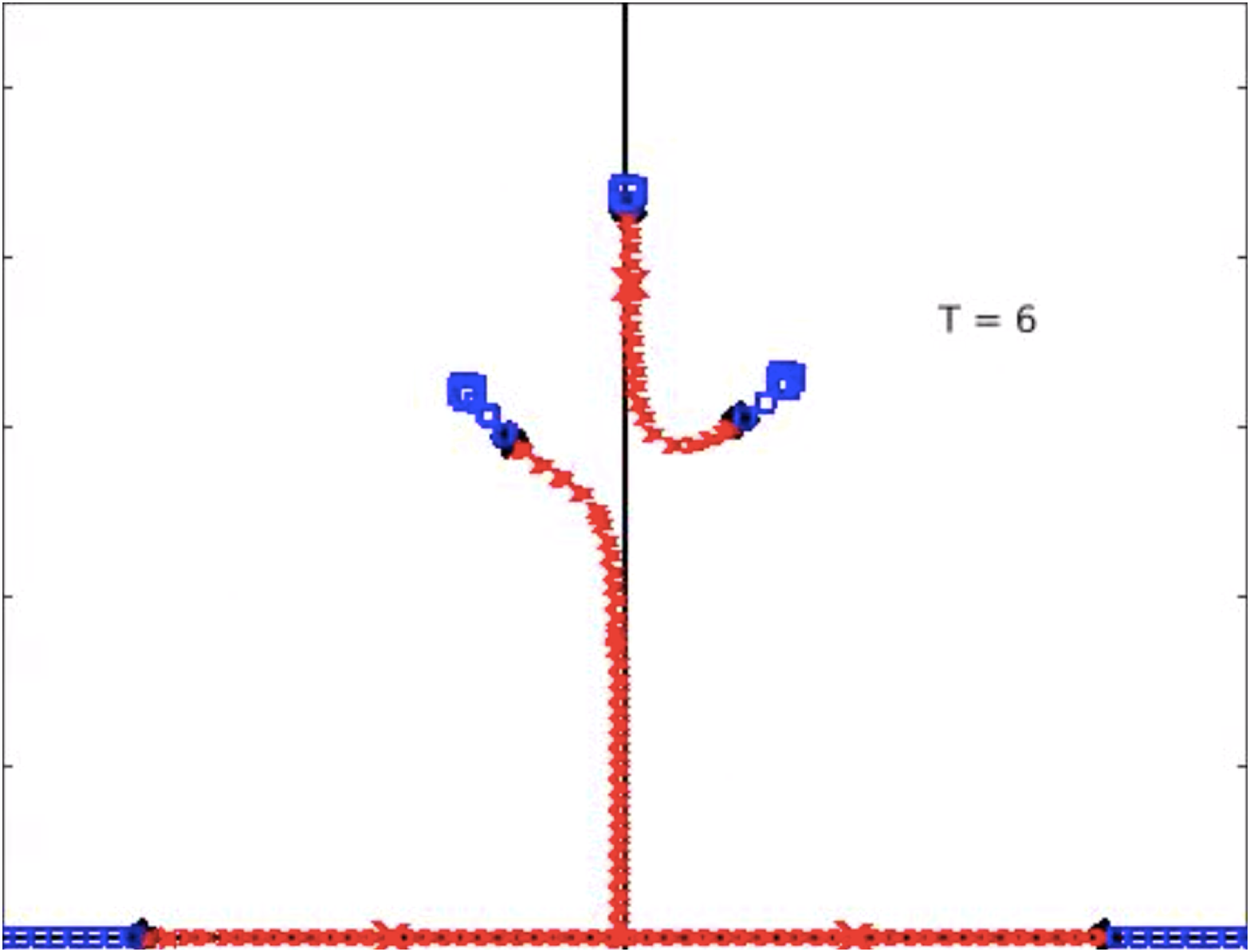}
}
  \centerline{\hspace{.1in} A\hspace{2.5in} B}
  \vspace{12pt}
  \centerline{
    \includegraphics[width=.33\textwidth,height=1.125in]{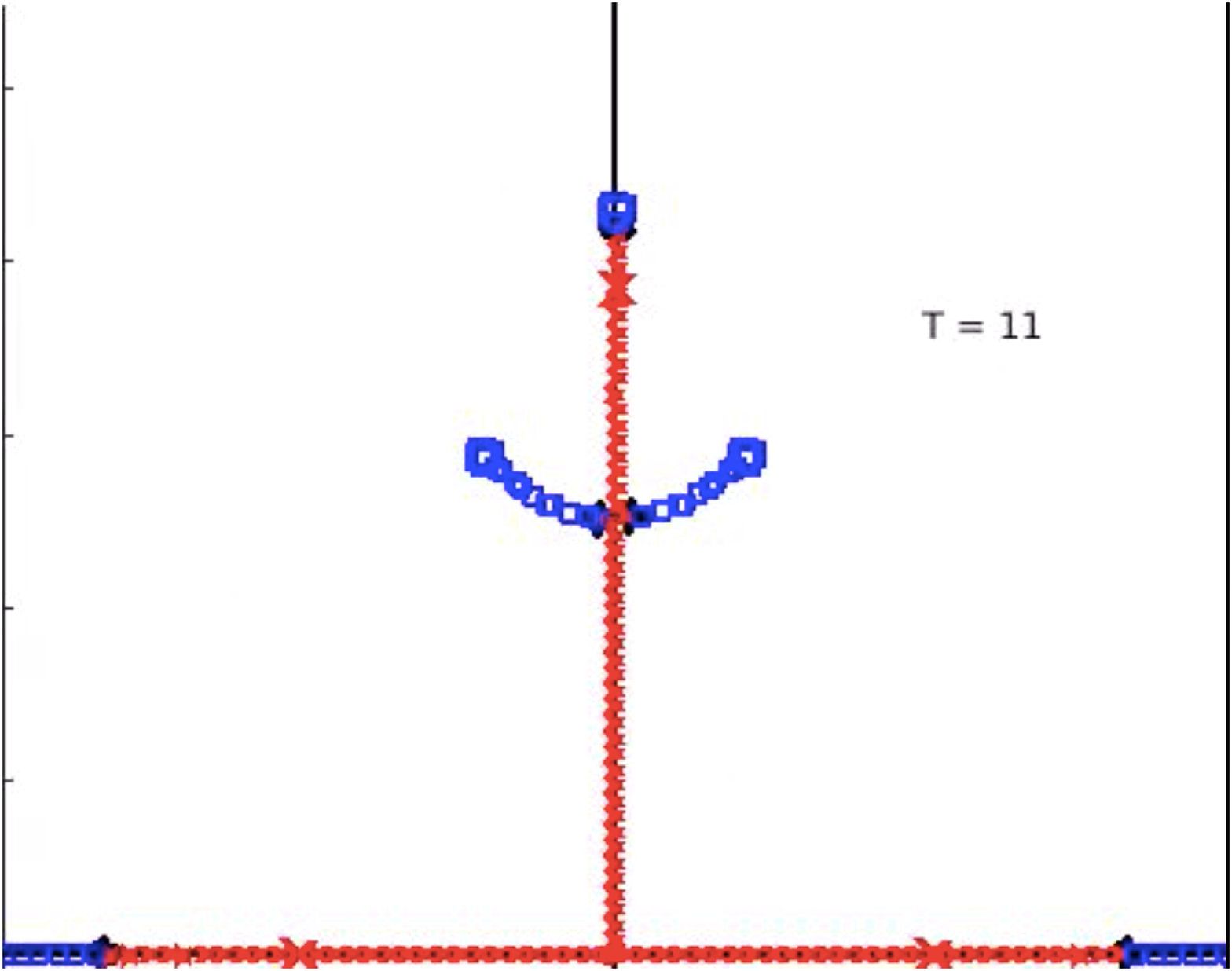}
      \includegraphics[width=.33\textwidth,height=1.125in]{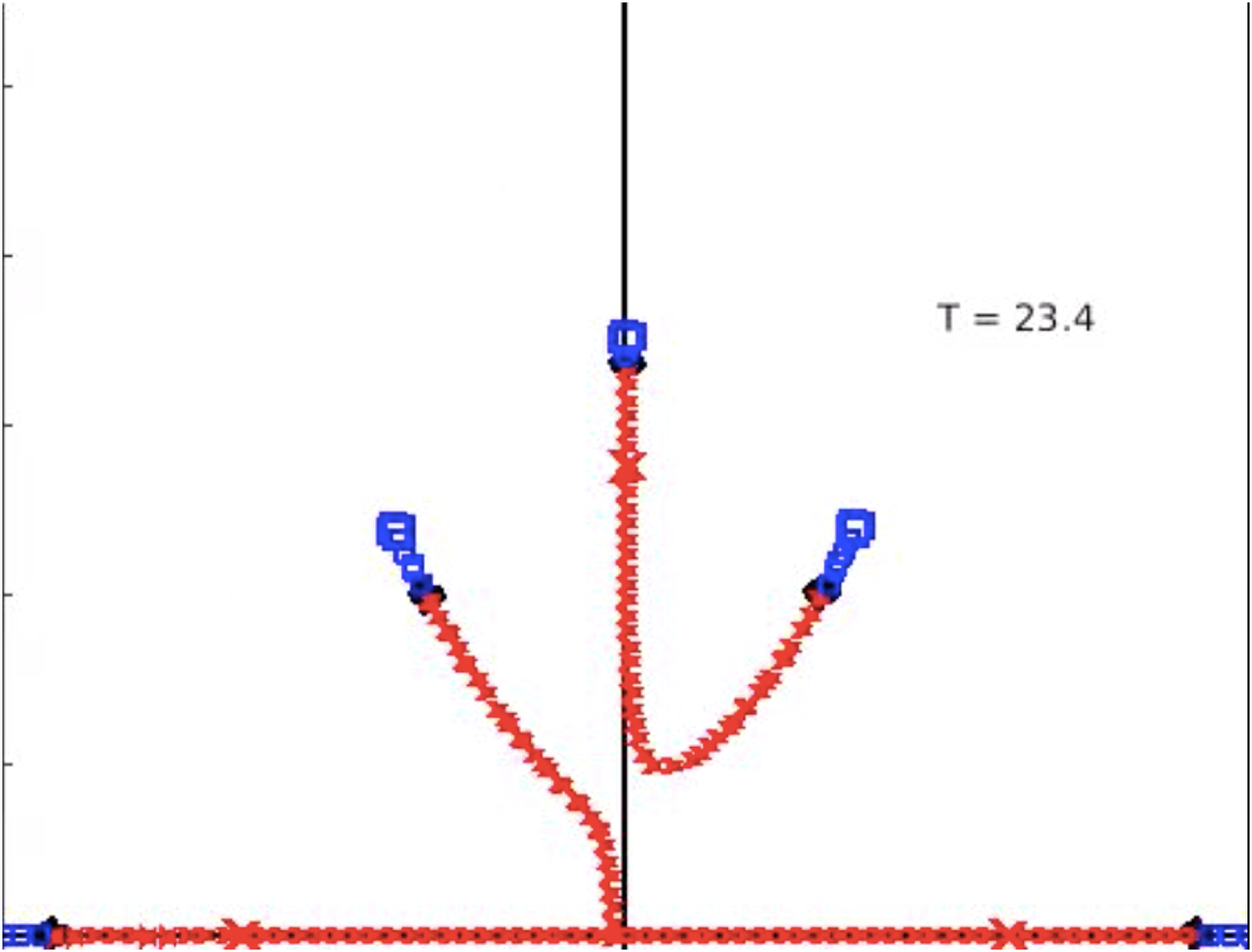}
      \includegraphics[width=.33\textwidth,height=1.125in]{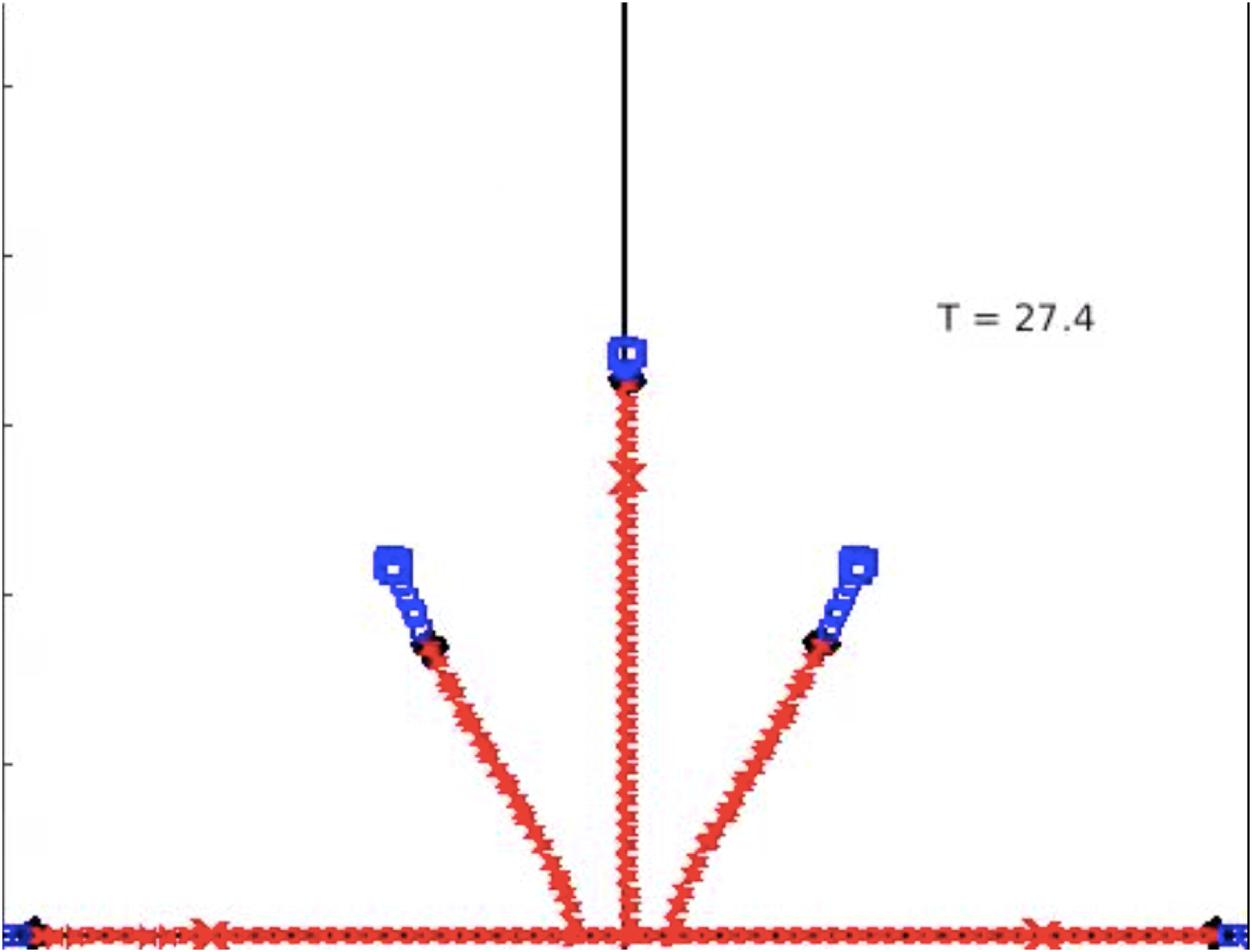}
}
 \centerline{C\hspace{2.5in} D\hspace{2.5in} E}
  \vspace{12pt}
  \centerline{
    \includegraphics[width=.33\textwidth,height=1.5in]{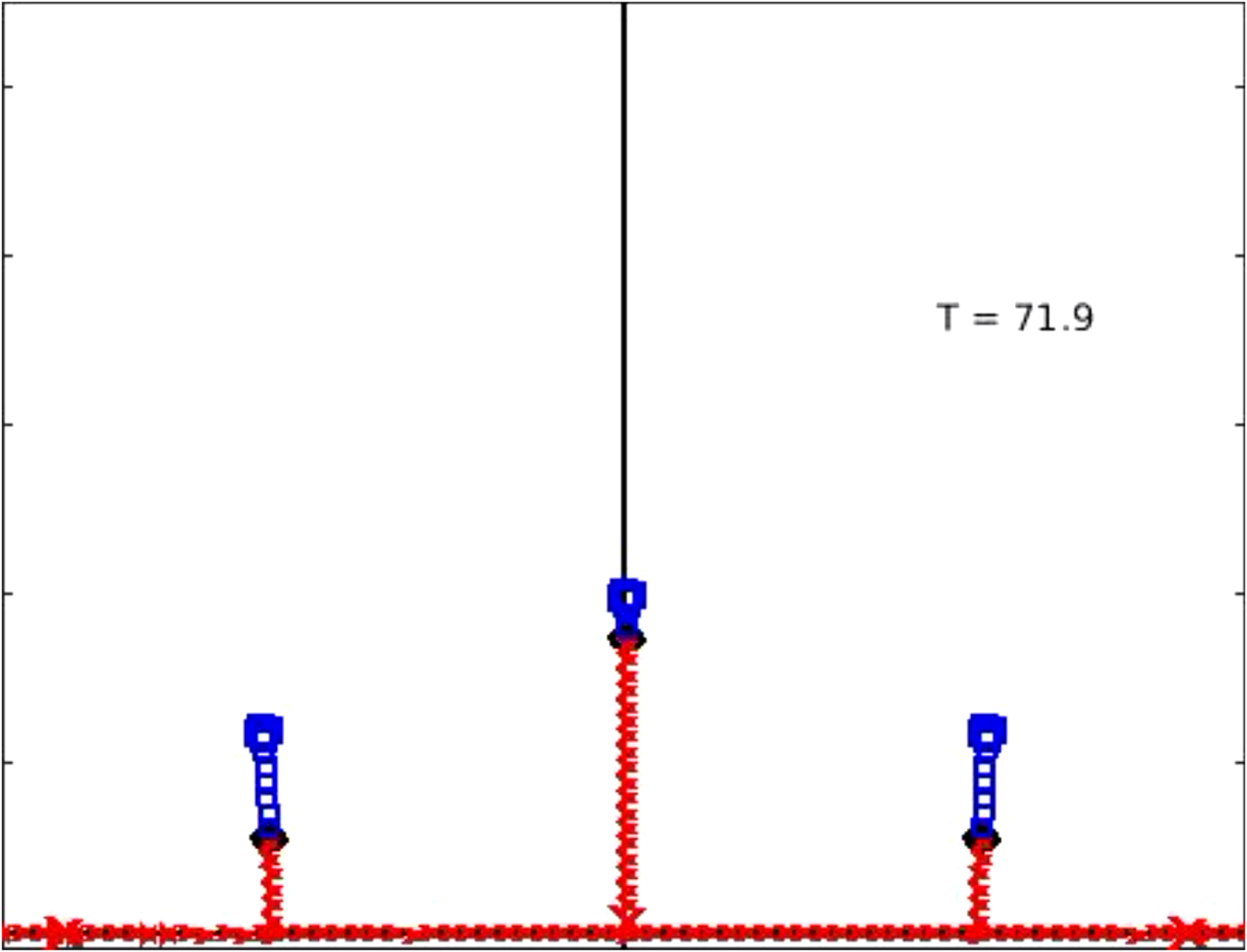}
    \includegraphics[width=.33\textwidth,height=1.5in]{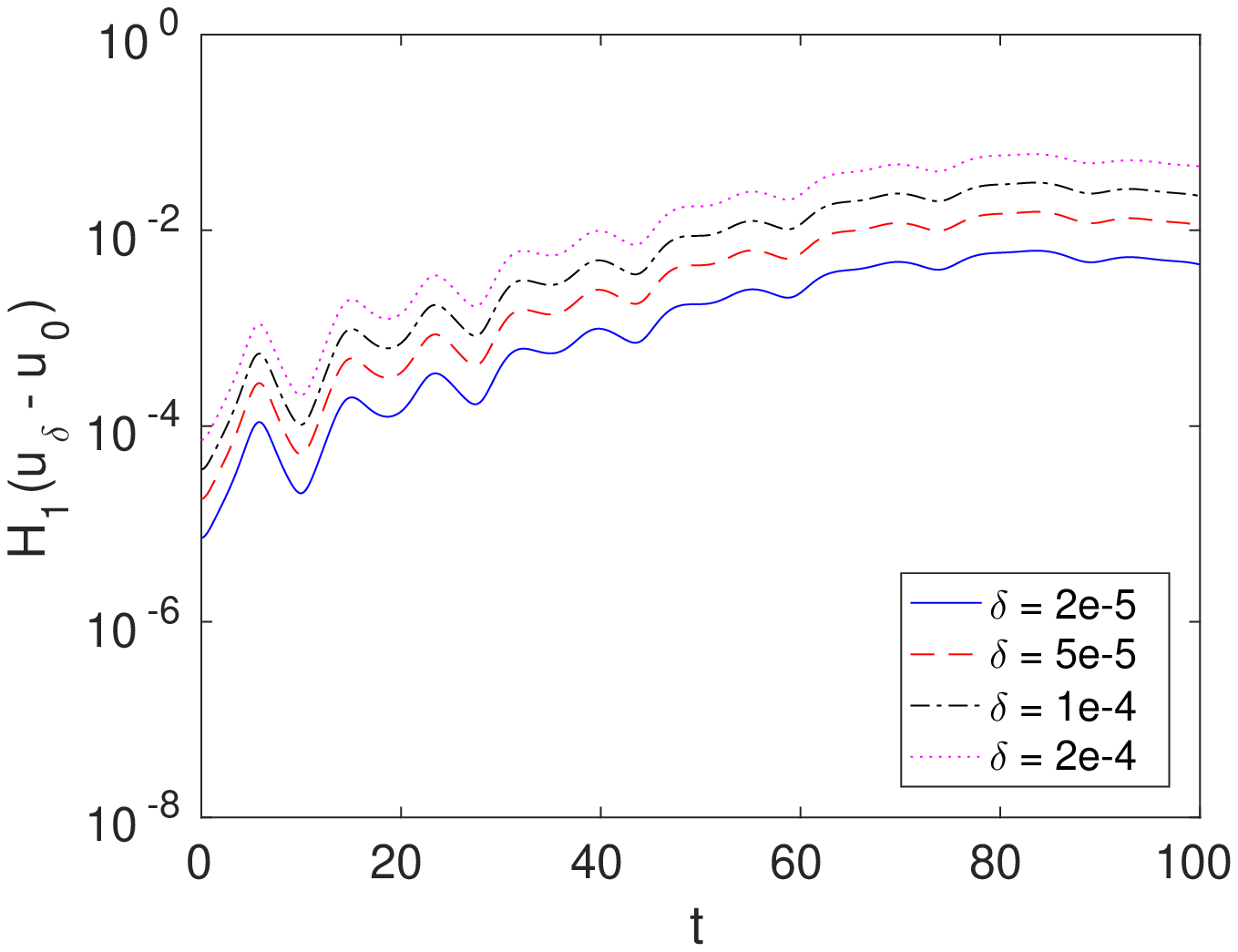}
  }
  \centerline{\hspace{.1in} F\hspace{2.5in} G}
  \caption{Two UM regime: A) $|U_{\epsilon,\gamma}^{(2)}(x,t)|$ for $0 \le t \le 100$ and Spectrum at B) $t=6$.
    C) $t=11$, D) $t= 23.4$, and E) $t= 27.4$,
  F) $t=71.9$, and
    G)  $\eta(t)$ for $f_1(x,t)$,  $\delta = 10^{-5},\dots,  10^{-4}$ and $\gamma = 0.01$.}
  \label{fig:7}
\end{figure}

{\bf 2.  $ \bf U_{\epsilon,\gamma}^{(2)}(x,t)$ in the two UM regime:}
We now consider $U^{(2)}_{\epsilon,\gamma}(x,t)$ whose
initial data is given by  Equation~\rf{SPB1} with $j =2$.
 Although $U^{(1)}(x,t)$ and $U^{(2)}(x,t)$ are both single mode SPBs over
the same Stokes wave, 
 their respective routes to stability under damping are quite different.
 Notice in Figure~\ref{fig:7}A the surface of $|U_{\epsilon,\gamma}^{(2)}(x,t)|$  for $0 \le t \le 100$ is a damped modulated traveling state,
 exhibiting regular behavior, 
 in contrast to the irregular
 behavior of $|U_{\epsilon,\gamma}^{(1)}(x,t)|$ in the two UM regime.

The spectrum of 
$U_{\epsilon,\gamma}^{(2)}$ at $t = 0$ is the same as in Figure~\ref{fig:6}B.
Under perturbation  $\lambda_2^d$ 
 immediately splits asymmetrically 
into $\lambda^{\pm}_2$ with the upper band in the right quadrant and the lower band in the left quadrant,
while the first double point $\lambda_1^d$ (indicated by the large ``$\times$'')  does not split.
Figure~\ref{fig:7}B clearly  shows that at $t = 6$ damping has only split $\lambda_2^d$,
i.e. the double point at which the SPB $U^{(2)}$ is constructed.
In fact $\lambda_1^d$ does not split for the duration of the damped HONLS evolution, $0 \le t \le 100$.
In Figure~\ref{fig:7}C, by $t = 11$, the two bands have  aligned forming a
cross state with a complex critical point at the transverse intersection of  the bands while $\lambda_1^d$ is still intact and  has simply
translated down the imaginary axis. 

The complex critical point subsequently splits  with the upper band of spectrum again in the right quadrant, Figure~\ref{fig:7}D. Complex critical points do not reappear in the spectrum.
At  $t = 27.4$ 
the vertex of the upper band of spectrum  touches the real axis
Figure~\ref{fig:7}E. As damping continues $\lambda_1^d$ moves
down the imaginary axis and the two bands move away from the imaginary axis with diminishing amplitude. In Figure~\ref{fig:7}F the complex double point
$\lambda_1^d$ has moved almost all the way down the imaginary axis. 
At $t \approx 76$, $\lambda_1^d =0$ and there are no 
complex double points in the subsequent spectral evolution.

We find $\eta(t)$ grows until $t = t_s \approx 80$,
Figure~\ref{fig:7}G,
consistent with
the expectation that $U_{\epsilon,\gamma}^{(2)}$ will stabilize once  all complex critical points and complex double points
vanish in the spectrum.   Until $\lambda^d_1$  moves onto the real axis,
  perturbations to the initial data $U_{\epsilon,\gamma,\delta}^{(2)}$ can excite 
 the first mode associated with $\lambda^d_1$  causing $U_{\epsilon,\gamma}^{(2)}$
 and $U_{\epsilon,\gamma,\delta}^{(2)}$ to grow apart.

 Why doesn't  the HONLS perturbation split  $\lambda_1^d$
 when given   $U^{(2)}(x,0)$ initial data?
 In Section~4, for short time, a suitable linearization of the  damped HONLS SPB data is found to be given by Equation~\rf{lic} for $j=1,2$, respectively where
$\tilde\epsilon = \tilde\epsilon(\epsilon, \gamma)$.
The perturbation analysis shows that at leading order
damping only asymmetrically splits the double point $\lambda_2^d$ associated with
$U^{(2)}(x,0)$.
The endpoint of spectrum, $\lambda_0^s$, decreases in amplitude
and the rest of the double points simply move along curves of continuous spectrum without splitting.
The resonant modes which correspond  to $\lambda_{2m}^d$ split asymmetrically at higher order $\mathcal{ O}(\tilde\epsilon^m)$.
The splitting of $\lambda_l^d$, $l\neq 2m$, is beyond all orders in
$\tilde\epsilon$ and is termed
``closed''.

The spectral evolution for $U_{\epsilon,\gamma}^{(2)}(x,t)$ in the 2 UM regime is reminiscent of the spectral evolution of 
$U_{\epsilon,\gamma}^{(1)}(x,t)$ in the one UM regime. There is an 
important difference though: The nearby cross state that appears in the
spectral decomposition of $U_{\epsilon,\gamma}^{(2)}(x,t)$
has two different types of instabilites: the instability associated with the complex critical point (potentially a phase instability)
and the exponential instability associated with the (nonresonant) complex double point $\lambda_1^d$.
The numerical results suggest that when  nonresonant modes are present in the damped HONLS,
their instabilities persist and organize the dynamics on a longer time scale.
 As $\lambda_1^d$ remains effectively closed for $t < t_s$, 
 $U_{\epsilon,\gamma}^{(2)}(x,t)$ can be characterized as a continuous deformation
 of a noneven generalization of the
 degenerate 3 phase solution given by Equation~\rf{3phase}
 (the parameter values change due to doubling the period $L$; i,e. a $j-$th mode excitation with period L becomes a $2j-$th mode excitation with period 2$L$.)

\begin{figure}[ht!]
  \centerline{
\includegraphics[width=.5\textwidth,height=1.5in]{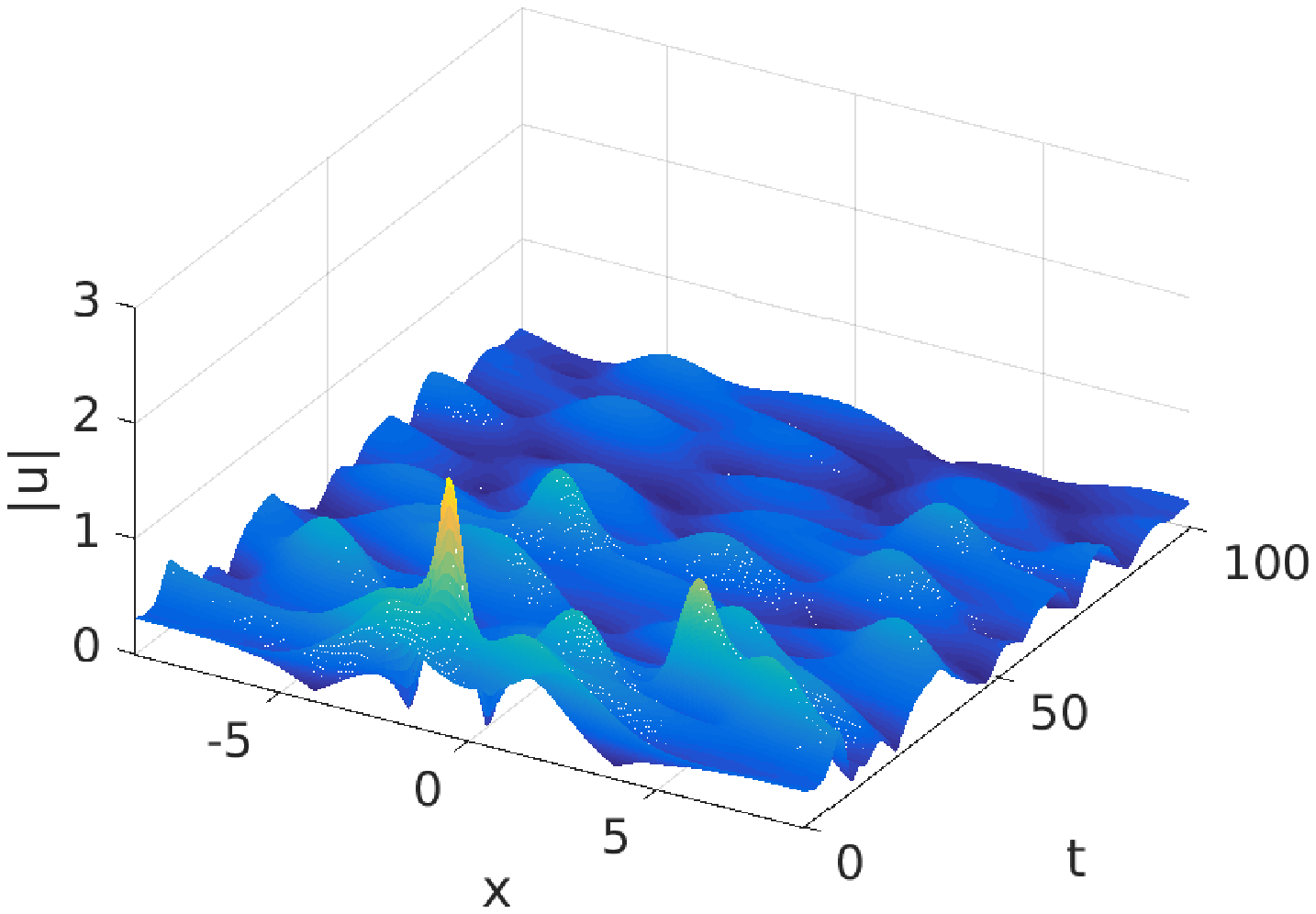}
\includegraphics[width=.33\textwidth,height=1.2in]{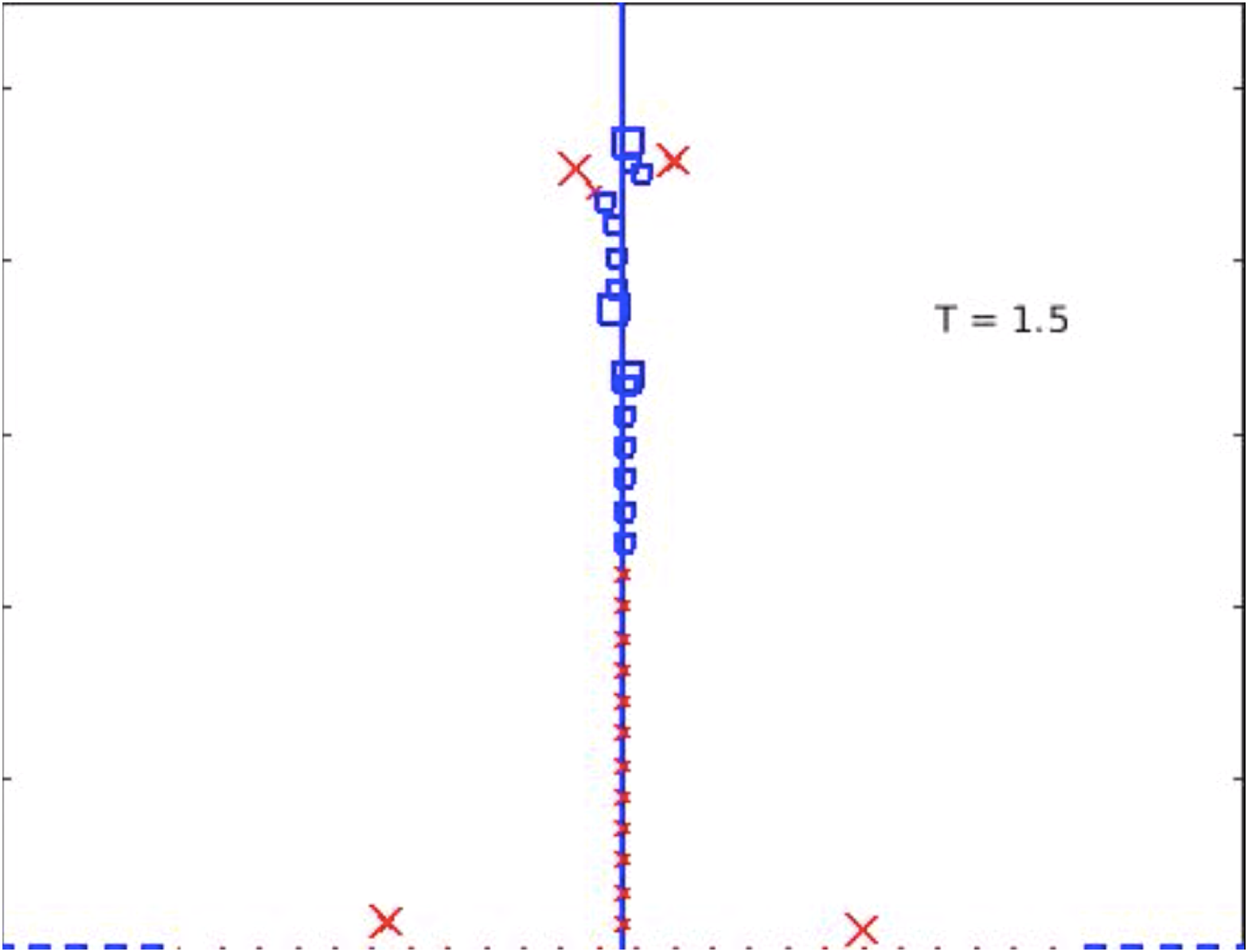}
  }
  \centerline{\hspace{.1in} A\hspace{2.5in} B}
  \vspace{12pt}
  \centerline{
\includegraphics[width=.33\textwidth,height=1.2in]{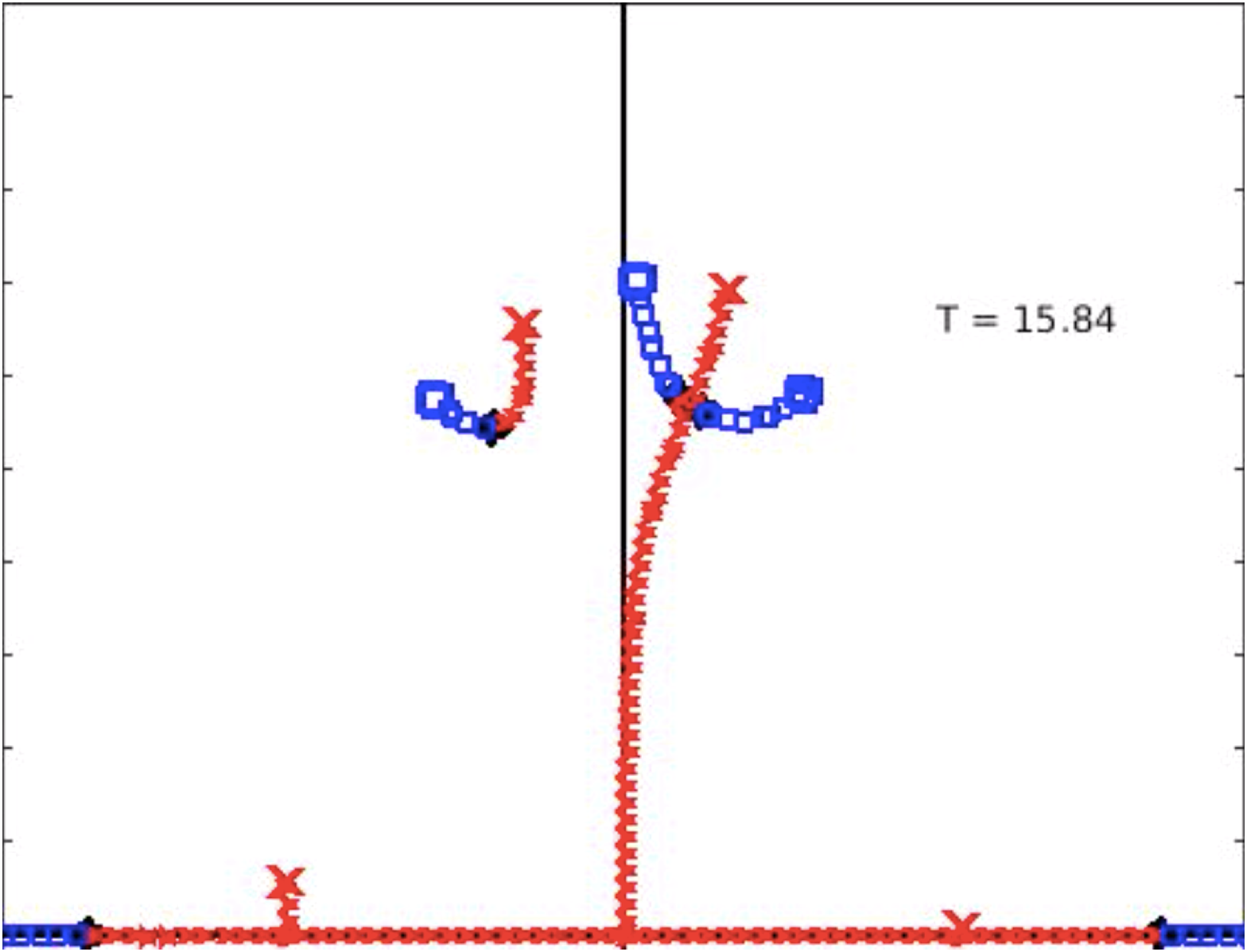}
\includegraphics[width=.33\textwidth,height=1.2in]{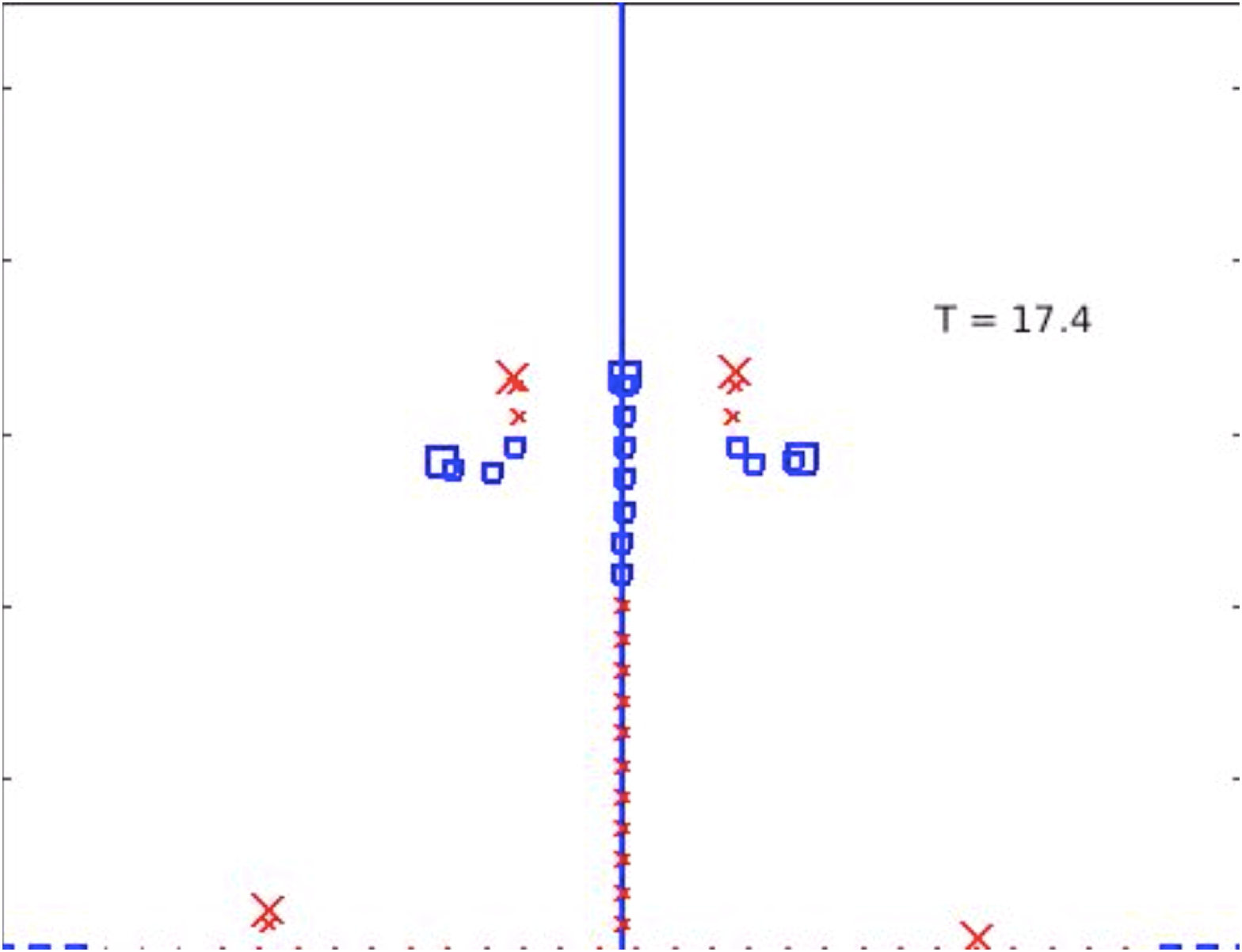}
\includegraphics[width=.33\textwidth,height=1.2in]{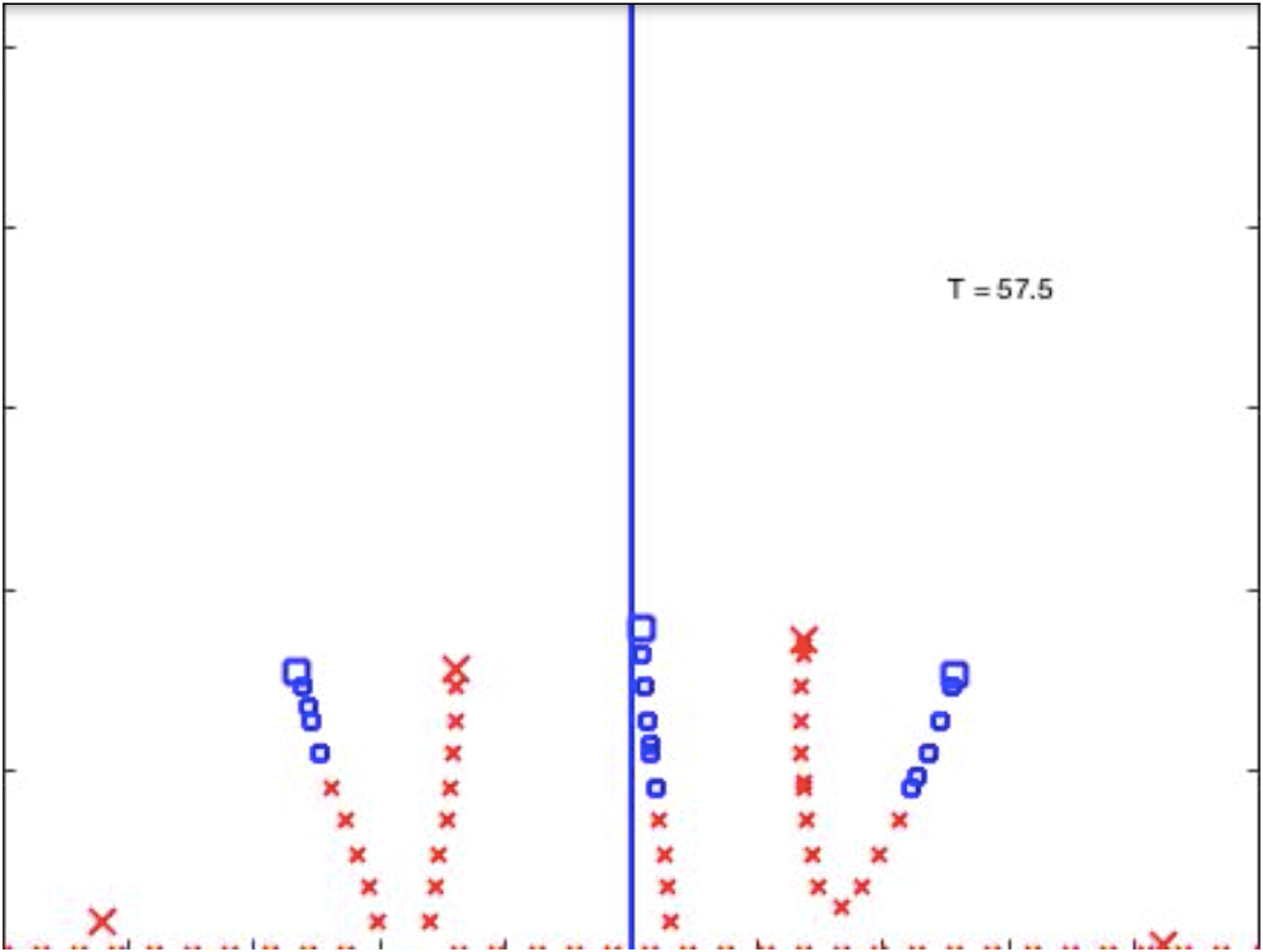}
}
 \centerline{C\hspace{2.5in} D\hspace{2.5in} E}
  \vspace{12pt}
  \centerline{
\includegraphics[width=.33\textwidth,height=1.2in]{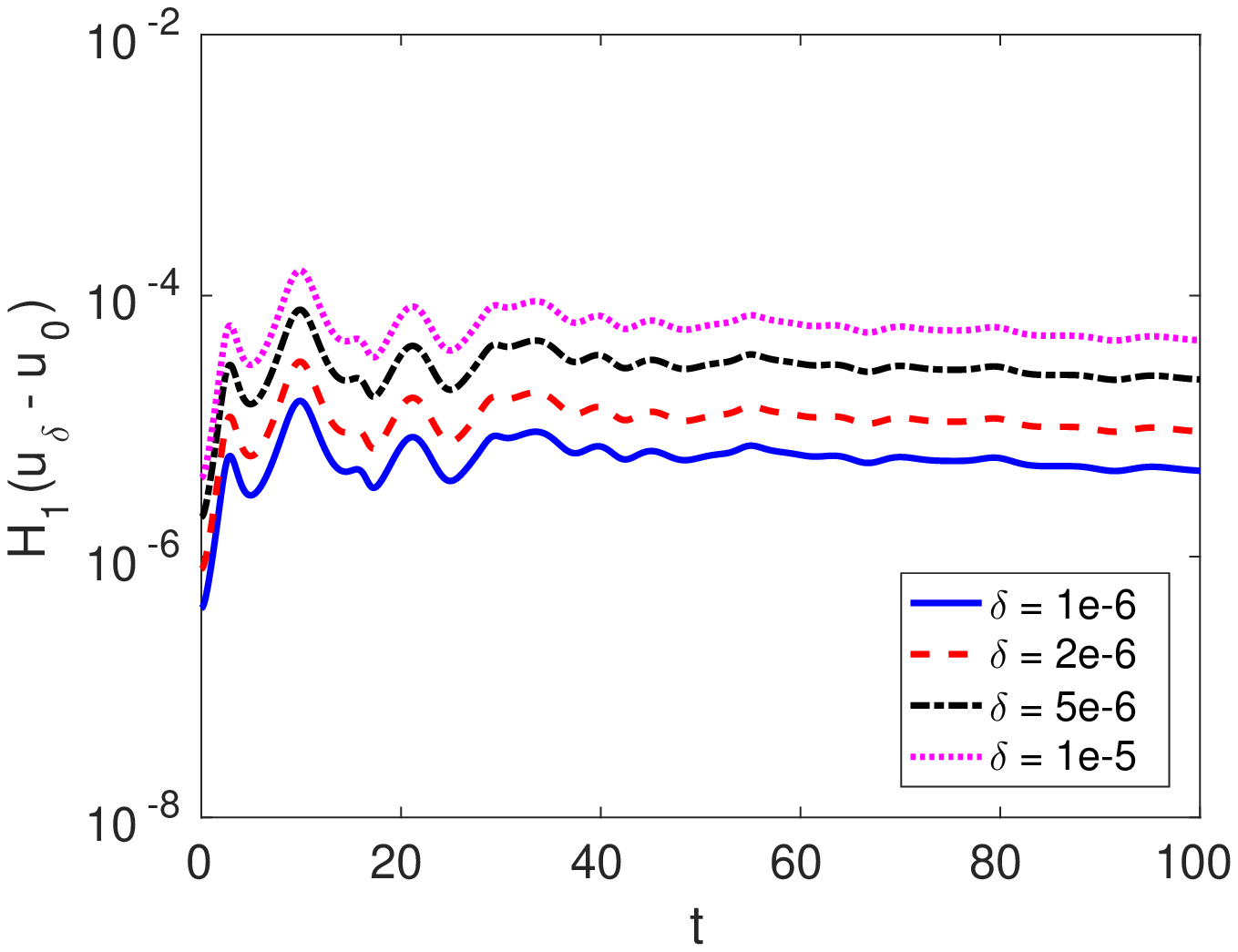}
\includegraphics[width=.33\textwidth,height=1.2in]{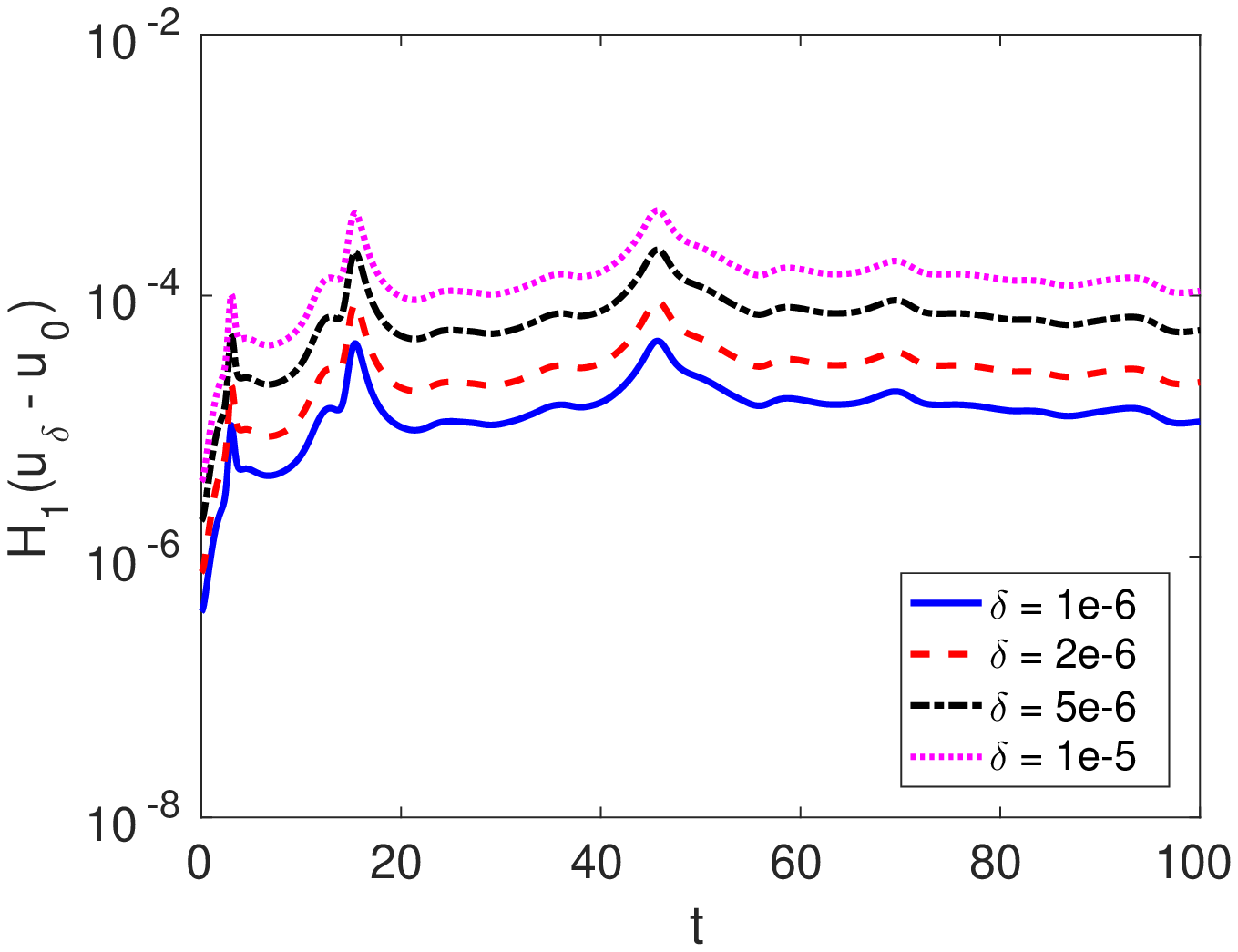}
  }
  \centerline{\hspace{.1in} F\hspace{2.5in} G}
  \caption{Two UM regime: A) $|U_{\epsilon,\gamma}^{(1,2)}(x,t)|$ for $0 \le t \le 100$ and spectrum at B) $t=1.5$,
    C) $t=15.9$, D) $t=17.4$, E) $t=55$, and
    $\eta(t)$ for $U_{\epsilon,\gamma}^{(1,2)}(x,t)$ with $f_1(x,t)$ 
      when 
    F)  $\rho = 0, \tau = -2$ (the uncoalesced SPB),
    G)  $\rho = 0.665, \tau = 1$ (the coalesced SPB).
  }
  \label{fig:8}
\end{figure}

{\bf 3. $\bf U^{(1,2)}_{\epsilon,\gamma}(x,t)$ in the two UM regime:}
Figure~\ref{fig:8}A  shows the surface
 $|U_{\epsilon,\gamma}^{(1,2)}(x,t)|$ for $ 0 < t < 100$
for initial data 
obtained from Equation~\rf{SPB2} by setting 
$\rho = 0$ and $\tau = -2$, $a = 0.5$.
The spectrum 
of $U_{\epsilon,\gamma}^{(1,2)}$ at $t = 0$ is given by  Figure~\ref{fig:6}B.
As we've seen in the previous examples, once complex double points split they
do not reform in the perturbed system; if they are present in the spectral decomposition of the damped HONLS, it is because the  modes corresponding
to $\lambda_j^d$ didn't resonate  under perturbation.
Here both  double points $\lambda^d_1$ and $\lambda^d_2$  split at leading order
under the damped HONLS perturbation.
Figure~\ref{fig:8}B shows   for short time ($t = 1.5$)
the spectrum 
 has a band gap structure indicating the first mode travels to the right and the second travels to the left.
 As time evolves the bands shift and align and Figure~\ref{fig:8}C 
 shows at $t \approx 15.9$ 
 two bands in the right quadrant intersect with an embedded complex critical point. 
 The critical point splits with one band moving back towards the imaginary axis.
 There are now two bands detached from the
primary band  and
moving downwards towards the real axis, Figures~\ref{fig:8}D-E. 
Complex critical points do not appear for $t > 15.9$.

Figure~\ref{fig:8}F shows $\eta(t)$  for the example under consideration
($\rho = 0, \tau = -2$) when $f_1(x,t)$,$\delta = 10^{-5},\dots,  10^{-4}$.
 We find $\eta(t)$ saturates at $ t_s \approx 20$
indicating $U_{\epsilon,\gamma}^{(1,2)}$ has  stabilized.
For $t>t_s$ $U_{\epsilon,\gamma}^{(1,2)}$ is characterized as a continuous deformation of a stable NLS five-phase solution.
The numerically observed initial splitting of complex double points $\lambda_1^d$ and $\lambda_2^d$ is
consistent with the perturbation analysis of damped HONLS  SPB data (\ref{lic}).

Among the  two mode SPBs in the 2 UM regime, the one of highest amplitude due to coalescence of the modes
 appeared to be more robust \cite{cs14}.
 An interesting observation is obtained if we examine the
 evolution of spectrum and of $\eta(t)$ using the initial data for
 the special coalesced two mode  SPB,
 generated from Equation~\rf{SPB2} by setting 
 $\rho = 0.665$,  $\tau = 1$, and $a = 0.5$. For the coalesced case,
 complex critical points form 4 times  for $0 < t < 45$
 in the damped HONLS system.
Figure~\ref{fig:8}G shows  $\eta(t)$ saturates for $t \approx 50$. A comparison
 with the results of the  non-coalesced two mode SPB given above indicate that
 remnants of the  coalesced $U^{(1,2)}$ and it's instabilities 
  influence the
 damped HONLS dynamics over a longer time period,  suggesting enhanced robustness with respect to perturbations of the NLS equation.

 \subsection{Damped SPBs in the three unstable mode regime}
The parameters  used for the three UM regime are $a = 0.7$ and 
$L = 4\sqrt{2}\pi$ 
($N = \displaystyle \lfloor {aL}/{\pi} \rfloor = 3$).
The damped HONLS perturbation parameters are $\epsilon = 0.05$ and
$\gamma = 0.01$.
We present the results of two damped HONLS SPBs, $U_{\epsilon,\gamma}^{(2)}(x,t)$ and 
$U_{\epsilon,\gamma}^{2,3}(x,t)$, which exhibit an interesting or  new feature. 
The evolutions of the other  damped HONLS SPBs in the
three UM regime  in the three UM regime are discussed in relation to these
cases.

\begin{figure}[ht!]
\centerline{
\includegraphics[width=.5\textwidth,height=1.5in]{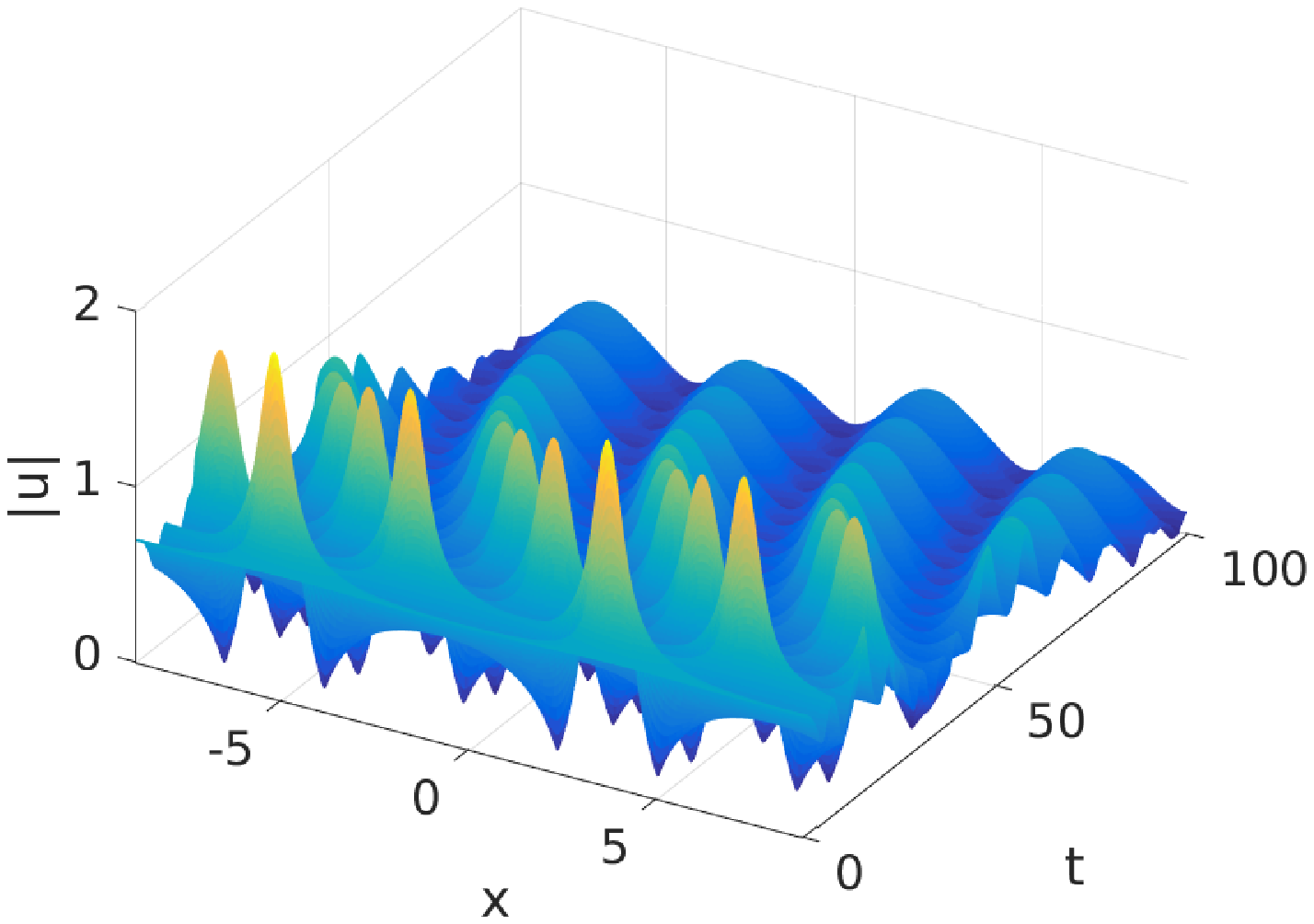}
\hspace{12pt}
\includegraphics[width=.33\textwidth,height=1.125in]{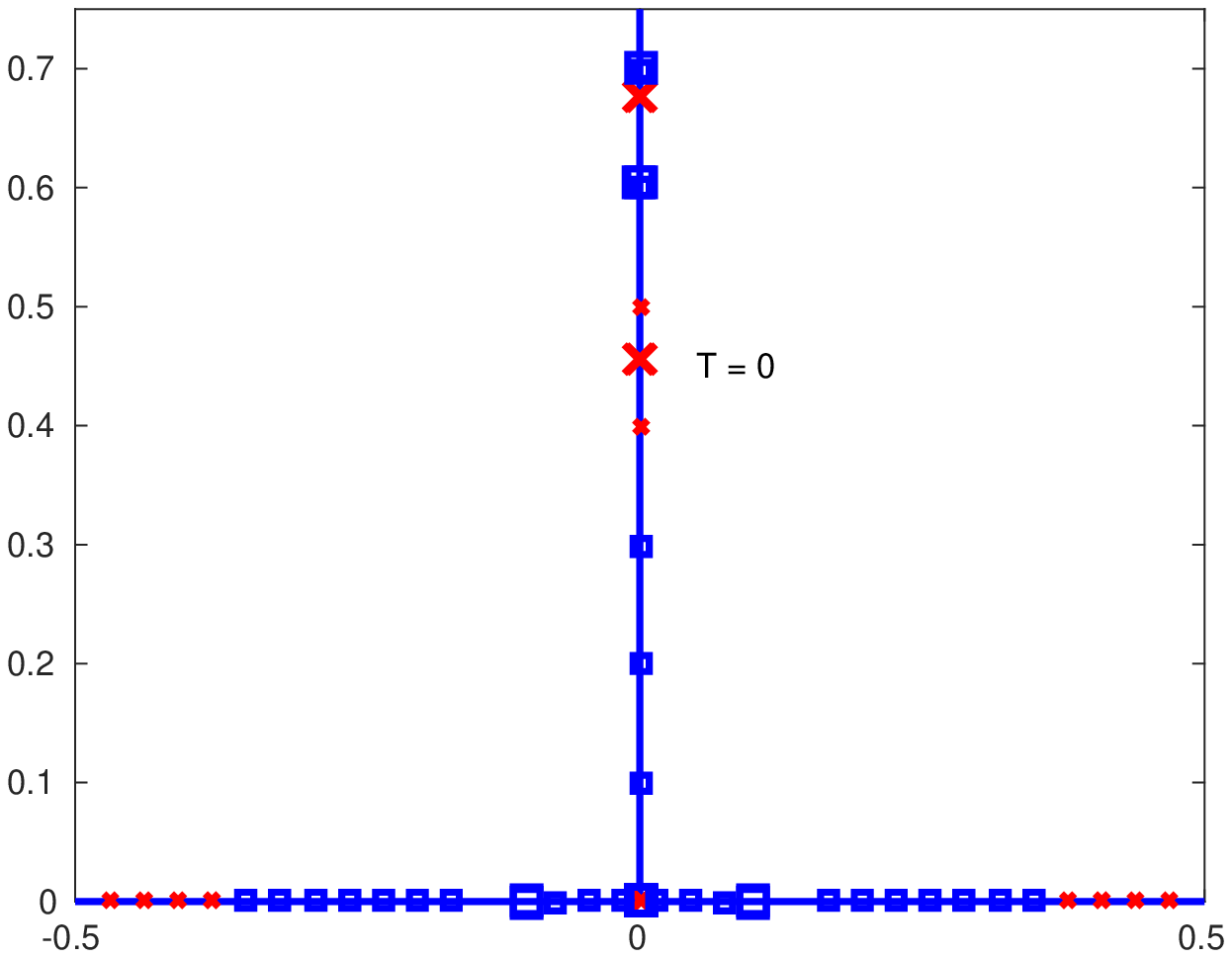}
}
  \centerline{\hspace{.1in} A\hspace{2.5in} B}
  \vspace{12pt}
\centerline{
\includegraphics[width=.33\textwidth,height=1.125in]{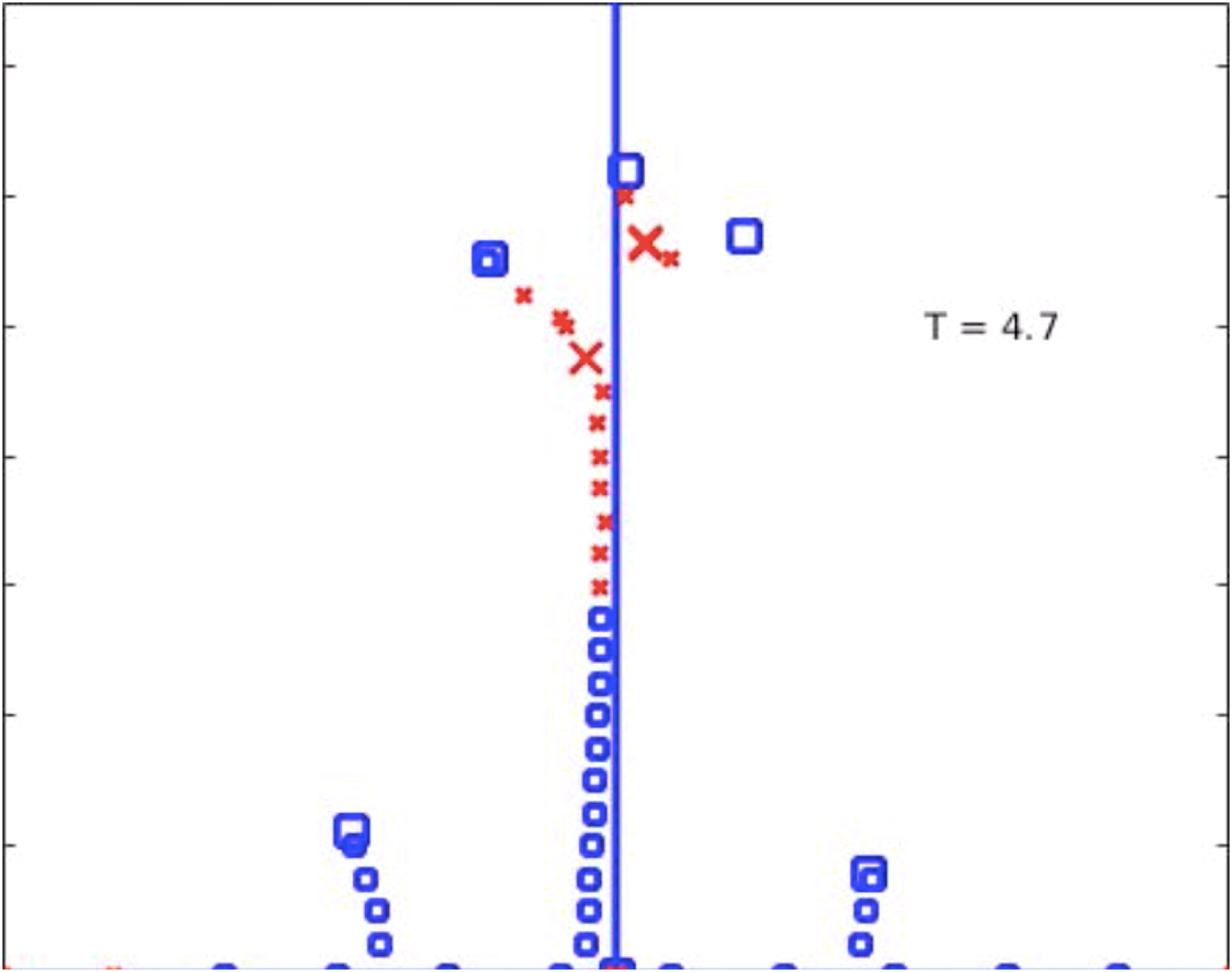}
\includegraphics[width=.33\textwidth,height=1.125in]{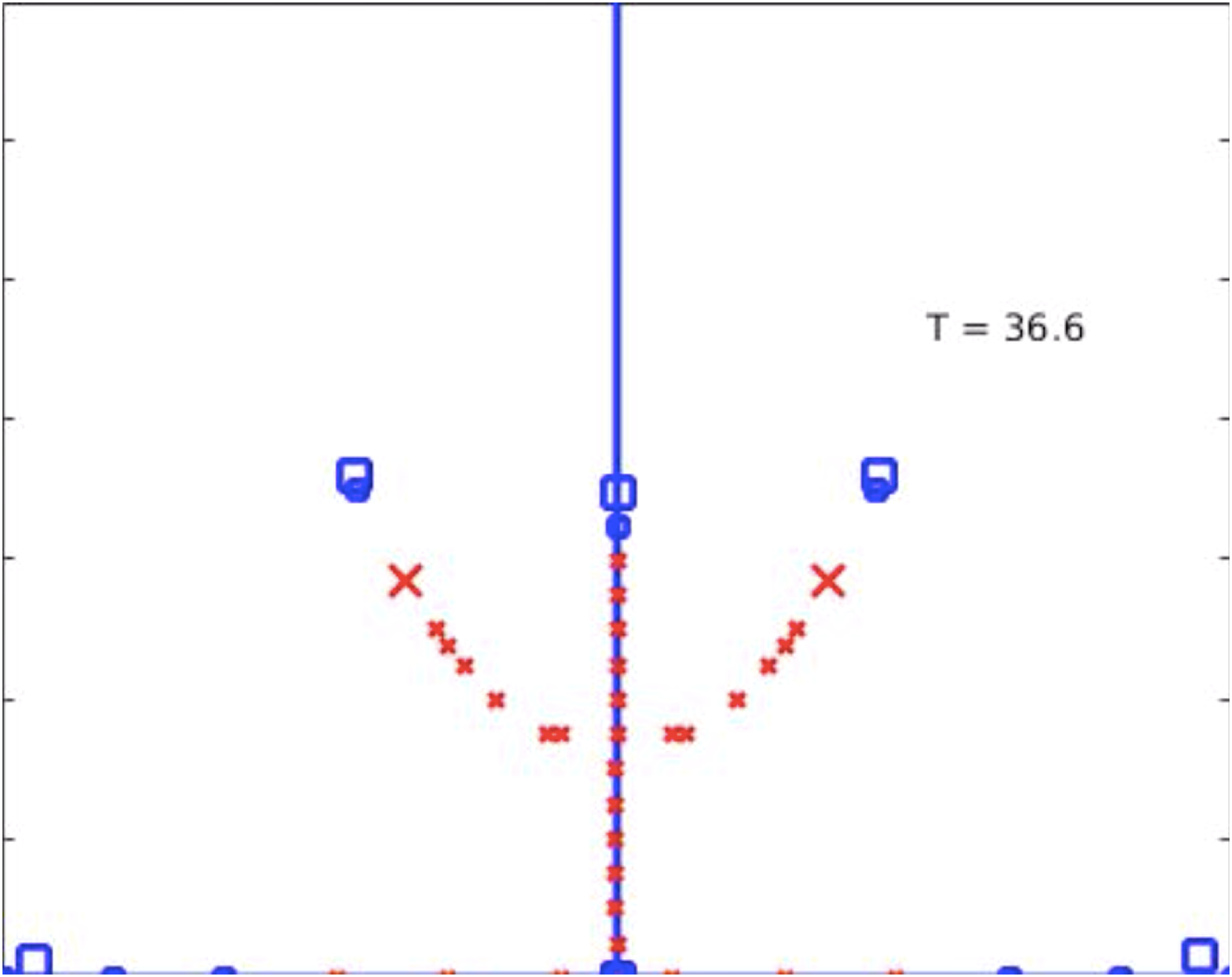}
}
  \centerline{\hspace{.1in} C\hspace{2.5in} D}
  \vspace{12pt}
\centerline{
\includegraphics[width=.33\textwidth,height=1.125in]{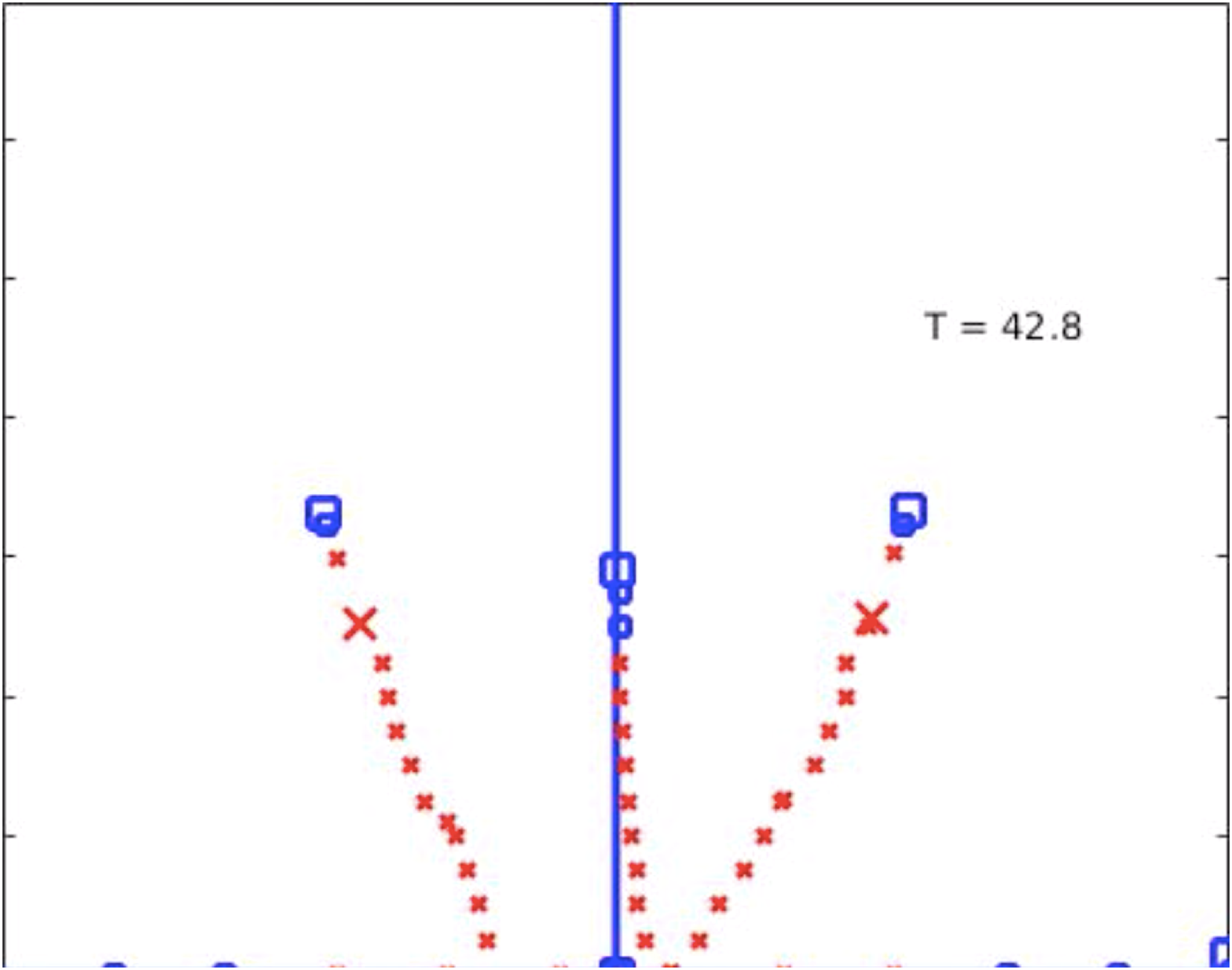}
\includegraphics[width=.33\textwidth,height=1.125in]{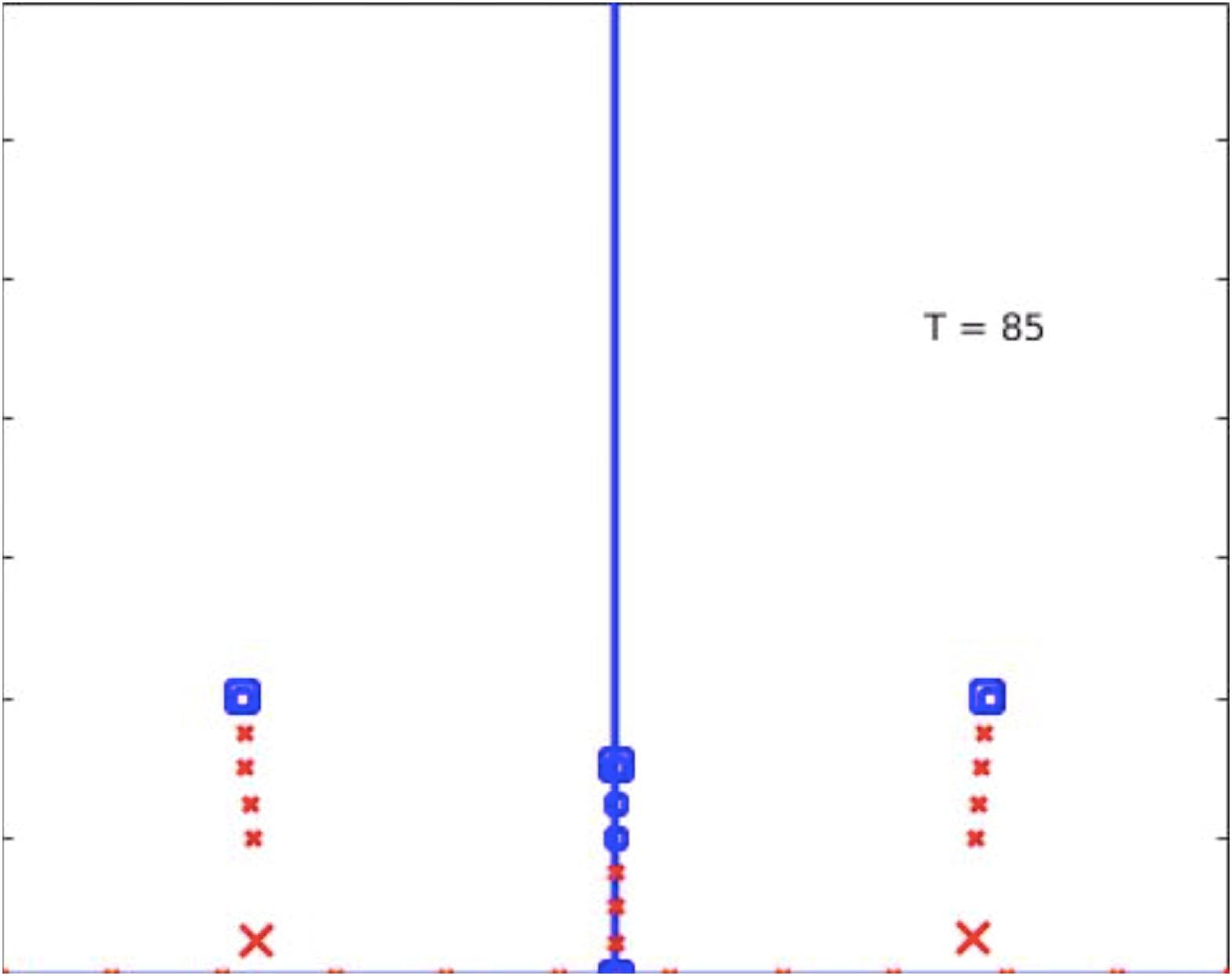}
\includegraphics[width=.33\textwidth,height=1.125in]{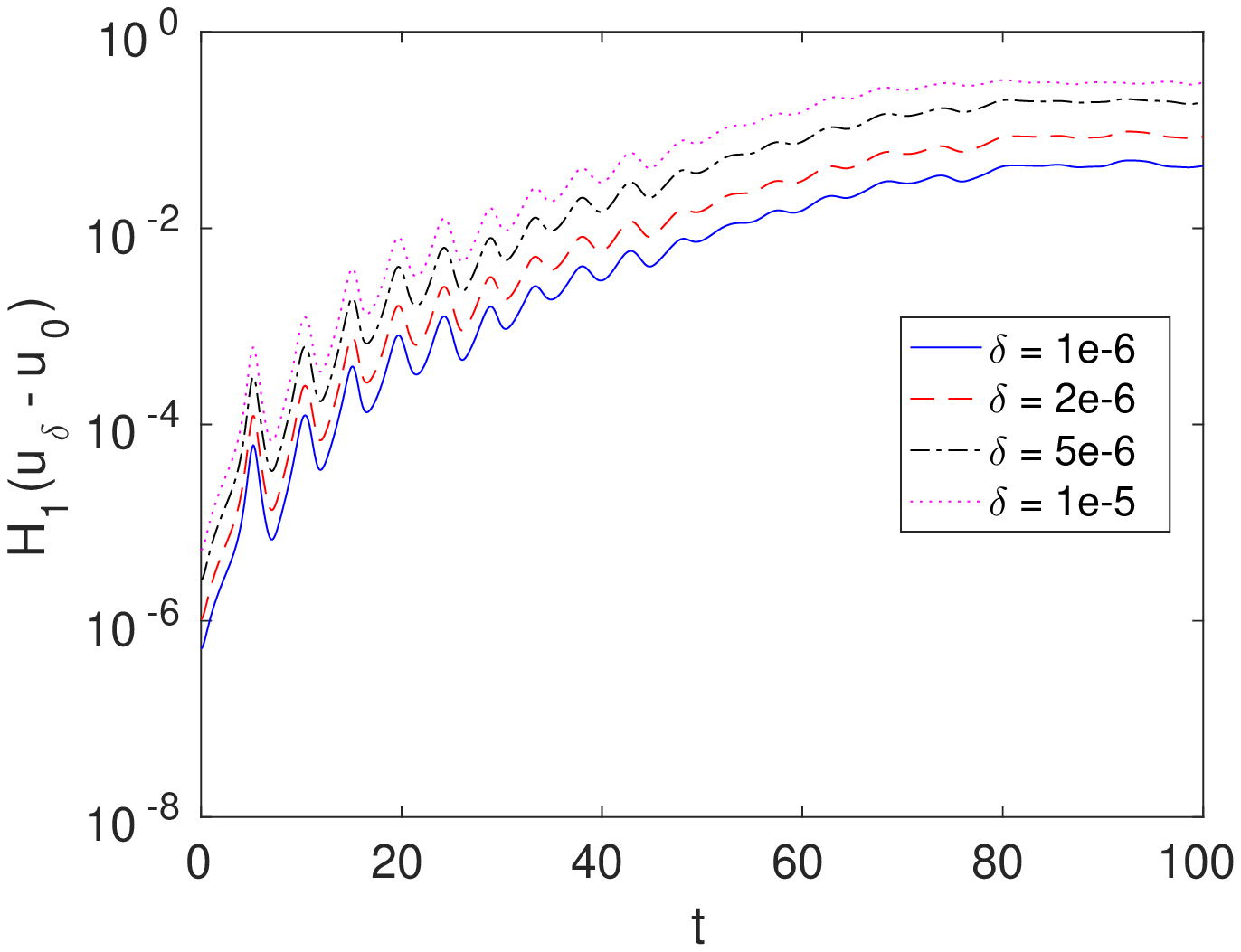}
}
 \centerline{E\hspace{2.5in} F\hspace{2.5in} G}
 \caption{Three  UM regime: A) $|U_{\epsilon,\gamma}^{(2)}(x,t)|$ for $0 \le t \le 100$ and Spectrum at B) $t=0$,
   C) $t = 4.7$,  D) $t=36.6$,
 E) $t=42.8$, F) $t = 85$ and G) $\eta(t)$ for $f_1(x,t)$,  $\delta = 10^{-6},\dots,  10^{-5}$.}
 \label{fig:9}
\end{figure}

{\bf 1. $\bf U_{\epsilon,\gamma}^{(2)}(x,t)$ in the three UM regime:}
The surface  $|U_{\epsilon,\gamma}^{(2)}(x,t)|$
 for initial data given by Equation~\rf{SPB1} with $j =2$ is shown in Figure~\ref{fig:9}A
 for $ 0 < t < 100$. Notice in the 3 UM regime  $U_{\epsilon,\gamma}^{(2)}(x,t)$ 
 exhibits regular behavior and is a damped  modulated right traveling wave
 as was $U_{\epsilon,\gamma}^{(2)}(x,t)$
 in the 2 UM
regime. 
The spectrum  at $t = 0$ is given in Figure~\ref{fig:9}B.
The end point of the band of spectrum, $\lambda_0^s = 0.7i$ is indicated by a ``box''.
There are three  complex double points  at $\lambda_1^d = 0.677i$,  $\lambda_2^d = 0.604 i$,
and  $\lambda_3^d = 0.456 i$
indicated by an  ``$\times$'', ``box'' and ``$\times$'', respectively.
Under the damped HONLS perturbation the complex double point
at which $U_{\epsilon,\gamma}^{(2)}(x,t)$ is constructed,
$\lambda_2^d$,  splits  into a right  state 
as shown in Figure~\ref{fig:9}C.
The complex double points $\lambda_1^d$ and
$\lambda_3^d$ remain closed; 
$\lambda_1^d$ lies on the upper band  in the right quadrant
and $\lambda_3^d$ lies on the lower  band.
Transverse cross states with embedded complex critical points form frequently
in the spectrum until $t \approx 68$, e.g. a cross state is shown
at $t = 36.6$ in Figure~\ref{fig:9}D. 
Figures~\ref{fig:9}E-F show the complex double points persist on the bands of spectrum until damping sufficiently diminishes the amplitude of the background
and the complex double points reach the real axis 
at $t  \approx 85 $. In Figure~\ref{fig:9}G $\eta(t)$ saturates at $t = t_s \approx 90$.
Due to the presence of
the complex double points for $t < t_s$,
$U_{\epsilon,\gamma}^{(2)}(x,t)$  can be viewed as a continuous deformation of an
unstable 3 phase solution (with two instabilities).
As discussed previously, the instabilities associated with the nonresonant modes persist longer than for the resonant modes. 

In the 3 UM regime the behavior of the SPB  $U_{\epsilon,\gamma}^3(x,t)$ is  similar to $U_{\epsilon,\gamma}^2(x,t)$.
In this case $\lambda_3^{d}$ initially splits asymmetrically  into the right state (which
we've now seen  frequently in the initial damped HONLS system when only one mode is activated).
The double points
$\lambda_1^d$ and $\lambda_2^d$ do not split,
they move along the 
band of spectrum created by $\lambda_0^s$ and $\lambda_3^{+}$.
As a result $U_{\epsilon,\gamma}^3(x,t)$ stabilizes only when
$\lambda_1^d$ and $\lambda_2^d$ become real,  at $t \approx 140$.
As in the previous cases, it is striking that 
the prediction from a short time perturbation analysis that certain double points remain closed,
holds for the duration of the experiments (even while the solution evolves
as  a perturbed degenerate  3-phase state (with two instabilities).
In contrast, for $U_{\epsilon,\gamma}^1(x,t)$
the higher order nonlinearities and damping excite all the modes. The solution is characterized by the formation of complex critical points and irregular
behavior before stabilizing at $t \approx 40$.

\begin{figure}[ht!]
 \centerline{
\includegraphics[width=.5\textwidth,height=1.5in]{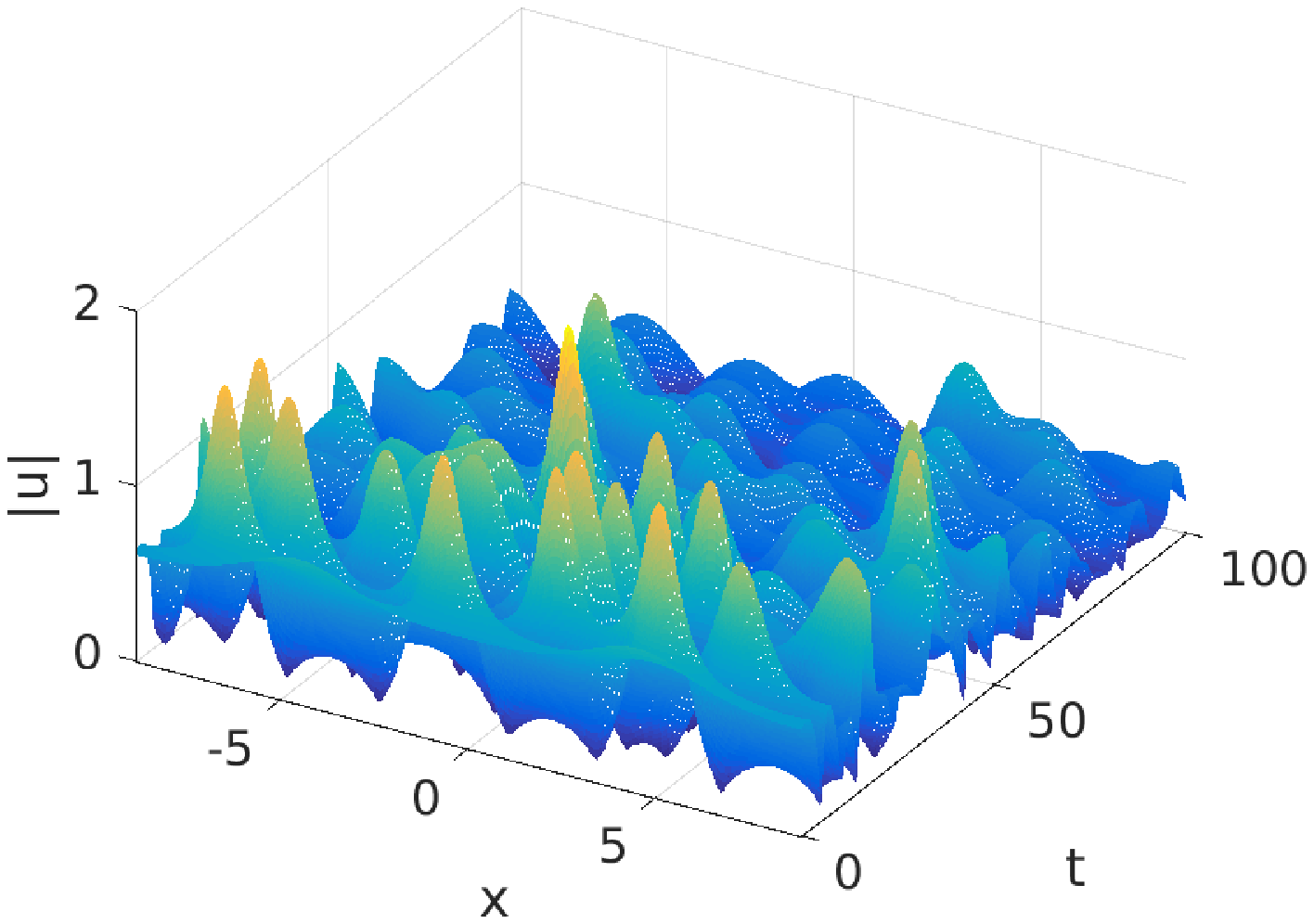}
\hspace{12pt}
   \includegraphics[width=.33\textwidth,height=1.125in]{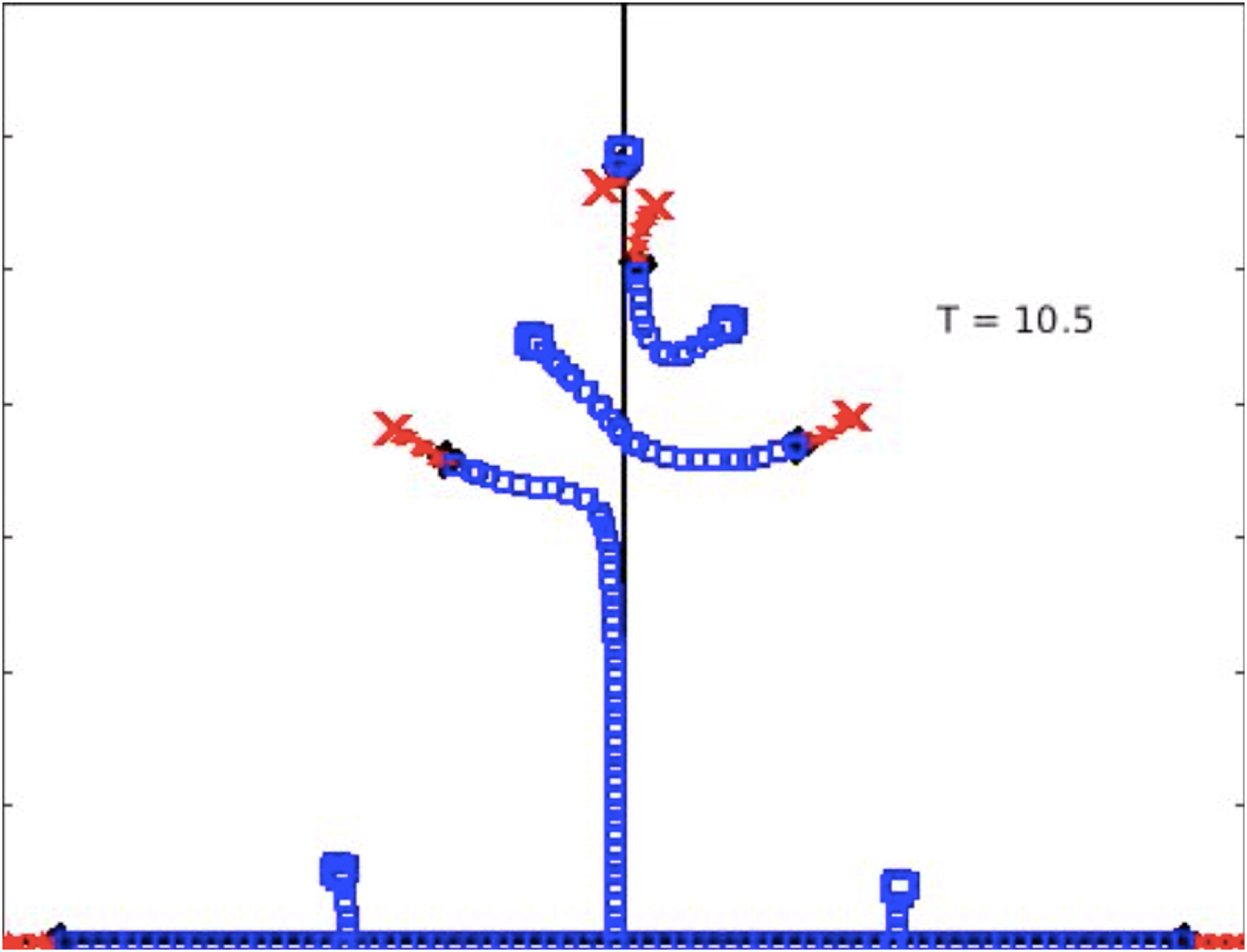}
}
  \centerline{\hspace{.1in} A\hspace{2.5in} B}
  \vspace{12pt}
 \centerline{
   \includegraphics[width=.33\textwidth,height=1.125in]{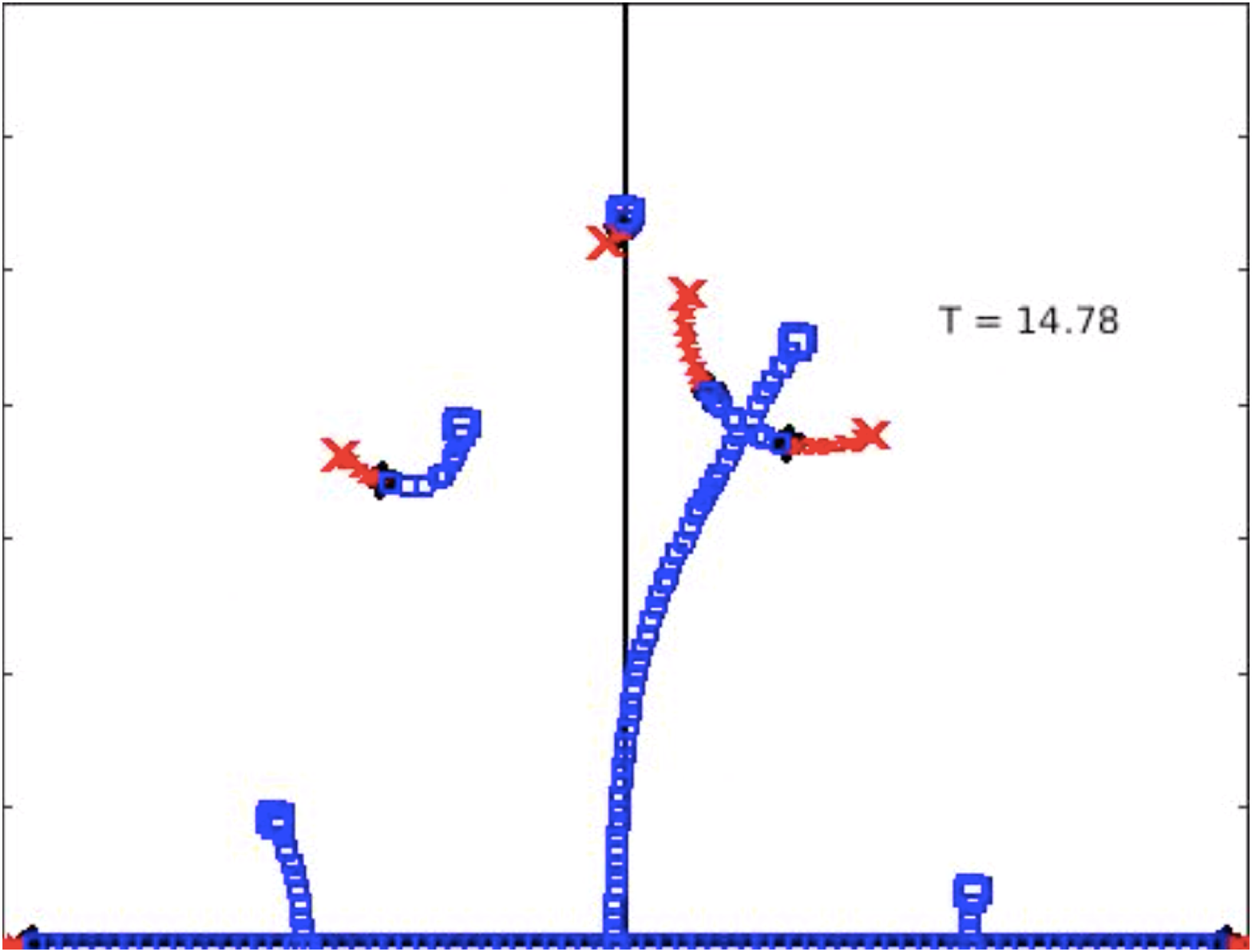}
   \includegraphics[width=.33\textwidth,height=1.125in]{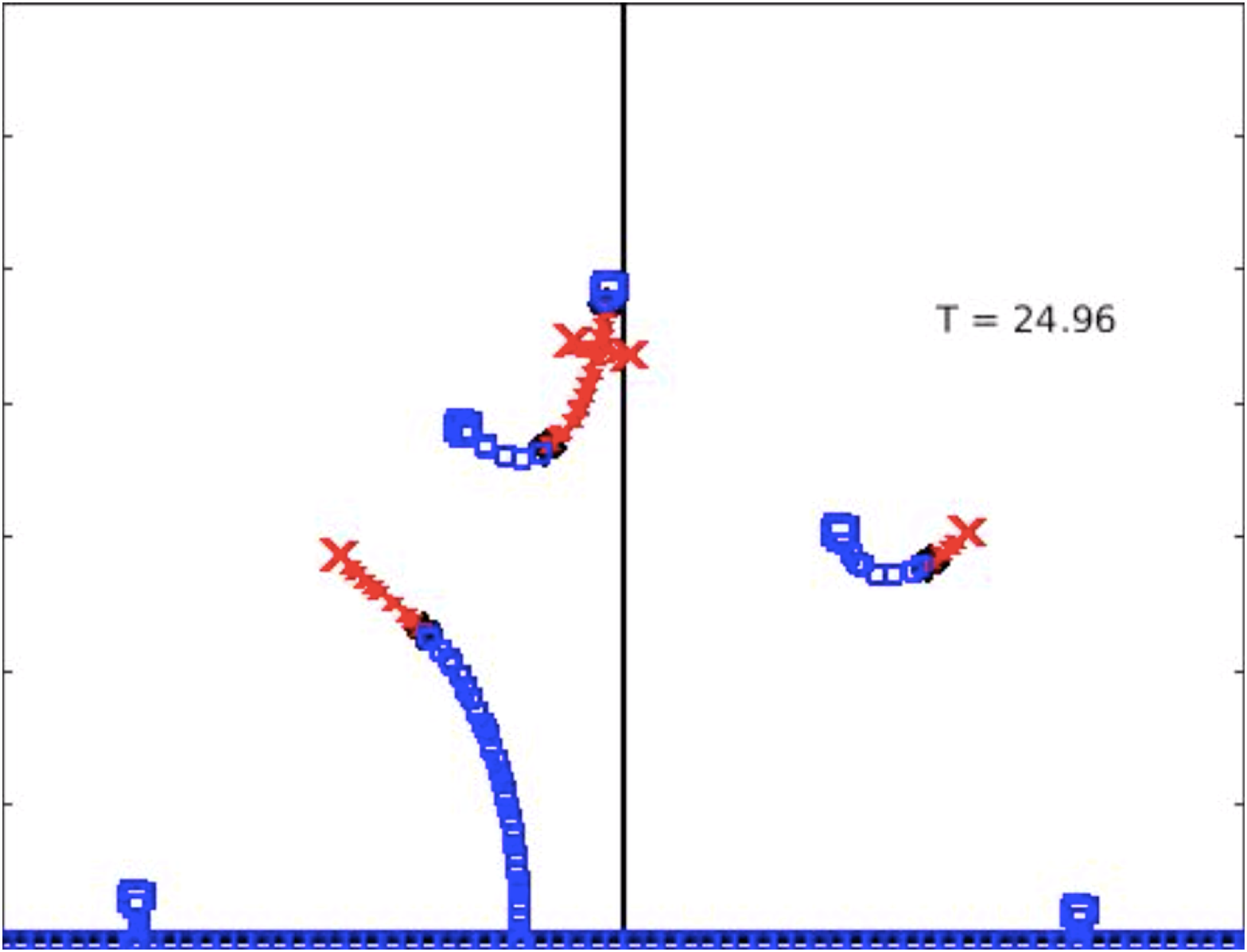}
   \includegraphics[width=.33\textwidth,height=1.125in]{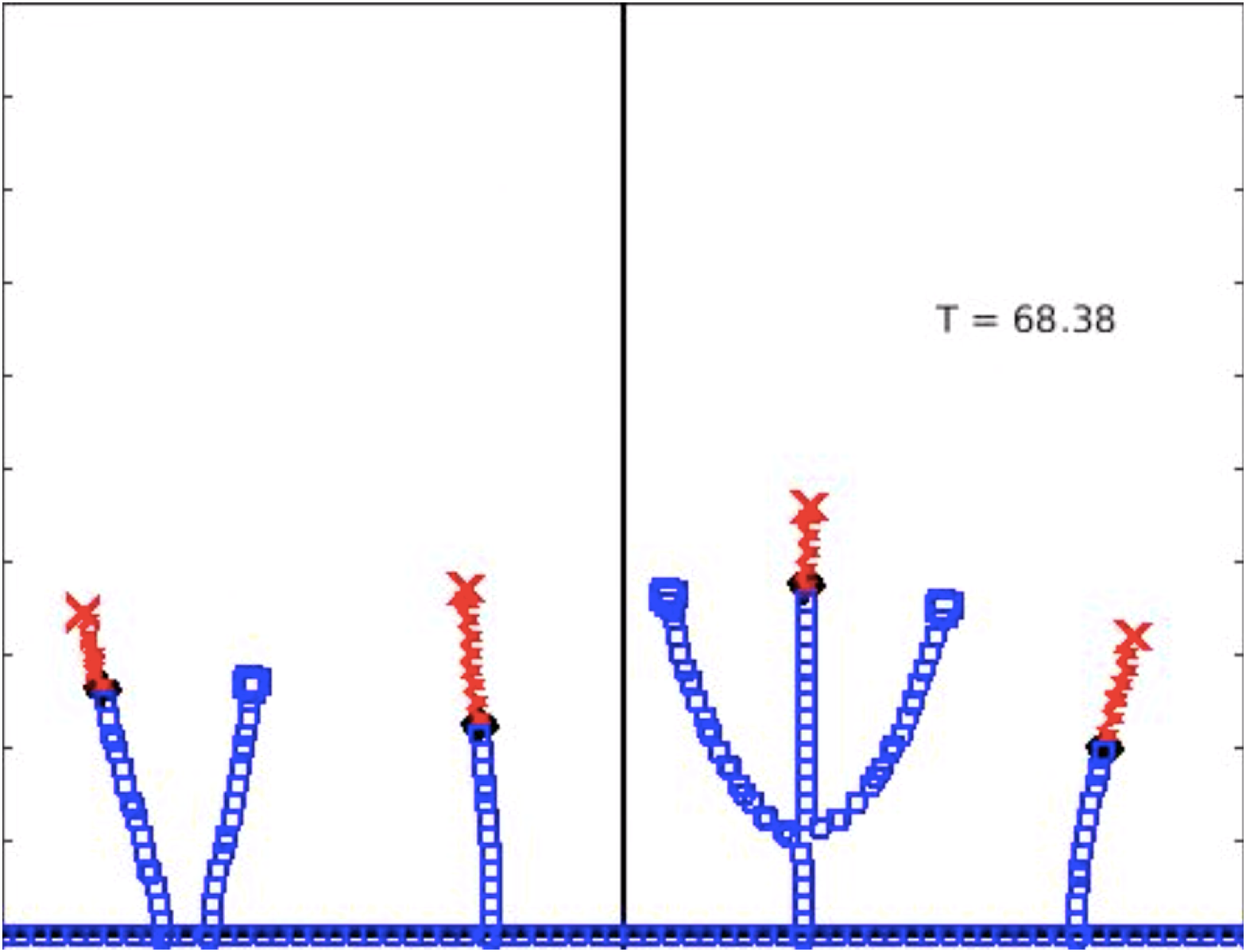}
 }
 \centerline{C\hspace{2.5in} D\hspace{2.5in} E}
  \vspace{12pt}
 \centerline{
\includegraphics[width=.45\textwidth,height=1.75in]{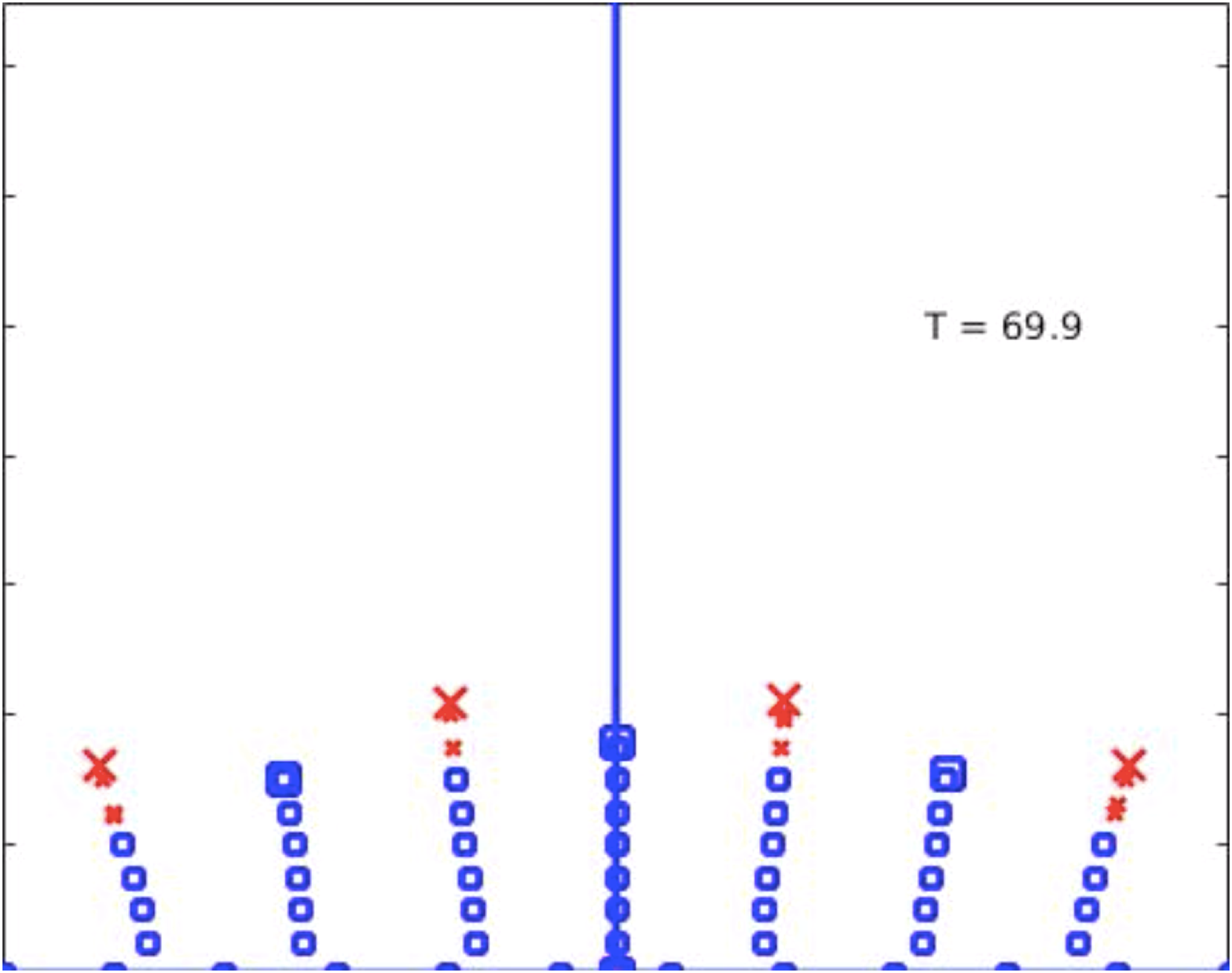}
\includegraphics[width=.45\textwidth,height=1.75in]{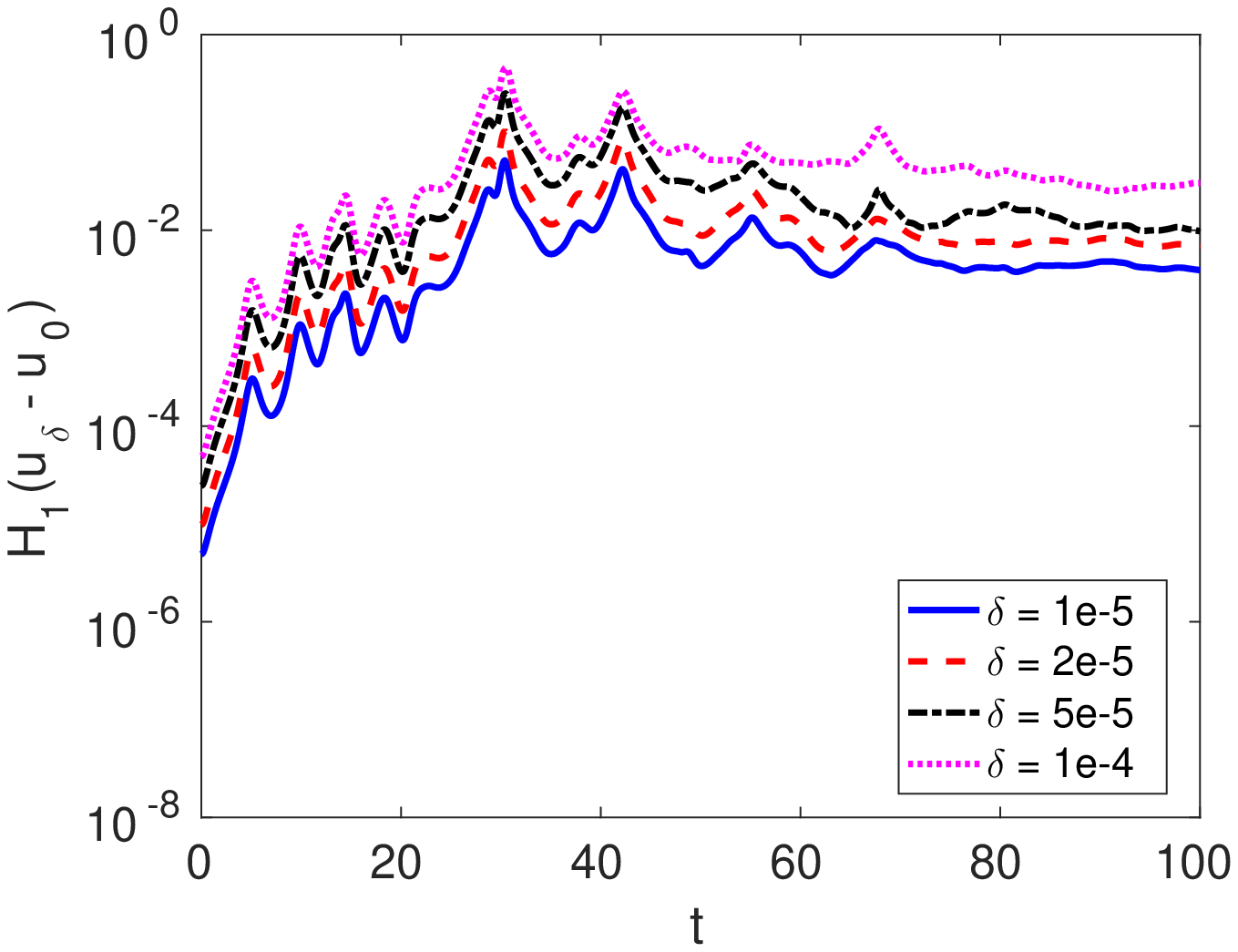}
 }
  \centerline{\hspace{.1in} F\hspace{2.5in} G}
 \caption{Three  UM regime: A) $|U^{(2,3)}(x,t)|$ and Spectrum at B) $t=10.5$,
   C) $t = 14.78$, D) $t = 24.96$, E) $t = 68.38$,
F) $t=69.9$, and G) $\eta(t)$ for $f_1(x,t)$,  $\delta = 10^{-5},\dots,  10^{-4}$ and $\gamma = 0.01$.}
 \label{fig:10}
 \end{figure}

{\bf 2. $\bf U_{\epsilon,\gamma}^{(2,3)}(x,t)$ in the three UM regime:}
 Figure~\ref{fig:10}A shows the surface  $|U_{\epsilon,\gamma}^{(2,3)}(x,t)|$
 for $ 0 < t < 100$ for initial data given by Equation~\rf{SPB2} with $i,j =2,3$. 
The spectrum  at $t = 0$ is as  in Figure~\ref{fig:9}B.
The perturbation initially splits the double points $\lambda_2^d$ and $\lambda_3^d$ into 
$\lambda_2^\pm$ and $\lambda_3^\pm$ that correspond to a left and right modulated traveling modes, respectively. The new feature here is that the first complex double point,  $\lambda_1^d$, splits at higher order into $\lambda_1^\pm$ (see the analysis in Section 4 showing that a
multi-mode perturbation in a higher UM regime introduces new resonances not seen with single mode perturbations).
The higher order splitting is  visible in Figure~\ref{fig:10}B at $t = 10.5$.

Complex double points are not observed in the spectral evolution for $t>0$.The formation of complex critical points in the spectrum occurs frequently  as shown, for example, in Figure~\ref{fig:10}C and Figure~\ref{fig:10}D. 
Since the amplitude of the background state at $a = 0.7$ is initially very close to the 4 UM regime, in this example we observe that nearby real double points are noticeably split by the perturbation.
Figure~\ref{fig:10}E shows the spectrum at $t = 68.4$ when the last  complex critical point  forms.  
This is
reflected in Figure~\ref{fig:10}G which shows  $\eta(t)$ saturates at
$t_s \approx 68$. Each of the bursts of growth in $\eta(t)$ can be correlated with a complex critical point crossing. As time evolves disspiation diminshes the strength of the instability captured by the complex critical points or complex double points.  $U_{\epsilon,\gamma}^{(2,3)}(x,t)$ exhibits quite rich and compex dynamics before damping saturates the instabilities and it's behavior is not easy to characterize as when dealing with the perturbed SPBs in the $N = 1,2$ UM regimes. For $t > t_s$ the evolution of $U_{\epsilon,\gamma}^{(2,3)}(x,t)$ may be characterized as a continuous deformation of  a stable 7-phase solution (Figure~\ref{fig:10}F).

As a  comparison, $U_{\epsilon,\gamma}^{(1,2)}(x,t)$ and  $ U_{\epsilon,\gamma}^{(1,3)}(x,t)$  exhibit shorter term  irregular
behavior with all the dominant modes excited and they stabilize at $t \approx 15, 18$. respectively.
$U_{\epsilon,\gamma}^{2,3}(x,t)$ was observed to  take longer to stabilize due  to the higher order splitting in $\lambda_1^d$.

The exact nature of the instability associated with complex critical points in under investigation. They may be weaker than the exponential instabilities associated with complex double points but the evolution of
$U_{\epsilon,\gamma}^{(2,3)}(x,t)$
illustrates the their cumulative impact can be significant.

\section{Perturbation Analysis}
While examining the route to stability of the SPBs  under the damped HONLS
several novel results arose. One  feature  was that the  instabilities of
nonresonant modes persist longer than the instabilities of the resonant modes.
We are interested in the fate of complex double points under noneven perturbations  induced by HONLS as they characterize the SPB. 
Following the  perturbation analysis in \cite{ahs96} used to
determine the $\mathcal{ O}(\eps)$ splitting of double points for single mode
perturbations, we carry the analysis to higher order for noneven multi mode
perturbations of the SPBs. We find  i) additional modes resonate with the perturbation and ii) complex double points associated with nonresonant modes remain efffectively closed.

To obtain linearized initial conditions for the one and two mode SPBs
we use the Hirota formulation of the SPBs \cite{h76}. For example for the one mode SPB  one obtains,
\be\label{one-homoclinic}
u^{(j)}(x,t) = a\, e^{2\ri a^2t}\,\frac{1 + 2\, \e^{2\ri\theta_j+\Omega_j  t + \gamma}\cos\mu_j x+ A_{12}\,  \e^{2(2\ri\theta_j+\Omega_j  t + \gamma)}}
{1 + 2\, \e^{\Omega_j  t + \gamma}\cos\mu_j x+ A_{12}\,  \e^{2(\Omega_j  t + \gamma)}}
\ee
where $\mu_j = 2\pi j/L$, $\Omega_j = \mu_j\sqrt{4a^2 - \mu_j^2}$, $\sin\theta_j= \mu_j/2a$, $A_{12} = \sec^2\theta_j$, and $\gamma$ is an arbitrary phase.

The appropriate  linearized initial conditions for the one and two
mode SPBs, $u^{(i)}(x,0)$ and $u^{(i,j)}(x,0)$ respectively, are  obtained by
choosing $t$ and $\gamma$ such that
$\tilde\eps_s = 4\ri\sin\theta_s\,\e^{\Omega_s t + \gamma}$, $s=i,j$, are small.
After neglecting second-order terms we obtain:
\bea
u^{(j)}(x,0) &=& a\,\left(1 + \tilde\eps_j\, \e^{\ri\theta_j} \cos\mu_j x\right),\\
u^{(i,j)}(x,0) &=& a\,\left(1 + \tilde\eps_i\, \e^{\ri\theta_i} \cos\mu_i x 
+ \tilde\eps_j\, \e^{\ri\theta_j} \cos\mu_j x\right).
\eea

The damped HONLS
yields the following  noneven  first order approximation for small time,
\bea
u^{(j)}(x,h) &=& a\left[1+
\tilde\eps_j\left(\e^{\ri\theta_j} \cos\mu_j x + r_j \,\e^{\ri\phi_j}\sin\mu_j x\right)
\right]\\
u^{(i,j)}(x,h) &=& a
\left[1+   \tilde\eps_i\,\left(
\e^{\ri\theta_i} \cos\mu_i x + r_i \,\e^{\ri\phi_i}\sin\mu_i x\right)
+  \tilde\eps_j\,\left( \e^{\ri\theta_j} \cos\mu_j x + r_j\e^{\ri\phi_j} \sin\mu_j x\right)
\right]\label{uij2}
\eea
where $\theta_s \neq \phi_s$ and $a, \tilde\eps_s, r_s$ are functions of $h$ and the damped HONLS parameters $\eps$ and $\gamma$, for $s = i,j$.
For simplicity we set $\eps = \tilde\eps_s$ and suppress their explicit dependence on  $\eps,\gamma$:
\be\label{lic}
u = a +  \eps\left[
\e^{\ri\theta_i} \cos\mu_i x + r_i \,\e^{\ri\phi_i}\sin\mu_i x
+  Q\,\left( \e^{\ri\theta_j} \cos\mu_j x + r_j\e^{\ri\phi_j} \sin\mu_j x\right)
\right] = a + \eps u^{(1)}
\ee
where $r_s \neq 0$ and $Q$ can be 0 or 1, depending on whether a one or two       mode SPB is under consideration.

Since  $\Delta(\lambda,u)$ and the eigenfunctions $v_n = \left[\ba{l} v_{n1} \\ v_{n2}\ea\right]$  are analytic functions of their arguments, at the double points $\lambda_n$  we  assume the following expansions:
\bea
v_n &=& v_n^{(0)} + \eps v_n^{(1)} + \eps^2 v_n^{(2)} +\cdots\\
\lambda_n &=& \lambda_n^{(0)} + \eps\lambda_n^{(1)} + \eps^2 \lambda_n^{(2)}+\cdots
. \eea

Substituting these expansions into Equation~\rf{pertL} we obtain the following:
\bea
\mathcal{ O}(\eps^0):\quad \mathcal{ L} v_n^{(0)} &=& 0\label{eps0}\\
\mathcal{ O}(\eps^1):\quad \mathcal{ L} v_n^{(1)} &=&  \left[\ba{l}-\ri\lambda_n^{(1)} v_{n1}^{(0)} + u^{(1)} v_{n2}^{(0)}\\
-\ri\lambda_n^{(1)} v_{n2}^{(0)} + u^{(1)*} v_{n1}^{(0)}\ea\right]
\equiv F\label{eps1}\\
\mathcal{ O}(\eps^2):\quad \mathcal{ L} v_n^{(2)} &=& \left[\ba{l}-\ri \lambda_n^{(1)} v_{n1}^{(1)} -\ri \lambda_n^{(2)} v_{n1}^{(0)} + u^{(1)} v_{n2}^{(1)}\\
- \ri\lambda_n^{(1)} v_{n2}^{(1)} - \ri\lambda_n^{(2)} v_{n2}^{(0)} + u^{(1)*} v_{n1}^{(1)}\ea\right]
\equiv G.\label{eps2}
\eea
where
\be
\label{pertL}
\mathcal{ L} = \left[\ba{cc} \partial/\partial x + \ri\lambda_n^{(0)} & -a\\-a & -\partial/\partial x + \ri\lambda_n^{(0)}\ea\right].
\ee

The leading order Equation~\rf{eps0} provides the spectrum for the Stokes wave.
At the double points $\lambda_n^{(0)}$ 
 the  two dimensional eigenspace is spanned by the eigenfunctions
 \be\label{phin}
 \phi_n^\pm = \e^{\pm \ri k_n x}\left(\ba{c} 1\\\frac{\ri}{a}\left(\pm k_n + \lambda_n\right)\ea\right),
 \ee
 where $(\lambda_n^{(0)})^2 = k_n^2 - a^2, k_n = n\pi/L$,
 and the general solution is given by
\be\label{S0}
v_n^{(0)} = A^+ \phi_n^+ + A^- \phi_n^-
\ee

\subsubsection{First order results}
For periodic $v$, the solvability condition for the system
$
\mathcal{ L} v = F = \left[\ba{l} F_1\\F_2\ea\right]
$
is given by the orthogonality condition
\[\int_0^L \left( F_1 w_1^* + F_2 w_2^*\right) = 0\]
for all $w$ in the nullspace of the Hermitian operator
\be
\mathcal{ L}^H = \left[\ba{cc} -\partial/\partial x - \ri\lambda_n^* & -a\\-a & \partial/\partial x - \ri\lambda_n^*\ea\right].
\ee
At the double points the nullspace of $\mathcal{ L}^H$ is spanned by the eigenfunctions
$\left(\ba{l} \phi^\pm_{n2}\\\phi^\pm_{n1}\ea\right)^*$
and the orthogonality condition becomes
\be\label{orth-cond}
\int_0^L \left(F_1\phi^\pm_{n2} + F_2\phi^\pm_{n1}\right) dx = 0.
\ee
Applying this orthogonality condition to Equation~\rf{eps1} yields the system of equations
\be\label{sys_eqn}
\left(\ba{ll} T_+ & T\\T & T_-\ea\right)\left(\ba{rr} A^+\\A^-\ea\right) = 0,
\ee
where
\bea
T &=& 2\lambda_n^{(0)}\lambda_n^{(1)}/a\\
T_\pm &=& -\half\left\{\ba{ll}\left(\frac{\pm k_n + \lambda_n}{a}\right)^2
\left(\e^{\ri\theta_n} \pm \ri r_n\e^{\ri\phi_n}\right) -
\left(\e^{-\ri\theta_n} \pm \ri r_n\e^{-\ri\phi_n}\right) & \quad n = i, j\\
0 & \quad n\neq i,j\ea\right. .
\eea
Non trivial solutions $A^\pm$  are obtained only at the complex double points $\lambda_n$, $n=i,j$ at which the SPB was constructed providing the first order correction
\be
\left(\lambda_n^{(1)}\right)^2 = \left\{\ba{ll} \frac{a^2}{4\lambda_n^2}
\left[\sin(\omega_n+\theta_n)\sin(\omega_n-\theta_n)\right. &\\
  \qquad+r_n^2\sin(\omega_n+\phi_n)\sin(\omega_n-\phi_n)& \quad n = i, j \\
\left.  \qquad+\ri r_n\sin(\phi_n-\theta_n)\sin 2\omega_n\right]\\
0 & \quad n \neq i, j\ea\right.
\ee
where $\tan\omega_n = Im\left(\lambda_n^{(0)}\right)/k_n$ and $\theta_s \neq \phi_s \pm n\pi$ for $s = i,j$.
As a result  $\lambda_{n}^{(1,\pm)}= \pm r^{1/2} \e^{ip/2}$ where $0 < p < 2\pi$
and the double point splits asymmetrically in any direction. Examining $\Delta$ in a neighborhood of $u^{(0)}$  we find that when $u^{(1)}$ resonates with a particular mode, the band of continuous spectrum along the imaginary axis  splits
asymmetrically  into two disjoint bands in the upper half plane.  
The other double points do not experience an $\mathcal{ O}(\eps)$ correction.

The spectral configuration is determind by the location of 
$\lambda_n^{(\pm)} = \lambda^{(0)} + \eps \lambda^{(1,\pm)}$.
$\lambda_n^{+}$ determines the speed and direction of the associated phase.
For example, in the one complex double point regime there are only two spectral configurations associated with noneven perturbation: i) For $0 < p < \pi$, Re $\lambda^{+} >0$ and the upper band of spctrum lies in the first quadrant. The
wave form is characterized by a single modulated mode traveling to the right.
ii) For $\pi < p < 2\pi$, Re $\lambda^+ <0$, the upper band of spectrum is in the second quadrant, and the wave form is characterized by a single modulated mode traveling to the left. 

As seen in the numerical experiments, for noneven perturbations under the damped HONLS, the evolution of spectrum between two distinct configurations  occurs when the continuous spectrum forms transverse bands with a complex ctitical point (not double point)  and then splits.

\subsubsection{Second order results}
Determining the  $\mathcal{ O}(\epsilon^2)$
 corrections to the double points $\lambda_n$ for 
 $n\neq i,j$,  requires determining
 the eigenfunctions at $\mathcal{ O}(\epsilon)$.
 When  $\lambda_n^{(1)} =0$ 
the right hand side of Equation~\rf{eps1} simplifies to
\[\mathcal{ L} v_n^{(1)} = \mathcal{ F} = \left[\ba{rr} 0 & u^{(1)}\\u^{(1)*} & 0 \ea\right] v_n^{(0)} = 
\left[\ba{r} u^{(1)}\left(A^+ \phi_{n2}^+ + A^- \phi_{n2}^-\right)\\u^{(1)*}\left(A^+ \phi_{n1}^+ + A^- \phi_{n1}^-\right)\ea\right]\]
where $\phi_n^\pm$ is given by Equation~\rf{phin}. 
We assume  $v_n^{(1)} = v_n^{(0)} + \sum_n^{(1)}$ where
\be\label{v_p}
\ba{rcl} \sum_n^{(1)} &=& \mathbf A_i \e^{\ri(k_n+\mu_i)x} + \mathbf B_i \e^{\ri(k_n+\mu_i)x}+ \mathbf C_i \e^{-\ri(k_n-\mu_i)x}+ \mathbf D_i \e^{-\ri(k_n+\mu_i)x}\\
&& + \mathbf A_j \e^{\ri(k_n+\mu_j)x} + \mathbf B_j \e^{\ri(k_n-\mu_j)x} + \mathbf C_j \e^{-\ri(k_n-\mu_j)x} + \mathbf D_j \e^{-\ri(k_n+\mu_j)x}.
\ea
\ee

Substituting $\sum_n^{(1)}$ into Equation~\rf{eps1} we find the coefficient vectors to be (with $n \neq s$, $s = i,j$)
\[
\mathbf A_s = \frac{A^+/2}{\mu_s^2+2k_n\mu_s}
\left[\ba{c} \frac{1}{a}\left[(2(\cos\theta_s-\sin\phi_s)a^2
    + \left(\e^{\ri\theta_s}+i\e^{\ri\phi_s}\right)\left(k_n+\lambda_n\right)\mu_s\right]\\
  \ri\left[2(\cos\theta_s-\sin\phi_s)(k_n+\lambda_n)
    + \left(\e^{\ri\theta_s}-i\e^{\ri\phi_s}\right)\mu_s\right]\ea\right]\\
\]
\[
\mathbf B_s = \frac{A^+/2}{\mu_s^2-2k_n\mu_s}
\left[\ba{c} \frac{1}{a}\left[(2(\cos\theta_s-\sin\phi_s)a^2
    - \left(\e^{\ri\theta_s}+i\e^{\ri\phi_s}\right)\left(k_n+\lambda_n\right)\mu_s\right]\\
  \ri\left[2(\cos\theta_s-\sin\phi_s)(k_n+\lambda_n)
    - \left(\e^{\ri\theta_s}-i\e^{\ri\phi_s}\right)\mu_s\right]\ea\right]\\
\]
\[
\mathbf C_s = \frac{A^-/2}{\mu_s^2-2k_n\mu_s}
\left[\ba{c} \frac{1}{a}\left[(2(\cos\theta_s-\sin\phi_s)a^2
    + \left(\e^{\ri\theta_s}+i\e^{\ri\phi_s}\right)\left(-k_n+\lambda_n\right)\mu_s\right]\\
  \ri\left[2(\cos\theta_s-\sin\phi_s)(-k_n+\lambda_n)
    + \left(\e^{\ri\theta_s}-i\e^{\ri\phi_s}\right)\mu_s\right]\ea\right]\\
\]
\[
\mathbf D_s = \frac{A^-/2}{\mu_s^2+2k_n\mu_s}
\left[\ba{c} \frac{1}{a}\left[(2(\cos\theta_s-\sin\phi_s)a^2
    - \left(\e^{\ri\theta_s}+i\e^{\ri\phi_s}\right)\left(-k_n+\lambda_n\right)\mu_s\right]\\
  \ri\left[2(\cos\theta_s-\sin\phi_s)(-k_n+\lambda_n)
    - \left(\e^{\ri\theta_s}-i\e^{\ri\phi_s}\right)\mu_s\right]\ea\right]\\
\]

With $v_n^{(1)}$ in hand, applying the orthogonality condition to 
Equation~\rf{eps2}  yields the system
\be
\left[\ba{cc} \alpha_n^+ & \lambda_n^{(2)} - \beta_n\\ \lambda_n^{(2)} - \beta_n& \alpha_n^-\ea\right]\left[\ba{c} A^+\\A^-\ea\right] = 0
\ee
giving an $\mathcal{ O}(\eps^2)$ correction of the form
\be
\left(\lambda_n^{(2)} - \beta_n\right)^2 =
\left\{\ba{ll} \alpha_n^+\alpha_n^- & \quad n=2i, 2j, i+j, j-i\\
&\\
0 & \quad \mbox{for all other cases}\ea\right.
\ee

Consequently only the double points $\lambda_n^{(0)}$ with $n=2i, 2j, i+j$, or $j-i$ experience an $\mathcal{ O}(\eps^2)$ splitting.  All other double points experience  an $\mathcal{ O}(\eps^2)$  translation.
This calculation can be carried to higher order $\mathcal{ O}(\eps^m)$. In the simpler case of a damped  single mode SPB, $U_{\epsilon,\gamma}^{(j)}(x,t)$, only
$\lambda_n^{(0)}$ corresponding to the resonant mode $n=mj$ will split at order  $\mathcal{O}(\eps^m)$
whereas the splitting is beyond all orders for $\lambda_n^{(0)}$, $n \neq mj$ \cite{as94}.

For the two mode damped SPB $U_{\epsilon,\gamma}^{(2,3)}$ in the 3 UM regime we find
$\lambda_2^d$ and $\lambda_3^d$ will split at $\mathcal{ O}(\eps)$.
The mode associated
with $\lambda_1^{(0)}$ resonates also with  $u^{(1)}$ at $\mathcal{ O}(\eps^2)$.
All 3 complex double points split, in contrast with one mode
$U_{\epsilon,\gamma}^{(2)}$ where $\lambda_1^{(0)}$ and $\lambda_3^{(0)}$ do not split.

\section{Conclusions}

In this paper we investigated the route to stability for single and multi-mode
SPBs (which are even solutions of the NLS equation) in the framework of a damped HONLS equation using the Floquet spectral  theory of the NLS equation.
We found  novel instabilities emerging in the symmetry broken solution space
of the damped HONLS which are not captured by complex double points in the Floquet spectrum. 
To develop a broadened Floquet characterization of instabilities we  examined
the stability of an even 3-phase solution of the NLS equation with respect to noneven perturbations. We found the transverse complex critical point in its spectrum is  associated with an instability arising from symmetry breaking which is not excited when evenness is imposed.

The association of instabilities due to symmetry breaking with complex critical points of the Floquet spectrum was corroborated by the numerical experiments.
If one of the complex double points present at $t = 0$  splits in the damped HONLS system, the subsequent spectral evolution involves repeated formation and splitting of complex critical points (not double points) which we correlated  with the observed  instabilities. 

In the numerical study we presented experiments using fixed values of the perturbation parameters $\eps$ and $\gamma$. As these parameters are varied fewer or
more critical points may form and the time the damped HONLS solution stabilizes
may vary but the following interesting results are independent of their specific value: i) Instabilities due to symmetry breaking are associated  with complex critical points. ii)  Solutions stabilize once damping eliminates all the  complex critical points and complex double points in the spectral deomposition of the damped HONLS data. iii) Only certain modes
resonate with the damped HONLS perturbation. Resonant modes
aid in stabilizing the solution. If nonresonant modes are present, their
instabilities persist and appear to organize the dynamics on a longer timescale.

Each burst of growth in $\eta(t)$ can be correlated with the emergence of a complex critical point. The numerics suggest the instabilities associated with complex critical points may  be weaker than  those associated with complex double points. Even so the exact nature of the instability is warrants further investigation.
As demonstrated by the evolution of $U_{\epsilon,\gamma}^{(2,3)}(x,t)$
their  cumulative impact can be significant.

A perturbation analysis is presented to confirm the splitting of the initial complex double points 
observed in the numerical experiments. We find that certain complex double points present in the SPB initial data do not split under the damped HONLS flow.  
This  short time analysis prediction  holds for the duration of the experiments,
even though the solution evolves as  a damped multi-phase state.

\section*{Funding}
This work was partially supported by Simons Foundation, Grant  \#527565

\bibliographystyle{plain}
%\bibliography{schober2020}

\end{document}